\documentclass[a4paper,11pt]{article}
\pdfoutput=1

\usepackage{jheppub}
\usepackage[utf8]{inputenc}
\usepackage{mathtools}
\usepackage{enumitem}
\usepackage{amsmath}
\usepackage{amsthm} 
\usepackage{amssymb}
\usepackage{amscd}
\usepackage{amsthm}
\usepackage{color}
\usepackage{textcomp}
\usepackage{array} 
\usepackage{graphicx}
\usepackage{adjustbox}
\usepackage{mathrsfs}
\usepackage[samesize]{cancel}
\usepackage{ytableau}
\usepackage{booktabs}
\usepackage{tikz-cd}
\usepackage{relsize}
\usepackage{here}
\usepackage{bm}
\usepackage[capitalize]{cleveref}
\usepackage{tabularx}
\usepackage{upgreek}

\usepackage{tikz}
\usepackage{adjustbox}
\usetikzlibrary{positioning,fit}
\usetikzlibrary{calc,intersections,through,backgrounds}
\usetikzlibrary{calc,arrows,decorations.markings,shapes.arrows,patterns, decorations.pathmorphing}
\usetikzlibrary{positioning,matrix,arrows,decorations.pathmorphing,decorations.pathreplacing}
\usetikzlibrary{arrows,matrix,positioning,shapes,arrows}
\usetikzlibrary{shapes.geometric}
\usetikzlibrary{decorations.text}
\usetikzlibrary{arrows.meta}
\usetikzlibrary{math}
\tikzset{
	->-/.style={decoration={markings, mark=at position 0.5 with {\arrow{to}}},
		postaction={decorate}},}
\tikzset{
	-<-/.style={decoration={markings, mark=at position 0.5 with {\arrow{to reversed}}},
		postaction={decorate}},}

\tikzset{
	pics/torus/.style n args={3}{
		code = {
			\providecolor{pgffillcolor}{rgb}{1,1,1}
			\begin{scope}[
				yscale=cos(#3),
				outer torus/.style = {draw,line width/.expanded={\the\dimexpr2\pgflinewidth+#2*2},line join=round},
				inner torus/.style = {draw=pgffillcolor,line width={#2*2}}
				]
				\draw[outer torus] circle(#1);\draw[inner torus] circle(#1);
				\draw[outer torus] (180:#1) arc (180:360:#1);\draw[inner torus,line cap=round] (180:#1) arc (180:360:#1);
			\end{scope}
		}
	}
}

\newcommand{\tikznode}[2]{\relax
	\ifmmode%
	\tikz[remember picture,baseline=(#1.base),inner sep=0pt] \node (#1) {$#2$};
	\else
	\tikz[remember picture,baseline=(#1.base),inner sep=0pt] \node (#1) {#2};%
	\fi
}

\newcommand{\dd}{{\rm d}}
\newcommand{\no}{\nonumber}

\newcommand{\cC}{\mathcal C}

\newcommand{\cO}{\mathcal O}\newcommand{\cP}{\mathcal P}

\newcommand{\sfz}{\mathsf z}

\newcommand{\la}{\lambda}
\newcommand{\Tr}{{\rm Tr}}

\newcommand{\Pexp}{\text{P}{\text{exp}}}

\newcommand{\ra}{\rightarrow}
\newcommand{\tn}[1]{\underline{\bm{#1}}}

\def\tRG{\mathfrak{t}_{\rm RG}}
\usepackage{slashed}
\hypersetup{
  pdftitle={
    Time-Dependent Integrability from Gauge Theory, I
  },
  pdfauthor={
    Shota Komatsu, Jun-ichi Sakamoto,
    Anders Wallberg, Masahito Yamazaki
  },
  pdfsubject={
    A unifying gauge-theoretic framework for
    time-dependent integrability and exactly
    solvable non-equilibrium dynamics.
    Generalizing four-dimensional Chern-Simons
    theory to spacetime-dependent one-forms,
    this work establishes a systematic construction
    of classically integrable field theories
    with explicitly time-dependent couplings.
    A central and surprising result is the
    universal emergence of one-loop
    renormalization-group flow as the consistency
    condition governing time dependence,
    revealing a deep connection between
    classical integrability and quantum
    renormalization.
    Despite their explicit time dependence,
    these theories retain Lax integrability
    and remain amenable to inverse scattering.
    The framework encompasses both ultralocal
    and non-ultralocal integrable models,
    including fermionic theories, sigma models,
    and their integrable deformations.
    It further establishes a connection to
    two-dimensional dilaton gravity, generating
    broad families of integrable gravity-matter
    systems and linking gauge theory,
    non-autonomous integrable systems,
    isomonodromic structures, and
    non-equilibrium physics.
    These results provide new theoretical tools
    for studying time-dependent dynamics
    and potential benchmarks for quantum
    simulation.
  },
  pdfkeywords={
    time-dependent integrability,
    time-dependent integrable field theories,
    non-autonomous integrable systems,
    four-dimensional Chern-Simons theory,
    4d Chern-Simons gauge theory,
    classical integrability,
    integrable non-equilibrium dynamics,
    time-dependent coupling constants,
    renormalization group flow,
    one-loop renormalization group,
    one-loop beta functions,
    integrability and renormalization,
    gauge theory and integrability,
    Lax pairs and Lax connections,
    inverse scattering method,
    isomonodromic deformations,
    variable spectral parameter,
    spacetime-dependent integrable systems,
    ultralocal integrable field theories,
    non-ultralocal integrable sigma models,
    integrable order and disorder defects,
    principal chiral model,
    Gross-Neveu model,
    Faddeev-Reshetikhin model,
    lambda-deformed sigma models,
    eta-deformed sigma models,
    two-dimensional dilaton gravity,
    integrable gravity-matter systems,
    dimensionally reduced Einstein gravity,
    exactly solvable dynamical systems,
    non-equilibrium theoretical physics,
    quantum simulation benchmarks
  }
}

\title{Time-Dependent Integrability from Gauge Theory, I}
\author[a]{Shota Komatsu,}
\author[b]{Jun-ichi Sakamoto,} 
\author[a,c]{Anders Wallberg,} 
\author[d,e,f]{Masahito Yamazaki}
\affiliation[a]{Department of Theoretical Physics, CERN, 1211 Meyrin, Switzerland}
\affiliation[b]{Department of Physics, The University of Osaka, Machikaneyama-Cho 1-1, Toyonaka 560-0043, Japan}
\affiliation[c]{Laboratory for Theoretical Fundamental Physics, EPFL
	1015 Lausanne, Switzerland}
\affiliation[d]{Department of Physics, Graduate School of Science, University of Tokyo,
	Hongo 7-3-1, Bunkyo-ku, Tokyo 113-0033, Japan}
\affiliation[e]{Kavli Institute for the Physics and Mathematics of the Universe,
	UTIAS, University of Tokyo, 5-1-5 Kashiwanoha, Chiba 277-8583, Japan}
\affiliation[f]{Trans-Scale Quantum Science Institute, University of Tokyo,
	Hongo 7-3-1, Bunkyo-ku, Tokyo 113-0033, Japan}
\abstract{
    Solvable time-dependent systems provide important settings for studying non-equilibrium physics, where exact results are rare. They are also useful for benchmarking quantum simulations, which can directly probe real-time dynamics beyond the reach of conventional numerical approaches. In this paper, we show that the four-dimensional Chern–Simons theory offers a natural and unifying framework for constructing such systems. 
	Focusing on classically integrable field theories, we consider a generalization of the four-dimensional Chern–Simons theory in which the usual holomorphic one-form is replaced by a more general, spacetime-dependent one-form.
    This yields a systematic procedure for generating time-dependent integrable field theories and establishes a universal relation: for every theory obtained in this way, the allowed time dependence coincides with the one-loop renormalization group flow.
     Despite the explicit time dependence, these theories retain Lax integrability and remain solvable by inverse scattering methods.
	Our construction applies to both ultralocal and non-ultralocal theories and extends previously known time-dependent sigma models to a much broader class of integrable systems. It also admits rewriting as dilaton gravity coupled to matter, producing a large family of classically integrable dilaton gravity theories in two dimensions. We also comment on connections to time-dependent integrable models studied recently in condensed matter physics and non-autonomous integrable systems arising from dimensionally-reduced Einstein gravity.
}
\begin{document}
\begin{flushright}
\texttt{OU-HET-1317}\\
\texttt{CERN-TH-2026-151}
\end{flushright}
\vspace*{0.5cm}
\maketitle

\section{Introduction}

	Understanding the dynamics of non-equilibrium, time-dependent, or externally driven systems is an important problem across many areas of physics, including condensed matter, statistical mechanics, high-energy physics, and cosmology. Such systems exhibit a wealth of novel phenomena, such as dynamical phase transitions, emergent scaling behavior, and intrinsically non-perturbative effects that have no counterpart in time-independent settings. 
	Yet, despite their importance, analytic tools for studying time-dependent dynamics remain limited.
	
	One strategy for making progress would be to focus on solvable models, often referred to as integrable models. Although such models are special, they have repeatedly served as laboratories for discovering and testing fundamental ideas. A classic example is the two-dimensional Ising model, whose exact solution by Onsager \cite{Onsager:1944zz} provided the first concrete demonstration of a phase transition in a microscopic model, helping to establish the modern theory of critical phenomena. More recently, integrable stochastic models such as the asymmetric simple exclusion process (ASEP) \cite{Derrida:1993,Derrida:1998,Corwin:2011,Mallick:2015} have played a key role in uncovering universal features of non-equilibrium steady states. In particular, they provided deep insight into the Kardar–Parisi–Zhang universality class in 1+1 dimensions \cite{Kardar:1986xt,Sasamoto:2010ac}, enabling exact computations of fluctuation statistics that were later confirmed experimentally in turbulent liquid crystal \cite{Takeuchi:2010zz}.
	
	These examples suggest that identifying solvable models of time-dependent systems could significantly advance our understanding of non-equilibrium dynamics. In addition, their importance has grown significantly with the rapid development of quantum simulation platforms, which can probe real-time dynamics directly. In this context, analytically tractable models play a crucial role: they provide reliable benchmarks for quantum simulators \cite{Maruyoshi:2022jnr} in regimes where conventional numerical methods, such as Monte Carlo simulations, are not applicable. While several such models have been constructed in specific contexts \cite{Belinsky:1971nt,Belinsky:1979gra,Hoare:2020fye,Cesaro:2025msv,Fioretto:2012caux,Sinitsyn:2017bmj,Yuzbashyan:2018,Gritsev:2017fyt,Gamayun:2020,Pasnoori:2025kondo,Pasnoori:2025wavefunction,Pasnoori:2025qkz,Pasnoori:2025rgflow,Pasnoori:2026tdgn,Chernyak:2020parallel,Chernyak:2021quadratic,Barik:2025rg}, there is currently no general framework or even a universally accepted definition of time-dependent integrability. A central obstacle is conceptual: integrability is traditionally associated with the existence of infinitely many conserved charges, but in time-dependent systems, even energy need not be conserved, making it unclear how to generalize this notion.
	
	This paper aims to propose a unifying framework for constructing time-dependent integrable systems in a systematic way. Our starting point is the four-dimensional Chern–Simons theory (4d CS) \cite{Costello:2013zra,Costello:2017dso,Costello:2018gyb,Costello:2019tri},\footnote{Precursors of this modern perspective already appeared in earlier works on higher-dimensional theories with holomorphic structures; see \cite{Nekrasov:1996fht,Gorsky:1994cx,Nair:1990bt,Losev:1995cr}.} which has recently emerged as a powerful organizing principle for integrable models. In its standard form, this theory naturally gives rise to a large class of integrable field theories and lattice systems. We show that it can also be adapted to describe time-dependent systems. In this paper, we focus on classical integrable field theories, but we expect that our construction can be generalized to other integrable systems.

    \paragraph{Key idea and results.}
	The key idea is simple. In the usual formulation, the theory depends on a holomorphic one-form defined on an auxiliary space. We generalize this structure by allowing the one-form to depend explicitly on time. While this may appear to be a modest modification, we find that, under natural consistency requirements, there is essentially a unique way to introduce time dependence within this framework. This rigidity indicates that the resulting systems are not arbitrary time-dependent deformations, but a distinguished class selected by the 4d CS structure. Importantly, despite the explicit time dependence, these systems preserve a key hallmark of integrability: their equations of motion can still be expressed in terms of a Lax connection, encoding the dynamics as a flatness condition. As a result, standard tools of integrable systems, such as the inverse scattering method, remain applicable, providing a controlled way to study time-dependent dynamics. Moreover, the flat connection also implies an {\it isomonodromic} structure: although explicit time dependence destroys standard conserved charges, certain {\it monodromy data} in the spectral-parameter plane remain invariant under time evolution. In this sense, isomonodromy replaces the usual {\it isospectrality} of time-independent integrable systems.\footnote{The connection with isomonodromy will be addressed in a sequel to this paper \cite{toappear}.}
    
    The structure becomes even richer when one allows the couplings to depend on both {\it space and time}. In that setting, we find a large family of admissible deformations, all preserving a Lax connection. The purely time-dependent theories arise as a specialized corner of this broader space. Precisely there, the rigidity of the four-dimensional construction becomes most pronounced.
	
	A particularly striking consequence is that the allowed time dependence is governed by the one-loop renormalization group (RG) equation
	\begin{align}
		\partial_{t}\, \hat{h}_i (t) \propto \beta_i \left(\hat{h}_j(t)\right)\,,
	\end{align}
	where $\hat{h}_i$'s are coupling constants and $\beta_i$'s are the corresponding one-loop beta functions. It is worth stressing that this relation is not imposed by hand; it instead follows from the 4d CS construction itself, and applies universally to any integrable field theories obtained in this way. Since our analysis is entirely classical, the appearance of the one-loop RG flow is at first sight surprising. The 4d CS theory also sheds new light on this feature: the same structural data that control the one-loop RG flow in the 4d CS description \cite{Delduc:2020vxy,levine_universal_2023,Levine:2023wvt,Lacroix:2024wrd,Lacroix:2025ias} reappear here as the consistency conditions governing time dependence.
	
	We also show that these theories admit an equivalent formulation as two-dimensional dilaton gravity coupled to integrable matter fields. Upon gauge fixing, the gravitational description reduces precisely to the time-dependent integrable field theories described above. In this way, our framework generates a broad family of classically integrable dilaton gravity theories, extending earlier examples obtained from dimensional reductions of higher-dimensional gravity \cite{Belinsky:1971nt,Belinsky:1979gra}.
	
	Our construction applies quite broadly. It encompasses both ultralocal theories, such as the Gross-Neveu model and its generalizations, and non-ultralocal theories, such as integrable sigma models and their generalizations. It therefore extends known time-dependent sigma models to a much wider class of integrable systems. In this sense, the 4d CS theory provides a natural framework in which time dependence and integrability coexist coherently and tractably.
	
	\paragraph{Relation to earlier works.} Time-dependent integrable systems are remarkable in part because they bring into contact a wide range of ideas: from diverse areas of theoretical physics to closely related structures in mathematics.
	
	Perhaps the earliest example of integrable structures with explicit spacetime dependence arose in the study of dimensional reductions of higher-dimensional gravitational systems. It was observed in \cite{Belinsky:1971nt,Belinsky:1979gra} that dimensional reduction of Einstein gravity gives rise to special coset sigma models coupled to two-dimensional dilaton gravity. These models admit a Lax connection and can be solved by inverse-scattering methods. They are often called {\it non-autonomous} integrable systems since, unlike standard integrable systems, the Lax connection contains explicit spacetime dependence. Their structure, as well as various generalizations, has since been explored \cite{BZM:1987,Breitenlohner:1986um,Hoare:2020fye,Penna:2020uky}. More recently, Cole and Weck \cite{Cole:2024skp}  beautifully explained how this system can be obtained from the 4d CS theory. (See also \cite{Cesaro:2025msv,Ashwinkumar:2026dwd,Bielli:2026vit} for related discussions of integrable deformations of this setup.) As in our construction, time dependence is introduced by modifying the meromorphic one-form. The main point of the present paper is to show that this mechanism is not special to systems that arise from dimensional reduction: it applies much more generally, including to ultralocal theories, which typically do not arise from dimensional reduction of gravitational systems.
	
	A connection between time dependence and the one-loop RG for a class of integrable sigma models was noticed in the pioneering work by Hoare, Levine, and Tseytlin \cite{Hoare:2020fye}. There, the analysis was carried out directly in two dimensions, and the requirement of a flat Lax connection was shown to select a particular time dependence, precisely matching the one-loop RG flow. Our approach extends this observation to a broader class of theories, again including ultralocal examples, by reformulating the problem in the 4d CS theory. As mentioned, in this formulation, the emergence of the one-loop RG flow is not a coincidence, but follows from universal structures of the 4d CS theory.
	
	At the same time, our analysis reveals an important distinction between purely time-dependent systems and more general {\it spacetime}-dependent deformations. The connection to dilaton gravity and to one-loop RG flow persists only in the former case. Once dependence on both space and time is allowed, the space of admissible integrable deformations becomes vastly larger. Most of these deformations cannot be directly interpreted in terms of RG or coupling to dilaton gravity. This broader structure does not appear to have been noticed previously, largely because it becomes visible only when one goes beyond the relatively special classes of models studied so far in the literature.
	
	A parallel line of development has appeared in condensed matter physics, where solvable quantum-mechanical systems with explicit time dependence have been found and studied; see, for example, \cite{Fioretto:2012caux,Sinitsyn:2017bmj,Yuzbashyan:2018,Gritsev:2017fyt,Gamayun:2020,Pasnoori:2025kondo,Pasnoori:2025wavefunction,Pasnoori:2025qkz,Pasnoori:2025rgflow,Pasnoori:2026tdgn,Chernyak:2020parallel,Chernyak:2021quadratic,Barik:2025rg}. The works closest in spirit to ours are \cite{Pasnoori:2025kondo,Pasnoori:2025wavefunction,Pasnoori:2025qkz,Pasnoori:2025rgflow,Pasnoori:2026tdgn}, which analysed the Kondo model and the UV-regulated chiral Gross-Neveu model, treating the latter effectively as a quantum-mechanical system, and uncovered a relation among solvability, time dependence and RG. Since these examples are primarily quantum-mechanical rather than field-theoretic, a detailed comparison lies beyond the scope of this paper. Nevertheless, it is natural to expect that the 4d CS theory can also generate and generalise such models; we postpone the analysis to future works.
	
	Finally, the non-autonomous integrable systems studied in this paper are closely tied to some of the important concepts in mathematics, including isomonodromic deformations and Painlevé equations \cite{Ueno:1981,Korotkin:1994hy,Korotkin:1995hf,Korotkin:1996xk,Korotkin:1996cz,Nicolai:1996he,Korotkin:1996cm,Korotkin:1998aa}. We expect the 4d CS theory to offer a natural and streamlined framework for these ideas. In particular, it should provide systematic generalisations, clarify their relations to other related concepts, and perhaps open a path toward quantum versions of these classical structures. This also raises a broader question: how do the familiar mathematical structures of integrable systems get modified once explicit time dependence is introduced? Examples include Yangian symmetry, the relation to affine Gaudin models, and the ODE/IM correspondence. We return to these questions in the conclusion.
	\paragraph{Structure of the paper.} The rest of this paper is structured as follows. In \textbf{\cref{sec:classicint}}, we review the basic concepts of classical integrable field theory: the Lax formulation, the classical $r$-matrix, and the distinction between ultralocal and non-ultralocal theories. These ideas are illustrated by two canonical examples, the Gross-Neveu model as an ultralocal theory and the principal chiral model as a non-ultralocal one.
	
	In \textbf{\cref{sec:timedep}}, we turn to time-dependent integrability from the viewpoint of the Lax formalism. We first introduce a general framework for describing spacetime-dependent integrable field theories, and then apply it to the two examples introduced earlier. This leads to a time-dependent principal chiral model, previously constructed in \cite{Belinsky:1971nt,Belinsky:1979gra}, and to a time-dependent Gross-Neveu model, which appears to be new. In both cases, we emphasize the relation between the allowed time dependence and the one-loop RG flow.
	In \textbf{\cref{sec:classicint4dCS}}, we review the construction of integrable field theories from the 4d CS theory, preparing for the time-dependent generalization. We also recall the role of order and disorder defects: order defects give rise to ultralocal theories, while disorder defects lead to non-ultralocal ones.
	
	The main results are presented in \textbf{\cref{sec:timedep4dCS}}. We give a systematic construction of time-dependent integrable field theories from the 4d CS theory. After reviewing how the one-loop RG flow arises in this framework, we show that the same structures reappear when the holomorphic one-form is replaced by a time-dependent one. This provides a general mechanism for time-dependent integrability and, at the same time, establishes a universal relation between time-dependent integrability and the one-loop RG flow.
	We then work out several examples in \textbf{\cref{sec:examples}}, reproducing and generalising known time-dependent integrable field theories. 
	
	In \textbf{\cref{sec:BMtype}}, we compare the time-dependent Lax connections obtained from the 4d CS theory with another formulation of non-autonomous integrability, widely used in dimensionally-reduced gravitational systems, where the spectral parameter itself depends on spacetime. We explain the map between the two descriptions and illustrate it explicitly in examples. In \textbf{\cref{sec:dilaton}}, we show that the systems discussed in this paper admit an equivalent formulation as two-dimensional dilaton gravity coupled to integrable matter fields.
	Finally, in \textbf{\cref{sec:conclusion}}, we summarize the results and discuss future directions, including brief comments on time-dependent quantum integrable systems studied in condensed matter physics. 
    
    In the appendices, we analyse the one-loop RG flows of general integrable field theories that arise from multiple chiral and antichiral order defects in the 4d CS theory. These analyses appear to be new, and the appendices may therefore be read as independent, self-contained results. The rational and trigonometric cases are
    treated in \textbf{Appendix \ref{app:ClRG}}, while the elliptic case is
    discussed both in \textbf{Appendix \ref{app:ClRG}} and \textbf{Appendix \ref{app:elliptic}}. For the rational cases, we prove in general that the positions of the order
    defects evolve linearly along the RG flow. Based on several examples, we conjecture that the same behaviour persists in the trigonometric and elliptic cases. 
	
	\section{Standard integrability: review}\label{sec:classicint}
	Before discussing time-dependent integrability, let us begin by reviewing the notion of standard (i.e.\ time-independent) integrability in two-dimensional classical field theories. 
	
	\subsection{Generalities} 
	For systems with finitely many degrees of freedom, such as classical mechanics of a particle moving in finite-dimensional space, there exists a mathematically precise definition of integrability, called {\it Liouville integrability} \cite{Liouville1855}. A system is Liouville-integrable if it possesses as many conserved charges as the degrees of freedom, which in turn implies the solvability of its dynamics.
	In field theories, however, the infinite number of degrees of freedom makes it difficult to define a universal notion of integrability. 
    
    In this paper, we adopt a practical perspective based on {\it Lax integrability},\footnote{See e.g.\ \cite{Babelon:2003qt} for pedagogical exposition on Lax integrability.} a well-established framework extensively used in the physics literature. As we see below, this approach naturally leads to two characteristic features of integrable systems: the existence of an infinite set of conserved charges and solvability via the so-called inverse scattering method.
	
	Consider a two-dimensional field theory on a surface $\Sigma$ containing fields $\phi^i:\Sigma\ra \mathcal{T}$ for some target space $\mathcal{T}$, satisfying a set of equations of motion.
	We say that the model is \textit{Lax integrable} if the equations of motion can be expressed alternatively as a one-parameter family of flat connections. Concretely, in Lax integrable systems,
	there exists a connection, called the \textit{Lax connection} $L_\mu:\Sigma\times C\ra \mathfrak{g}$ with $\mu=1,2$, where $C$ is a Riemann surface\footnote{Typically taken to be either the Riemann sphere $\mathbb{CP}^1$, the complex cylinder $\mathbb{C}^\times$ or the complex torus $\mathbb{T}$.} and $\mathfrak{g}$ is a Lie algebra, which satisfies the \textit{Lax equation}
	\begin{equation}
		\forall z\in C:\qquad \partial_\mu L_\nu(z)-\partial_\nu L_\mu(z)+[L_\mu(z),L_\nu(z)]=0\,.\label{eq:LaxEq}
	\end{equation}
	Here $z \in C$ parametrizes the family of connections and is called the \textit{spectral parameter}. Importantly, imposing the Lax equation for any $z$ is equivalent to imposing the equations of motion, and $L_\mu$ contains all the dynamical information contained in the fields $\phi^i$. As alluded to above, the Lax equation is equivalent to the flatness condition of the connection, built from covariant derivatives $D_\mu=\partial_\mu+L_\mu$,
	\begin{equation}
		[D_\mu(z),D_\nu(z)]=0\,.\label{eq:LaxFlat_independent}
	\end{equation}
	The flatness condition guarantees the existence of a simultaneous eigenvector $\Phi$ of $D_{\mu}$'s,
	\begin{equation}\label{eq:ALP}
		D_\mu(z)\,\Phi(z)=0\,\qquad \qquad \mu=1,2\,.
	\end{equation}
	The equation \eqref{eq:ALP} is called the {\it auxiliary linear problem}, and plays an important role in classical integrable field theories, allowing one to construct solutions to the equations of motion, which are generally a system of non-linear PDEs, from the analysis of the linear problem \eqref{eq:ALP}. This method is called the {\it inverse scattering method} and is summarized, e.g. in \cite{Babelon:2003qt}, although we will not study this aspect much in this paper. 
	To see why the Lax equation is a natural definition of integrability, consider the \textit{monodromy matrix} $M(z)$, given by\footnote{Here we assumed that the fields $\phi^i$ satisfy appropriate boundary conditions as $x\ra\pm \infty$.}
	\begin{equation}
		M(z)=\Pexp\left(-\int_{-\infty}^{\infty}\dd x\,L_x(x,z)\right)\,,\label{eq:monodromy}
	\end{equation}
	where $\Pexp$ is the path-ordered exponential. From the monodromy matrix, one can construct the \textit{transfer matrix} 
    \begin{align}
    t(z)=\Tr \,M(z)\,,
    \end{align} where the trace is taken in some representation of $\mathfrak{g}$. Note that $t(z)$ is essentially a holonomy of the connection $D_{\mu}$. Due to the flatness condition \eqref{eq:LaxFlat_independent}, $t(z)$ is conserved for all values of $z$. Thus, by expanding it in a power series
	\begin{equation}
		t(z)=\sum_{n=0}^\infty Q_{n}\,z^n\,,
	\end{equation}
	we obtain, as the coefficients, an infinite set of conserved charges $\{Q_n\}_{n=0}^\infty$, a hallmark of integrability. 
	
	One can also verify that the infinite set of charges $Q_n$ is Poisson-commuting. This is ensured if the Lax connection satisfies the following Poisson bracket, often called \textit{Maillet bracket} in the literature;
	\begin{align}
		\Big\{L_{x}(x_1,z_1)_{\tn{1}},L_{x}(x_2,z_2)_{\tn{2}}\Big\}=&\Big([r(z_1,z_2)_{\tn{12}},L_{x}(x_1,z_1)_{\tn{1}}]-[r(z_2,z_1)_{\tn{21}},L_{x}(x_2,z_2)_{\tn{2}}]\Big)\delta_{x_1x_2}\nonumber
		\\
		&-\Big(r(z_1,z_2)_{\tn{12}}+r(z_2,z_1)_{\tn{21}}\Big)\delta'_{x_1x_2}\,.\label{eq:Maillet}
	\end{align}
	Let us unpack the notation. The Poisson bracket is naturally valued in $\mathfrak{g}\otimes\mathfrak{g}$. For $X\in\mathfrak{g}$, $X_{\tn{i}}$ denotes that $X$ is taken to live in the $i$'th copy of $\mathfrak{g}$, \textit{e.g.} $X_{\tn{1}}=X\otimes\mathbb{I}$; $\delta_{x_1x_2}$ is the Dirac $\delta$-function which vanishes whenever $x_1\neq x_2$, while $\delta'$ is its derivative. The object $r:C\otimes C\ra\mathfrak{g}\otimes\mathfrak{g}$ is called the \textit{classical $r$-matrix}. In order for the Maillet bracket to satisfy the Jacobi-identity, $r$ needs to satisfy the classical Yang-Baxter equation (\textit{cYBE})
	\begin{equation}
		\left[r(z_1,z_2)_{\tn{12}},r(z_1,z_3)_{\tn{13}}\right]+\left[r(z_1,z_2)_{\tn{12}},r(z_2,z_3)_{\tn{23}}\right]+\left[r(z_3,z_2)_{\tn{32}},r(z_1,z_3)_{\tn{13}}\right]=0\,. \label{eq:cybe}
	\end{equation}
	The cYBE is a classical analog of the famous Yang-Baxter equation governing quantum integrable systems, and the classical $r$-matrix provides the leading semi-classical approximation to the quantum $R$-matrix (see e.g.~\cite{Korepin:1993kvr} for an introduction to quantum integrability and the Yang-Baxter equation),
	\begin{align}
		R(z_1,z_2) \propto \mathbb{I}\otimes \mathbb{I} +\hbar\,r(z_1,z_2)+\cdots\,.
	\end{align}
	
	Note that, if $r$ is skew-symmetric, i.e.\ $r(z_1,z_2)_{\tn{12}}=-r(z_2,z_1)_{\tn{21}}$, the Maillet bracket \eqref{eq:Maillet} simplifies since the $\delta'$ term vanishes. A system with such a simplified bracket structure is called \textit{ultralocal}, while a general system with a non-skew-symmetric $r$-matrix is called {\it non-ultralocal}. As we see later, the time-dependent generalisations of these two classes are slightly different, and they have different realizations in the four-dimensional Chern-Simons theory.
	
	As shown in \cite{Belavin:1982rtj} (see also \cite{Costello:2017dso}), under some technical assumptions, ultralocal solutions of the cYBE have been classified into three classes: rational, trigonometric, and elliptic. These correspond  to taking the Riemann surface $C$, on which $z$ lives, to be the Riemann sphere $\mathbb{CP}^1=\mathbb{C}\cup\{\infty\}$, the complex cylinder $\mathbb{C}^*=\mathbb{C}\setminus\{0\}$, or the complex torus $\mathbb{T}=\mathbb{C}/(\mathbb{Z}+\tau\mathbb{Z})$, respectively. Given such an ultralocal solution $r_0(z_1,z_2)$ as a seed solution, one can find another solution
	\begin{align}\label{eq:twistfunction}
		r(z_1,z_2)=r_0(z_1,z_2)\varphi^{-1}(z_2)\,,
	\end{align}
	for any scalar function $\varphi$,
	thus generating many non-ultralocal solutions from a given ultralocal one. The function $\varphi(z)$ is called the {\it twist function} and plays an important role in the 4d CS construction. 
	
	Below, we will see examples of both of these types of solutions.
	
	\subsection{Examples}
	For illustration, let us discuss two examples, the \textit{Gross-Neveu Model} (GN) and the \textit{Principal Chiral Model} (PCM), well-known both in condensed matter and high-energy physics. The former being ultralocal and the latter being non-ultralocal, these examples allow us to highlight key differences between different classes of theories.
	\subsubsection*{Lie algebra conventions}
	Before presenting examples, here we summarize conventions and notations for the Lie algebra used throughout this paper.
	
	Let $\mathfrak{g}$ be a semi-simple Lie algebra. Let  $\{T_a\}\,(a=1,\dots, \dim\mathfrak{g})$ be a set of corresponding generators, which should satisfy the commutation relations 
	\begin{align}
		[T_a,T_b]=f_{ab}{}^{c}T_c\,,		\label{eq:lie-structure}
	\end{align}
	where $f_{ab}{}^{c}$ is the structure constant.
	The canonical choice of inner product on $\mathfrak{g}$ is the invariant bilinear form, which we denote $\langle A, B\rangle$ ($A, B\in \mathfrak{g}$). In terms of a trace in the defining representation ${\rm Tr} (T_a T_b)=\kappa_{ab}$, it is given by\footnote{The sign of (\ref{eq:bilineark}) is chosen such that $\left<\cdot,\cdot\right>$ is positive definite for a compact Lie algebra $\mathfrak{g}$.}
	\begin{equation}
		\left<T_a,T_b\right>=-\kappa_{ab}\,.
		\label{eq:bilineark}
	\end{equation}
	We use raised indices $\kappa^{ab}$ to denote the inverse of $\kappa_{ab}$, from where we can define the dual generators $T^a=\kappa^{ab}T_b$.
	
	The structure constants $f_{ab}{}^{c}$ define an alternative inner product by the trace in the adjoint representation, $f_{ac}{}^{d}f_{bd}{}^c$. The two inner products are proportional, and the constant of proportionality is called the dual Coxeter number $c_G$
	\begin{equation}\label{eq:cGdef}
		c_{G}\, \kappa_{ab}=f_{ac}{}^{d}f_{bd}{}^c
		\,.
	\end{equation}
	\subsubsection*{Gross-Neveu model}
	Consider a 2d field theory consisting of $N$ interacting massless Majorana fermions $\psi_n$, $n=1, \dots, N$. The action is given by  
    \begin{equation}
        \text{S}_{\text{GN}}[\psi^n]=\int_\Sigma\dd^2x\,i\bar{\psi}^n\slashed{\partial}\psi^n+\frac{h}{2}(\bar{\psi}^n\psi^n)^2\,.\label{eq:SGN0}
    \end{equation}
    This action has an $O(N)$ flavour symmetry, acting by $\psi^n\mapsto O_{nm}\psi^m$. The theory further has a chiral symmetry, which makes it natural to decompose the fermion into its chiral and antichiral components $\psi_+^n$ and $\psi_-^n$. Denoting by $T^a$, $a=1,\dots N(N-1)/2$, the generators of the Lie algebra $\mathfrak{so}(N)$, corresponding to the $O(N)$ symmetry in the vector representation,  the action can be rewritten as
	\begin{equation}
		\text{S}_{\text{GN}}[\psi_\pm^n]=\int_\Sigma\dd ^2x\,i\psi^n_+\partial_-\psi^n_++i\psi^n_-\partial_+\psi^n_-+h\left<J_+,J_-\right>\,,\label{eq:SGN} \quad J^a_{\pm}=-i\psi^n_\pm T^a_{nm}\psi^m_\pm\in\mathfrak{so}(N)
	\end{equation}
	where $x^{\pm}=t\pm x$ are light-cone coordinates and $h\in\mathbb{R}$ is a coupling constant.
	The equations of motion are
	\begin{equation}	\partial_\pm\psi_{\mp}^n+h\,J_{\pm,a}\,T^a_{nm}\,\psi^m_\mp=0\,.\label{eq:GNEom}
	\end{equation}
	To see that the model is indeed integrable, we need to express these equations as the flatness of a Lax connection. It is straightforward to check that 
	\begin{equation}
		L_\pm(z)=\mp \frac{J_\pm}{z-z^\pm}:\Sigma\times C\rightarrow\mathfrak{so}(N)\label{eq:LGN}
	\end{equation}
	is indeed flat upon imposing the equations of motion, provided that $z^\pm$ satisfy
	\begin{equation}
		h=\frac{1}{z^+-z^-}\,.\label{eq:hfromzGN}
	\end{equation}
	It also satisfies the Maillet bracket \eqref{eq:Maillet}, with the rational  classical $r$-matrix $r(z_1,z_2)=r_Y(z_1-z_2)$ given by \cite{Yang:1967bm,Belavin:1982rtj} 
	\begin{equation}
		r_{\text{Y}}(z)=\frac{\kappa_{ab}\,T^a\otimes T^{b}}{z}\,,\label{eq:Yang}
	\end{equation}
	where $\kappa_{ab}$ is the bilinear form \eqref{eq:bilineark}.
	Since both the Lax connection and the $r$-matrix are rational functions of the spectral parameter $z$, the Gross-Neveu model is said to be a rational integrable system, corresponding to taking $C=\mathbb{CP}^1$ in the discussion of section \ref{sec:classicint}. Also, since this $r$-matrix is skew-symmetric $r(z)_{\tn{12}}=-r(-z)_{\tn{21}}$, the Gross-Neveu model is ultralocal.
	\subsubsection*{Principal chiral model}
	A canonical example of a non-ultralocal theory\footnote{Most known integrable $\sigma$-models are non-ultralocal. See \cite{bykov_complex_2016,Affleck:2021vzo} for counterexamples.} is the principal chiral model. It is a $\sigma$-model consisting of a group valued field $g:\Sigma\ra G$, whose action is given by
	\begin{equation}
		\text{S}_{\text{PCM}}[g]=\frac{h}{2}\int\dd^2x\,\left<j_{+},j_{-}\right>\qquad \text{with} \quad j_\pm= g^{-1}\partial_\pm g\,.
		\label{eq:SCPM}
	\end{equation}
	This theory has a $G\times G$ flavour-symmetry under which the field transforms as $g\ra k_Lgk_R$, $k_L, k_R\in G$. The equations of motion are given by
	\begin{equation}
		\partial_\mu j^\mu=\partial_+j_-+\partial_-j_+=0\,.\label{eq:PCMEOM}
	\end{equation}
	The corresponding Lax connection is given by
	\begin{equation}
		L_{\pm}(z)=\frac{h\,j_{\pm}}{h\mp z}\label{eq:LPCM}
	\end{equation}
	which is flat upon imposing the equations of motion \eqref{eq:PCMEOM} and the Maurer-Cartan equation $\partial_\mu j_\nu-\partial_\nu j_\mu+[j_\mu,j_\nu]=0$, which is a Bianchi identity and thus true also off-shell. The similarity to \eqref{eq:LGN} is striking, though there is a crucial difference. The Poisson bracket of $L_x$ with itself still takes the form of a Maillet bracket \eqref{eq:Maillet}, but now the $r$-matrix is non-ultralocal and is given by
	\begin{equation}
		r(z_1,z_2)=r_Y(z_2-z_1)\varphi^{-1}(z_2)\quad \text{ with }\quad \varphi(z)=\frac{h^2-z^2}{2\,z^2}\,,\label{eq:rPCM}
	\end{equation}
	where $r_Y$ is identical to the $r$-matrix of the GN model \eqref{eq:Yang} and $\varphi:\mathbb{CP}^1\ra \mathbb{C}$ is the twist function \eqref{eq:twistfunction}. Like the GN model, the PCM is also a rational integrable system, but the presence of a non-trivial $\varphi$ spoils the ultralocality of the Maillet bracket. As we will see later, $\varphi$ plays a crucial role in the construction of the theory from the four-dimensional Chern-Simons theory.
	
	\section{Time-dependent integrability}\label{sec:timedep}
	
	\subsection{Generalities}
	We now discuss the notion of \textit{time-dependent} integrability. Consider an integrable field theory with coupling constants $h_i$. We try to upgrade these to functions of time $\hat{h}_i(t)$, in such a way that we still have some notion of integrability. To distinguish time-independent and time-dependent quantities, in what follows we add hats $\hat{\quad}$ to the latter. 
	
	Obviously, the first question to ask is: {\it How do we define time-dependent integrability?} It clearly cannot be the existence of an infinite tower of conserved charges, since not even the energy will be conserved once we have time-dependent couplings. Instead, we require that we still have a Lax formulation of the equations of motion. As we see later from the analysis of the 4d CS theory, it turns out that the time-dependent generalisation of the Lax connection depends on whether the theory is ultralocal or not. If the model is ultralocal, the structure of the Lax equation remains intact \eqref{eq:LaxEq}. In contrast, for non-ultralocal theories, we need to modify the derivatives so that they include a derivative along the spectral parameter direction, $\partial_\pm\ra\hat{\partial}_\pm(z)\equiv \partial_\pm\pm \hat{f}_\pm(z)\partial_z$,  and impose the equation
	\begin{equation}
		\hat{\partial}_{\mu}(z)\hat{L}_\nu(z)-\hat{\partial}_{\nu}(z)\hat{L}_\mu(z)+[\hat{L}_\mu(z),\hat{L}_\nu(z)]=0\,,\label{eq:Laxtime}
	\end{equation}
	where $\hat{L}$ will generically have explicit time dependence (i.e.\ not only through its dependence on the fields).
	We further require that these derivatives commute $[\hat{\partial}_\mu(z),\hat{\partial}_\nu(z)]=0$, such that \eqref{eq:Laxtime} can be written in the form
	\begin{equation}
		[\hat{D}_\mu(z),\hat{D}_\nu(z)]=0\,,\qquad \hat{D}_\mu(z)\equiv\hat{\partial}_\mu(z)+\hat{L}_\mu(z)\,.\label{eq:Flattime}
	\end{equation}
	
	While the modification of the Lax equation \eqref{eq:Laxtime} does not allow for an infinite tower of conserved charges, \eqref{eq:Flattime} still has an associated auxiliary linear system and hence can be studied by a generalization of the inverse scattering method. The Lax connection also endows the system with an {\it isomonodromic} structure: certain monodromy data in the spectral-parameter plane remain invariant under time evolution, so that isomonodromy replaces the usual isospectrality of autonomous integrable systems. We will not pursue this connection in detail here, but return to it in a forthcoming work \cite{toappear}. Closely related Lax systems have appeared in non-autonomous integrable models arising from dimensionally-reduced Einstein gravity, where they are known as the {\it Belinsky--Zakharov linear system} \cite{Belinsky:1971nt,Belinsky:1979gra}. Their relation to isomonodromic deformation was identified in a subsector of these theories \cite{Ueno:1981,Korotkin:1994hy,Korotkin:1995hf,Korotkin:1996xk,Korotkin:1996cz,Nicolai:1996he,Korotkin:1996cm,Korotkin:1998aa}. More recently, similar structures also appeared in the studies of time-dependent integrable sigma models in \cite{Hoare:2020fye}. The key result of this paper is the generalisation of these results to a much broader class of integrable theories.
	
	In passing, let us mention that there is an alternative (but physically equivalent) formulation in which one does not modify the form of the Lax equation but instead promotes the spectral parameter to a spacetime-dependent function $z\ra z(t,x)$. In the context of dimensionally-reduced gravity, such a formulation is known as the \textit{Breitenlohner-Mason linear system} \cite{Breitenlohner:1986um}. In this paper, we distinguish the two by calling the former the \textit{fixed-spectral formulation} $\hat{L}^{\text{fix}}$ and the latter the \textit{variable-spectral formulation} $\hat{L}^{\text{var}}$. In most of this paper, we focus on the constant-spectral-parameter formulation and drop the superscript for simplicity, but we will discuss the connection to the variable-spectral-parameter formulation in \cref{sec:BMtype} and work out the corresponding function $z(t,x)$ in a number of examples.
	
	The goal of this paper is to systematically construct theories satisfying the requirements \eqref{eq:Laxtime} and \eqref{eq:Flattime} via the 4d CS theory. We will see that there exists an essentially unique way to introduce time dependence in the 4d CS theory, and \eqref{eq:Laxtime} and \eqref{eq:Flattime} follow automatically.
	Curiously, we find that the time dependence must coincide with the one-loop RG flow, i.e.\
	\begin{equation}
		\partial_t\, \hat{h}_i(t)\propto\beta_i \left(\hat{h}_j(t)\right)\,,\label{eq:timeisRG}
	\end{equation}
	where $\beta_i$ is the one-loop $\beta$-function for the coupling $h_i$. We stress that we do not impose this by hand; instead, it naturally follows from requiring the 4d CS theory to be well-defined in the presence of time dependence. This connection to RG has been observed for a class of integrable $\sigma$-models in \cite{Hoare:2020fye}, while \cite{Pasnoori:2025rgflow} found a similar relation in a quantum integrable system. We will find that it holds for any integrable field theory obtained from the 4d CS theory. 
	
	To illustrate these points, in the rest of this section, we discuss how the time dependence can be introduced for our familiar examples of the GN model and the PCM, postponing a systematic construction from the 4d CS theory to \cref{sec:timedep}.
	
	\subsection{Examples}
	\subsubsection*{Gross-Neveu model}
	We find it instructive to initially allow the coupling constant $\hat{h}$ of the GN model to depend on both $x$ and $t$, such that the action takes the form
	\begin{equation}
		\hat{\text{S}}[\psi^n_\pm]=\int_{\Sigma}\dd^2 x\,i\psi^n_+\partial_-\psi^n_++i\psi^n_-\partial_+\psi^n_-+\,\hat{h}(t,x)\,\left<J_+,J_-\right>\,.\label{eq:SGNtime}
	\end{equation}
	The equations of motion take the same form as in the time-independent case, except that the coupling constant now introduces explicit spacetime dependence
	\begin{equation}    \partial_\pm\psi_{\mp}^n+\hat{h}(t,x)\,J^a_{\pm}\,T^a_{nm}\,\psi^m_\mp=0\,.\label{eq:GNEomtime}
	\end{equation}
	For which choices of $\hat{h}(t,x)$ is this model still integrable?  We assume\footnote{As we see later, this naturally follows from the 4d CS theory.} that the dependence of the Lax connection on the fields and couplings takes the same form as in the time-independent case
	\begin{equation}
		\hat{L}_\pm=\mp\frac{J_\pm}{z-\hat{z}^\pm(t,x)}\,,\qquad \hat{h}(t,x)=\frac{1}{\hat{z}^+(t,x)-\hat{z}^-(t,x)}\,,\label{eq:hfromzGN2}
	\end{equation}
	where we now allow for the locations of the poles to depend on both time and space,
	and impose that \eqref{eq:GNEomtime} follows from the flatness condition for \eqref{eq:hfromzGN2}. Given the similarity to the time-independent case, it is easy to find that this holds provided that the derivatives $\partial_\pm$ only act on $J_\mp$ and not on $\hat{z}^\mp$, i.e.\ $\partial_\pm \hat{z}^\mp=0$. Hence, the most general deformation admitting the Lax connection is
	\begin{align}
		\hat{z}^{+}=\hat Z^+(x^{+})\,, \qquad \hat{z}^{-}=-\hat Z^-(x^{-})\quad \Rightarrow \quad \hat h(t,x)=\frac{1}{\hat Z^+(x^+)+\hat Z^-(x^-)}\label{eq:spacetime}
	\end{align}
	where $\hat Z^\pm$ are arbitrary chiral/antichiral functions.
	
	If we further require that $\hat{h}$ depends only on time and not on space, the only possible such function is
	\begin{equation}
		\hat{h}(t)=\frac{1}{a\,t+b}\,,\label{eq:hGNtime}
	\end{equation}
	for some constants $a,b$. Thus, given our notion of time-dependent integrability, the time-dependent GN model \eqref{eq:SGNtime} is integrable if and only if the coupling constant satisfies \eqref{eq:hGNtime}. 
	As mentioned earlier, this time dependence is identical to the one-loop RG flow. For the time dependence, we have
	\begin{equation}
		\partial_t\,\hat{h}(t)=-a\,\hat{h}(t)^2\,,
	\end{equation}
	while for the RG flow of the time-independent theory, we have instead \cite{Kutasov:1989dt,Gerganov:2000mt}
	\begin{equation}
		\frac{\dd h}{\dd \tRG}=-\frac{c_{SO(N)}}{2}h^2\label{eq:RGGN}
	\end{equation}
	where $\tRG\equiv \log\mu/4\pi$ is the \textit{RG-time} and $c_{SO(N)}=N-2$ is the dual Coxeter number of $SO(N)$. Thus, we see that the two flows are identical under the identification $t=\frac{c_{SO(N)}}{2a}\tRG$. 
	
	One may worry that the time-dependent coupling constant blows up at finite time $t=-a/b$ and changes sign beyond that, making the theory ill-defined. This can be remedied by changing the coordinates. For instance, after a holomorphic change of coordinates
	\begin{align}\label{eq:holochangecor}
		x^{\pm}=\cosh\tilde{x}^{\pm}-\frac{b}{a}\,,
	\end{align}
	the time dependence in the new coordinates $\tilde{x}^{\pm}=\tilde{t}\pm \tilde{x}$ reads
	\begin{align}
		\hat{h}(\tilde{t},\tilde{x})=\frac{1}{a \cosh \tilde{x} \cosh \tilde{t}}\,.\label{eq:GNnonsingular}
	\end{align}
	In the new coordinates, the coupling constant vanishes at $\tilde{t},\tilde{x}\to \pm \infty$ and never blows up.
	Thus, at the expense of introducing the additional space-dependence, one can make the theory better defined.\footnote{Alternatively, one can obtain \eqref{eq:GNnonsingular} directly from \eqref{eq:spacetime} by setting $\hat{Z}^+(x^{+})=\cosh (x^{+})/2$ and $\hat{Z}^-(x^{-})=-\cosh(x^{-})/2$.}   A similar change of the coordinates can be performed in any of the models that we discuss in this paper. 
	
	Note that recently, a time-dependent generalization of the Gross-Neveu model was discussed in \cite{Pasnoori:2025qkz}. It will be interesting to understand the precise connection with our results.
	\subsubsection*{Principal chiral model}
	Next, let us discuss the time-dependent PCM. Just like for the GN model, we start by considering a general spacetime-dependent coupling
	\begin{equation}
		\hat{\text{S}}_{\text{PCM}}[g]=\int\dd^2x\,\frac{\hat{h}(t,x)}{2}\,\left<j_\mu,j^\mu\right>\,.\label{eq:SCPMtime}
	\end{equation}
	The corresponding equations of motion take the form
	\begin{equation}
		\partial_+\big(\hat{h}(t,x)j_-\big)+\partial_-\big(\hat{h}(t,x)j_+\big)=0\,,\label{eq:PCMEOMtime}
	\end{equation}
	and we also have the Maurer-Cartan equation $\partial_+j_--\partial_-j_++[j_+,j_-]=0$. Again, assuming that the dependence of the Lax connection on the fields and couplings takes the same form as in the time-independent case \eqref{eq:LPCM}, we obtain 
	\begin{equation}\label{eq:time-pcm-lax}
		\hat{L}_\pm(z)=\frac{\hat{h}\,j_\pm}{\hat{h}\mp z}\,.
	\end{equation} 
	However, unlike the GN model, if we further require that the Lax equation is not modified, we cannot obtain a nontrivial time dependence. We thus need to modify the Lax equation by replacing the derivatives as $\partial_\pm\ra\hat{\partial}_\pm(z)=\partial_\pm\pm \hat f_\pm(z)\partial_z$ for some functions $\hat f_\pm(z)$. The modified Lax equation $\hat{\partial}_+\hat{L}_--\hat{\partial}_-\hat{L}_++[\hat{L}_+,\hat{L}_-]=0$ then takes the form 
	\begin{equation}
		\frac{\partial_+\big(\hat{h}j_-\big)}{\hat{h}+z}-\frac{\hat{h}j_-\big(\partial_+\hat{h}+\hat{f}_+)}{(\hat{h}+z)^2}-\frac{\partial_-\big(\hat{h}j_+\big)}{\hat{h}-z}+\frac{\hat{h}j_+\big(\partial_-\hat{h}+\hat{f}_-)}{(\hat{h}-z)^2}+\frac{\hat{h}^2[j_+,j_-]}{\hat{h}^2-z^2}=0\,.
	\end{equation}
	This equation follows from the EOM and the Maurer-Cartan equation if we identify
	\begin{equation}
		\hat{f}_\pm(z)=\pm\frac{2\partial_\pm\hat{h}\,z}{\hat{h}\mp z}\,.\label{eq:fpmPCM}
	\end{equation}
	Further requiring that $\hat{\partial}_+$ and $\hat{\partial}_-$ commute, we find $\hat{h}$ must satisfy $\partial_+\partial_-\hat{h}(t,x)=0$ and thus that $\hat{h}$ is a sum of a chiral and an ,
    \begin{equation}
        \hat h(x^+,x^-)=\hat Z^+(x^+)+\hat Z^-(x^-)\label{eq:hfromZ}
    \end{equation}
	
	If we focus on solutions for which $\hat{h}$ only depends on time but not on space, \eqref{eq:hfromZ} simplifies to
	\begin{equation}
		\hat{h}(t)=a\,t+b\,.\label{eq:hPCMtime}
	\end{equation}
	This again matches, after an appropriate rescaling of time, the one-loop RG flow of the time-independent PCM, which takes the form \cite{Brezin:1975sq}
	\begin{equation}
		\frac{\dd h}{\dd\tRG}=c_G\,,
	\end{equation}
	where $\tRG \equiv \log \mu/4\pi$  
	is again the RG-time and $c_G$ is the dual Coxeter number \eqref{eq:cGdef}.
	
	Note that $\hat{h}(t)$ \eqref{eq:hPCMtime} vanishes at a finite time, but this can once again be remedied by changing the coordinates. For instance, after the holomorphic change of the coordinates \eqref{eq:holochangecor}, we obtain
	\begin{align}
		\hat{h}(\tilde{t},\tilde{x})=a\cosh\tilde{t}\cosh\tilde{x}\,.
	\end{align}
	
	\section{Standard integrability from the 4d CS theory: review}\label{sec:classicint4dCS}
	
	We now come back to standard (time-independent) integrable field theories and review how they can be systematically constructed from the 4d CS theory \cite{Costello:2017dso,Costello:2018gyb,Costello:2019tri}, as a preparation for the generalisation to the time-dependent cases discussed in the next section.
	
	We will not attempt to give a comprehensive review and refer the reader to the original articles, in particular \cite{Costello:2019tri} (see also \cite{Lacroix:2021iit,Yamazaki:2025yan} for recent reviews).
	\subsection{Generalities}
	Traditionally, integrable systems have often been found by “trial and error'', by attempting to rewrite equations of motion into the Lax connection form. Without a guiding principle, however, this approach makes systematic construction difficult. The 4d CS theory overcomes this limitation by providing a unified framework for generating a wide class of integrable systems. In this approach, the Lax connection arises naturally, enabling a systematic construction of integrable structures. 
	
	Perhaps the best way to motivate this approach is through the Lax equation \eqref{eq:LaxEq}. Being a flatness condition, it closely resembles the equation of motion $F=0$ of the 3d Chern-Simons theory. Indeed, the similarity between the 3d Chern–Simons theory and integrable systems was recognized early on: the dynamics of Wilson lines describe non-Abelian anyons whose braiding relations resemble the Yang–Baxter equation. The crucial difference, however, is the absence of a spectral parameter in the 3d Chern–Simons theory. In contrast to 3d Chern-Simons theory, the Lax connection $L$ of 2d integrable systems depends on both the spacetime and the spectral parameter and thus naturally lives on $\Sigma\times C$, suggesting that the relevant theory may be a four-dimensional gauge theory. At the same time, $L_\mu$ has components only along $\Sigma$, not along all four directions of $\Sigma \times C$. The question is then: What kind of four-dimensional gauge theory produces precisely such an object?
	\begin{figure}[t]
	    \centering
			\begin{tikzpicture}
		
		\newcommand{\R}{0.125\textwidth} 
		\filldraw[color=gray!70, fill=gray!15, very thick] (-4*\R,-\R) rectangle (-2*\R,\R);
		\node[gray!70,font=\Huge] at (-1.5*\R,0){$\times$};
		\filldraw[color=gray!70,fill=gray!15,thick] (0,0) circle (\R);

		\draw[gray!70, dashed] (-\R,0) arc [start angle=180, end angle=0, x radius=\R, y radius=0.3*\R];
		
		\draw[gray!70] (-\R,0) arc [start angle=180, end angle=360, x radius=\R, y radius=0.3*\R];
		
		
		
		\node[] at (-3*\R,1.5*\R) {\LARGE $\Sigma:(t,x)$};
		\node[] at (0,1.5*\R) {\LARGE $C:(z,\bar{z})$};
		\node[] at (-1.5*\R,1.5*\R) {\LARGE $\times$};
	\end{tikzpicture}
        \caption{The four-dimensional manifold on which the 4d CS theory is defined. It is the product of the 2d spacetime $\Sigma$, on which dynamical degrees of freedom live, and the Riemann surface $C$, on which the spectral parameter $z$ takes values.}
        \label{fig:4dCSspacetime}
	\end{figure}
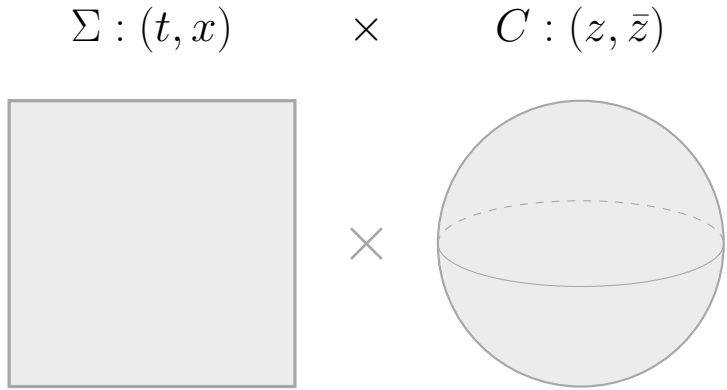    
	This question led the authors of \cite{Costello:2017dso,Costello:2018gyb,Costello:2019tri} to introduce what is now called \emph{the four-dimensional Chern-Simons theory}.\footnote{See \cite{Costello:2013zra} for an earlier construction of a closely related gauge theory in a different context.} The theory is defined on a product $\Sigma\times C$, where $\Sigma$ is a two-dimensional surface, often referred to as the topological directions, and $C$ is a complex curve, often referred to as the holomorphic direction. See Figure \ref{fig:4dCSspacetime}. The fundamental field is a $\mathfrak{g}$-valued one-form $A$, where $\mathfrak{g}$ is a complexified semisimple Lie algebra. We denote light-cone coordinates on $\Sigma$\footnote{For the most part of this paper, we assume  $\Sigma$ is a Lorentzian manifold.} by
	$(x^+,x^-)$, and holomorphic/anti-holomorphic coordinates on $C$ by
	$(z,\bar z)$.
	Given an invariant bilinear form $\langle \cdot,\cdot\rangle$ on $\mathfrak g$, the Chern-Simons three-form is
	\begin{align}
		\text{CS}[A]=\left<A,\dd A+\frac{2}{3}A\wedge A\right>\,.
	\end{align}
	The four-dimensional Chern-Simons action is then given by\cite{Costello:2013zra,Costello:2017dso,Costello:2018gyb,Costello:2019tri}
	\begin{equation}
		\text{S}_{\text{CS}_4}[A]=\frac{i}{4\pi}\int_{\Sigma \times C}\omega\wedge\text{CS}[A]\,.\label{eq:S4DCS}
	\end{equation}
	Here, $\omega$ is a meromorphic one-form on $C$. We write it as
	\begin{align}
		\omega=\varphi(z) \dd z ,
	\end{align}
	for some meromorphic function $\varphi(z)$. The resemblance between $\varphi(z)$ and the twist function appearing in \eqref{eq:rPCM} is not accidental; as we will see, they are in fact the same object in this framework. Note that $\omega$ depends only on the holomorphic coordinate on $C$, and not on the topological directions $\Sigma$.
	
	To extract the equations of motion, we take the variation of the action with respect to the gauge field ($\omega$ is not a dynamical field and thus should not be varied)
	\begin{equation}
		\label{eq:delta4DCS}
		\delta \text{S}_{\text{CS}_4}[A]=\frac{i}{2\pi}\int_{ \Sigma \times C}\omega\wedge \left<\text{F}[A],\delta A\right>-\frac{i}{4\pi}\int_{ \Sigma \times C}\dd\omega\wedge\left<A,\delta A\right>\,,
	\end{equation}
	where $\text{F}[A]=\dd A+A\wedge A$ is the field strength of $A$. Since the 1-form $\omega$ is meromorphic, the integral of the second term is localised at the poles of $\omega$, and hence can be regarded as a boundary term. Thus, for the variational problem to be well-defined, we need to set a boundary condition so that the contribution from this term vanishes.
	Once a boundary condition is imposed, the equations of motion take the form of a flatness condition
	\begin{equation}
		\omega\wedge \text{F}[A]=0\,.\label{eq:4DCSEOM}
	\end{equation}
	This already has the structure anticipated above, but one more step is still needed to make contact with the Lax connection $L$. The four-dimensional gauge field has components not only along $\Sigma$, but also along $C$, namely $A_z$ and $A_{\bar z}$. By contrast, the Lax connection should have only the two components along $\Sigma$, and we must therefore remove the components along $C$. The first simplification comes from the special form of the action \eqref{eq:S4DCS}. Since $\omega$ is proportional to $\dd z$, the component $A_z$ does not appear in the action. As a result, the theory has a shift gauge symmetry $A\mapsto A +\chi(x,z)\,\dd z$, where $\chi(x,z)$ is an arbitrary smooth $\mathfrak{g}$-valued function on $\Sigma\times C$. This allows us to set $A_{z}=0$.
	The remaining component $A_{\bar z}$ can then be removed by an ordinary four-dimensional gauge transformation,
	\begin{align}\label{eq:4Dgaugetr}
		A\rightarrow A^u:= uAu^{-1}-\dd uu^{-1}\,,
	\end{align}
	which leaves the equation of motion invariant.\footnote{The gauge invariance of the action itself is more subtle, see \cite{Benini:2020skc} for a discussion.} More precisely, because of the flatness condition \eqref{eq:4DCSEOM}, $A_{\bar{z}}$ can always be written in a ``pure gauge form'' $A_{\bar{z}}=-\partial_{\bar{z}}g g^{-1}$. Hence setting $u=g^{-1}$ gives $A^{u}_{\bar{z}}=0$.
	
	Ignoring for the moment the subtleties that arise at the poles of $\omega$, which will be discussed later, we may therefore choose a gauge in which $A_z=A_{\bar{z}}=0$. The remaining components are then identified with the Lax connection $A_{\pm} =L_{\pm}$. In this gauge, the four-dimensional Chern-Simons equations of motion \eqref{eq:4DCSEOM} reduce to
	\begin{subequations}
		\label{Lax4d}
		\begin{align}
			\partial_+L_--\partial_-L_++[L_+,L_-]=0\,,\label{eq:LaxFlat}\\
			\varphi\,\partial_{\bar{z}}L_{\pm}=0\,.\label{eq:LaxMero}
		\end{align}
	\end{subequations}
	The first equation is precisely the flatness condition. The second says that the connection is holomorphic in the spectral parameter $z$. These are the characteristic features of a Lax connection $L$ of a 2d integrable field theory! This is the power of the 4d CS theory: instead of guessing the Lax connection for a given 2d integrable field theory, one obtains it directly from the 4d gauge field.
	
	There is still an important point to clarify. If one chooses the simplest one-form, $\omega=\dd z$, then \eqref{eq:LaxMero} is too restrictive: it forbids poles in the spectral parameter and therefore leads only to a trivial Lax connection. Moreover, the discussion so far has not yet explained where the two-dimensional degrees of freedom come from. Both issues are resolved by introducing surface defects extended along $\Sigma$ and localized at points of $C$. These defects supply the two-dimensional dynamics and, at the same time, allow the Lax connection to acquire the prescribed singularities in the spectral parameter. In general, there are two types of such surface defects \cite{Costello:2019tri}:
	
	\begin{itemize}
		\item \textit{Order Defect}:\\
		Additional local degrees of freedom are introduced on 2d defects along $\Sigma\times \{z_i\}$ for some points $z_i\in C$. These defects are coupled to the 4d CS gauge field in a gauge invariant way, giving rise to delta function sources $\delta^{(2)}(z-z_i)$ in (\ref{eq:LaxMero}), and the associated Lax connection develops poles at $z=z_i$. This gives rise to ultralocal theories such as the Gross-Neveu model.
		\item \textit{Disorder Defect}:\\
		Instead of introducing additional degrees of freedom, we consider a meromorphic 1-form $\omega$ that has zeros and poles. At the zeros of $\omega$, \eqref{eq:LaxMero} allows the Lax connection to have poles. Because of the relation between $L$ and $A$, this corresponds to imposing singular boundary conditions on the gauge fields $A$, and can be interpreted as {\it disorder defects}. In addition, the four-dimensional gauge fields give rise to \textit{edge modes} at the pole of $\omega$ due to the boundary condition imposed there. These edge modes become dynamical degrees of freedom in the two dimensions of $\Sigma$. Disorder defects in general give rise to non-ultralocal theories such as the principal chiral model.
	\end{itemize}
	The Lax connection is determined by specifying the insertions of these surface defects and the boundary conditions at the poles of $\omega$. Once the Lax connection is obtained, integrating out the four-dimensional action over the spectral parameter space $C$ around the classical solution gives the two-dimensional action of the corresponding 2d classical integrable field theory. Importantly, it is conjectured in \cite{Costello:2019tri} that \textit{all} 2d integrable field theories can be obtained from some 4d Chern-Simons construction.
	
	In what follows, we explain more details of each construction and how they are realised in concrete examples.
	
	\subsection{Order defects}\label{sec:orderdefects}
	For the order defects, we consider a meromorphic 1-form $\omega$ without zeros. Namely, we take $\omega=\dd z$ for $C=\mathbb{CP}^1$, $\omega=\dd z/z$ for $C=\mathbb{C}^*$, and $\omega=\dd z$ for $C=\mathbb{T}$. In what follows, we mainly discuss the case for $C=\mathbb{CP}^{1}$, i.e.\ $\omega=\dd z$.
	
\begin{figure}[t]
	    \centering
   \begin{tikzpicture}
    
    \newcommand{\R}{0.125\textwidth} 
    
    \filldraw[color=gray!70, fill=gray!15, thick] ({-4*\R}, -\R) rectangle ({-2*\R}, \R);
    
    \filldraw[color=gray!70, fill=gray!15, thick] ({2*\R}, -\R) rectangle ({4*\R}, \R);
    
    \filldraw[color=gray!70, fill=gray!15, thick] (0,0) circle (\R);
    
    \draw[gray!70, dashed] (-\R,0) arc [start angle=180, end angle=0, x radius=\R, y radius=0.3cm];
    
    
\foreach \x/\multiplier/\label in {
        0/1%
    } {
        \fill[blue] (\x, {\multiplier*\R}) circle (1.5pt); 
    }
    
    \node[left, orange] at (-\R, 0) {$z^-$};
    \node[right, orange] at (\R, 0) {$z^+$};
    \node[above, blue] at (0,\R) {$\infty$};
    
    \draw[orange, ->, thick] (-1.04*\R, -0.04*\R) .. controls ({-1.25*\R}, {-0.6*\R}) and ({-1.75*\R}, {0.6*\R}) .. ({-1.96*\R}, 0.04*\R);
    
    \draw[orange, ->, thick] (1.04*\R, -0.04*\R) .. controls ({1.25*\R}, {-0.6*\R}) and ({1.75*\R}, {0.6*\R}) .. ({1.96*\R}, 0.04*\R);
    
    \node[] at ({-3*\R}, 0) {\Large$J_-$};
    \node[] at ({3*\R}, 0) {\Large$J_+$};

    \draw[gray!70] (-\R,0) arc [start angle=180, end angle=360, x radius=\R, y radius=0.3*\R];

    \draw[orange, thick](-\R-0.03*\R, -0.03*\R) -- (-\R+0.03*\R, 0.03*\R);
    \draw[orange, thick] (-\R-0.03*\R, 0.03*\R) -- (-\R+0.03*\R,-0.03*\R);

    \draw[orange, thick](\R-0.03*\R, -0.03*\R) -- (\R+0.03*\R, 0.03*\R);
    \draw[orange, thick] (\R-0.03*\R, 0.03*\R) -- (\R+0.03*\R,-0.03*\R);

    \end{tikzpicture}
	    \caption{The 4d CS theory with a pair of chiral and antichiral order defects, inserted at points $z^+$ and $z^-$ respectively. The defect at $z^{+}$ hosts the chiral current $J_+$ and couples to the bulk gauge field $ A_{-}$ while the defect at $z^{-}$ hosts $J_{-}$ and couples to the bulk gauge field $A_{+}$. Integrating out the bulk gauge fields leads to an interacting integrable field theory in two dimensions. The 1-form $\omega =\dd z$ has a pole at $z=\infty$, where we impose the boundary condition \eqref{eq:ordbound}.}
	    \label{fig:4dcsorder}
	\end{figure}
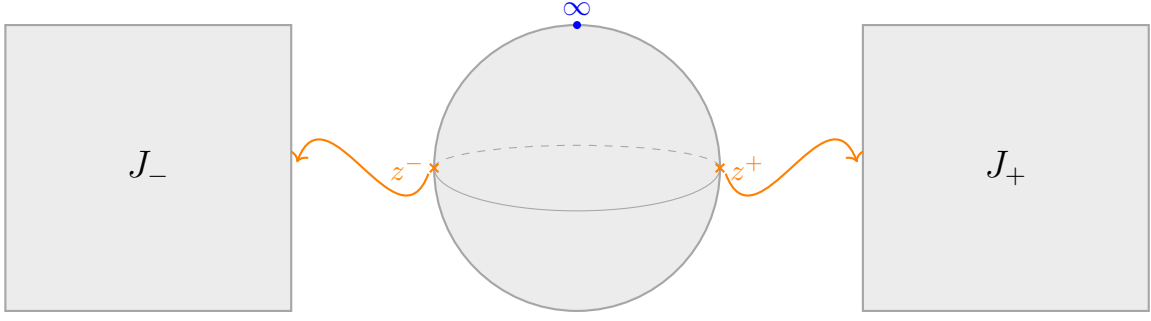

	We then introduce $N$ surface defects, each localised to a different point $z_i\in C$, $i=1,\dots, N$. The defect at $z_i$ consists of a collection of fields $\phi^{i,n}$ governed by a two-dimensional Lagrangian $\mathcal{L}^i[\phi^{i,n}]$. In principle, the choice of defects is quite broad. Following much of the existing literature, however, we restrict attention to defects that are either purely chiral, depending only on $x^{+}$, or purely antichiral, depending only on $x^{-}$.\footnote{Note that some nonchiral order defects can be realized in a scaling limit where a chiral and an antichiral defect collide.}
	Additionally, we assume that the defect field theories have global $\mathfrak{g}$ symmetry with conserved currents $J_{\pm}^i$ and can be consistently coupled to the 4d gauge fields $A$. See Figure \ref{fig:4dcsorder}.
	
	We therefore consider $N_+$ chiral surface defects at $z_1^+,\dots,z_{N_+}^+$ and $N_-$ antichiral defects at $z_1^-,\dots, z_{N_-}^-$ with $N=N_++N_-$.
	The action for the coupled 4d--2d system is then obtained by gauging the global symmetry $\mathfrak{g}$ by coupling the defect currents directly to the 4d gauge field \cite{Costello:2019tri}\begin{align} 
		\begin{split}\label{eq:orddefS}
			\text{S}_{\textrm{4d--2d}}\Big[A,\phi_{\pm}^i\Big]=
			\text{S}_{\text{CS}_4}[A] & + 
			\sum_{i=1}^{N_+} 
			\int_{\Sigma \times \{z_j^+\} }\Big(\mathcal{L}^i_+\left[\phi^i_{+}\right] 
			+\left< A_{-} , J_{+}^i\,\right>
			\Big)\dd x^+\wedge \dd x^- \\
			&+\sum_{i=1}^{N_-} 
			\int_{\Sigma \times \{z_i^-\} } 
			\Big( \mathcal{L}^{i}_- 
			\left[ \phi^{i}_{-} \right]
			+ \left<A_{+}, J_-^{i}\,\right>
			\Big)\dd x^+\wedge \dd x^-\,,  
		\end{split}
	\end{align}
	The equations of motion of $\phi$'s imply that the defect currents are separately conserved
	\begin{align}\label{eq:eom-order}
		\mathcal{L}^i_{+}:\quad \partial_{-}J_{+}^i=0\,,\qquad  \mathcal{L}^i_{-}:\quad \partial_{+}J_{-}^i=0\,.
	\end{align}

	Coupling to these defects adds additional terms on the right-hand side of the equations of motion for the gauge-field \eqref{eq:4DCSEOM}. While the flatness condition \eqref{eq:LaxFlat} for the Lax connection remains the same, the variation of the action with respect to $\delta A_\pm$ has additional contributions at $z=z_i^{\pm}$, and we get instead
	\begin{equation}
		\partial_{\bar{z}}L_{\pm}(z)=\pm 2\pi i\sum_{i=1}^{N_{\pm}}\delta^{(2)}(z-z_i^{\pm})J_{\pm}^i\,,
		\label{disordmod}
	\end{equation}
	where we again made the gauge choice $A_{\bar{z}}=0$ and we replaced $A_\pm$ by $L_\pm$. Being inhomogeneous differential equations, these equations alone do not specify the Lax connection uniquely; to any solution to \eqref{disordmod}, one can add solutions to the homogeneous differential equations $\partial_{\bar{z}} L_{\pm}(z)=0$. To remove this ambiguity, we need to take into account the boundary conditions of $A$ at the poles of the meromorphic one-form $\omega$. In the rational case, where $C=\mathbb{CP}^1$ and $\omega=\dd z$, there exists a pole at $z=\infty$ and a natural choice of the boundary condition is 
	\begin{align}
		A_{\pm}\lvert_{\infty}=L_{\pm}\lvert_{\infty}=0\,.\label{eq:ordbound}
	\end{align}
	Then, using the identity
	\begin{align}
		-\frac{1}{2\pi i}\partial_{\bar{z}}\left(\frac{1}{z-z_0}\right)=\delta^{(2)}(z-z_0)\,.\label{eq:dzbonz}
	\end{align}
	one obtains
	\begin{equation}
		L_\pm=\mp\sum_{i=1}^{N_\pm} \frac{J^i_\pm}{z-z^\pm_i}\,,\label{LdisformJ}
	\end{equation}
	Therefore, the Lax connection has simple poles at the defect locations, with residues determined by the corresponding defect currents.
	One can similarly construct the Lax connection for trigonometric and elliptic cases by solving the equations of motion (\ref{disordmod}) with appropriate boundary conditions at the poles of the 1-form $\omega$ \cite{Costello:2017dso}. The result can be written uniformly in terms of the classical $r$-matrix $r\in \mathfrak{g}\otimes \mathfrak{g}$:
	\begin{align}\label{eq:lax-order}
		L_{\pm,a}=\mp\sum_{i=1}^{N_\pm}J_{\pm}^{i,b}r_{ab}(z-z^\pm_i)\,,
	\end{align}
	where $a$ and $b$ labels generators $T^{a,b}$ of $\mathfrak{g}$.
	For rational models with the classical $r$-matrix (\ref{eq:Yang}), this reduces to \eqref{LdisformJ}.
	
	The effective two-dimensional theory is obtained by evaluating the 4d--2d action \eqref{eq:orddefS} on the classical solution $A=L$ and integrating over the auxiliary curve $C$. The result is a current-current deformation of the collection of chiral and antichiral defect theories \cite{Costello:2019tri}
	\begin{align}\label{eq:2daction-order}
		\begin{aligned}
			&\text{S}_{\text{2d}}[\phi_i^{\pm}]=\\
			&\int_{\Sigma}~\biggl[~\sum_{i=1}^{N_+}\mathcal{L}_+^i\left[\phi^i_{+}\right]+\sum_{i=1}^{N_-}\mathcal{L}^i_-\left[\phi^i_{-}\right]-\sum_{i=1}^{N_+}\sum_{j=1}^{N_-}J_+^{i,a}r_{ab}(z^+_i-z^-_j)J_-^{j,b}\biggr]\dd x^+\wedge \dd x^-\,.
		\end{aligned}
	\end{align}
	Therefore, each chiral defect interacts with each antichiral defect, and the strength of the interaction is fixed by the classical $r$-matrix evaluated at their separation on $C$, i.e.\ for the rational $r$-matrix \eqref{eq:Yang}, the potential goes like $1/(z^+_i-z^-_j)$. Classical integrability follows from the equivalence between the equations of motion of this two-dimensional theory and the flatness condition for the Lax connection (\ref{eq:lax-order}). 
	For details, see Appendix \ref{sec:int-order}.
	
	Note that the resulting 2d action depends only on the relative positions of the defects on $C$, not on their absolute locations. This is expected since $z$ is a spectral parameter rather than a physical coordinate, and the locations of the defects can be shifted by a coordinate transformation of $z$. More precisely, one can perform a global biholomorphism of $C$ (i.e.\ a holomorphic coordinate transformation that is bijective). In the rational case, this amounts to performing a Möbius transformation
	\begin{equation}
		z\mapsto\frac{az+b}{cz+d}\,,\quad ad-bc=1\,.\label{eq:Mobius}
	\end{equation}
	Since the boundary condition singles out $z=\infty$, it is natural to restrict to transformations preserving this point. These are the constant shift $z\ra z +b$, which indeed leave the 2d action \eqref{eq:2daction-order} invariant.
	\subsubsection*{Example: Gross-Neveu model}
	As an example of this machinery, let us derive the GN model that we encountered in \eqref{eq:SGN}. We take $C=\mathbb{CP}^1$ and $\omega=\dd z$, take $\mathfrak{g}=\mathfrak{so}(N)$ and consider one chiral and one antichiral defect. The corresponding gauged defect actions are free charged chiral fermions, i.e.
	\begin{align}\label{fermaction1}	
		\begin{aligned}
			\text{S}_{\textrm{4d--2d}}[A,\psi_\pm]=\text{S}_{\text{CS}_4}[A]
            &+\int_{\Sigma\times \{z^{+}\}} \, \psi_{+} i(\partial_-+ A_-)\psi_{+}\,\dd x^+\wedge \dd x^-
			\\
			&+\int_{\Sigma\times \{z^{-}\}}\psi_{-} i(\partial_{+}+A_+)\psi_{-}\,\dd x^+\wedge \dd x^-\,.
		\end{aligned}
	\end{align}
	The $\text{SO}(N)$ currents are given by $J_\pm^a=-i\psi^i_\pm T^a_{ij}\psi^j_\pm$ and the Lax connection \eqref{LdisformJ} reads
	\begin{equation}
		L_\pm=\mp\frac{J_\pm}{z-z^\pm}\,,
	\end{equation}
	which agrees with \eqref{eq:LGN}. To obtain the two-dimensional action, we evaluate the 4d--2d action on $L=A$ and integrate over $\mathbb{CP}^1$. Using the general expression \eqref{eq:2daction-order} and the $r$-matrix \eqref{eq:Yang}, we obtain
	\begin{equation}
		\text{S}[\psi_\pm]=\int_{\Sigma}i\psi_+\partial_-\psi_++i\psi_-\partial_+\psi_-+\,\frac{1}{z^+-z^-}\left<J_+,J_-\right>\dd x^+\wedge \dd x^-\,,\label{eq:SGrossNeveu4DCS}
	\end{equation}
	which reproduces the action of the GN model \eqref{eq:SGN} upon the identification between the coupling constant and $z^{\pm}$ \eqref{eq:hfromzGN}.
	
	One advantage of the 4D CS construction is that it is straightforward to identify integrable generalizations and deformations of the theory. Some of these are discussed in Section \ref{sec:examples}. For example, one can obtain trigonometric and elliptic deformations simply by replacing $\mathbb{CP}^1$ by $\mathbb{C}^*$ or $\mathbb{T}$, or one can replace the symmetry $SO(N)$ by any other semi-simple Lie group. Importantly, all these theories are integrable by construction.

	\subsection{Disorder defects}\label{sec:ti-disorder}
\begin{figure}[t]
    \centering
    	\begin{tikzpicture}
        
    \newcommand{\R}{0.125\textwidth}

    \filldraw[color=gray!70, fill=gray!15, thick] ({-4*\R}, -\R) rectangle ({-2*\R}, \R);
    
    \filldraw[color=gray!70, fill=gray!15, thick] ({2*\R}, -\R) rectangle ({4*\R}, \R);
    
    \filldraw[color=gray!70,fill=gray!15,thick] (0,0) circle (\R);
    
    \draw[gray!70, dashed] (-\R,0) arc [start angle=180, end angle=0, x radius=\R, y radius={0.3*\R}];
    
    \draw[gray!70] (-\R,0) arc [start angle=180, end angle=360, x radius=\R, y radius={0.3*\R}];
   
    
    \foreach \x/\multiplier/\label in {
        0/1,
        0/-1%
    } {
        \fill[blue] (\x, {\multiplier*\R}) circle (1.5pt); 
    }
    
    \draw[orange, thick](-\R-0.03*\R, -0.03*\R) -- (-\R+0.03*\R, 0.03*\R);
    \draw[orange, thick] (-\R-0.03*\R, 0.03*\R) -- (-\R+0.03*\R,-0.03*\R);

    \draw[orange, thick](\R-0.03*\R, -0.03*\R) -- (\R+0.03*\R, 0.03*\R);
    \draw[orange, thick] (\R-0.03*\R, 0.03*\R) -- (\R+0.03*\R,-0.03*\R);

    \node[left, orange] at (-\R, 0) {$z^-$};
    \node[right, orange] at (\R, 0) {$z^+$};
    \node[below, blue] at (0,-\R) {$p_1$};
    \node[above, blue] at (0,\R) {$p_2$};

     \draw[blue, ->, thick] (-0.04*\R, -1.04*\R) .. controls ({-\R}, {-1.5*\R}) and ({-1.5*\R}, {0*\R}) .. ({-1.96*\R}, -0.04*\R);
    
    \draw[blue, ->, thick] (0.04*\R, 1.04*\R) .. controls ({\R}, {1.5*\R}) and ({1.5*\R}, {0*\R}) .. ({1.96*\R}, 0.04*\R);
    
    \node[] at ({-3*\R}, 0) {\Large$g_1$};
    \node[] at ({3*\R}, 0) {\Large$g_2$};
    
    \end{tikzpicture}
     \caption{The 4d CS theory for a 1-form $\omega$ with poles at $p_1$ and $p_2$ and zeros at $z^+$ and $z^-$. At the poles $p_i$, one must impose boundary conditions for the gauge field $A$, such as \eqref{eq:bcgauge}, leading to edge modes localised at those points. At the zeros $z^{\pm}$, the Lax connection \eqref{LfromJ} can develop poles.
     For $C=\mathbb{CP}^{1}$, there exists an additional gauge symmetry \eqref{eq:Lv}, which allows one to set one of the edge modes to $0$.}\label{fig:4dcsdisorder}
    \end{figure}
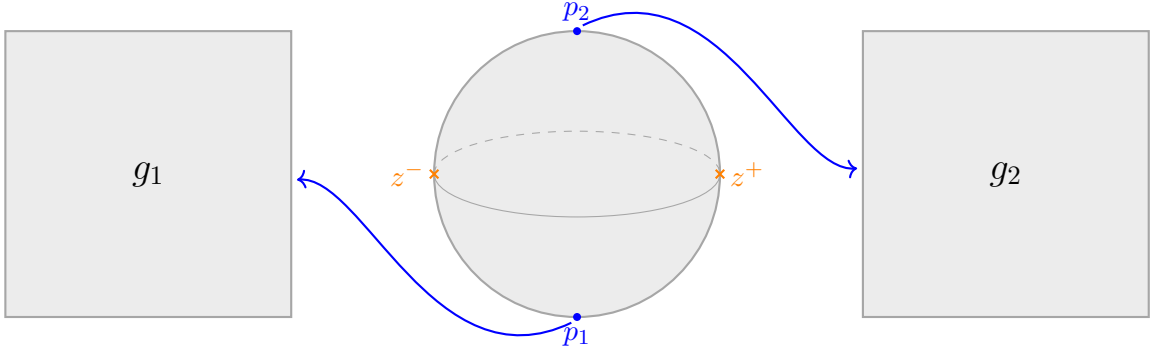

	We next turn to disorder defects. 
	Unlike order defects, we do not add new degrees of freedom by hand. Instead, the resulting two-dimensional theory is determined entirely by the meromorphic one-form $\omega$; more precisely, by the positions and multiplicities of the poles and zeros of $\omega$. See Figure \ref{fig:4dcsdisorder}.
    
	Let us again take $C=\mathbb{CP}^1$ for simplicity and write $\omega =\varphi(z)\dd z$. The function $\varphi(z)$ will later be identified with the twist function \eqref{eq:twistfunction}, so we already refer to it by that name.
	We denote the poles of $\varphi(z)$ by ${\mathbb{P}}=\left\{p_r\right\}_{r=1}^N$ and split the set of zeros into two groups ${\mathbb{Z}}^+=\left\{z^+_i\right\}_{i=1}^{N^+}$ and ${\mathbb{Z}}^-=\left\{z^-_i\right\}_{i=1}^{N^-}$. The reason for this split will become clear below: the two sets will correspond to the points at which the Lax connection $L_{\pm}$ becomes singular, respectively. The poles and the zeros need not be simple, and we denote their multiplicities (i.e.\ order) by $m_r,m_i^+,m_i^-\in\mathbb{N}$. Assuming that none of the zeros and poles of $\omega$ lie at $z=\infty$, which can always be achieved by a Möbius transformation of $z$, the twist function is fixed up to an overall constant $K$
	\begin{equation}\label{eq:disorder-omega}
		\omega=\varphi(z)\dd z=K\frac{\prod_{i=1}^{N^+}(z-z_i^+)^{m_i^+}\prod_{i=1}^{N^-}(z-z_i^-)^{m_i^-}}{\prod_{r=1}^N(z-p_r)^{m_r}}\dd z\,.
	\end{equation}
	Thus, once the numbers of poles and zeros and their multiplicities are specified, $\varphi(z)$ is parametrized by $(K,p_r,z_i^\pm)$. 
	Alternatively, $\varphi(z)$ can be characterised by its poles and the higher-order residues $\ell_{r,k}$ at these poles
	\begin{equation}\label{eq:principalpart}
		\varphi(z)=\sum_{r=1}^N\sum_{k=0}^{m_r-1}\frac{\ell_{r,k}}{(z-p_r)^{k+1}}\,.
	\end{equation}
	Below, we restrict to the case in which ${\mathbb Z}^+$ and
	${\mathbb Z}^-$ contain an equal number of zeros, counted with
	multiplicity. This condition ensures Lorentz invariance of the resulting
	two-dimensional theory, although it can be relaxed in more general
	constructions. Since we assume that $z=\infty$ is neither a pole nor a zero of
	$\omega$, this also fixes the total pole multiplicity
	\begin{equation}
		\sum_{i=1}^{N_+}m_i^+=\sum_{i=1}^{N_-}m_i^-=-1+\frac{1}{2}\sum_{r=1}^{N}m_r\,.\label{sumM}
	\end{equation}

	Let us next explain how the Lax connection and the effective 2d action arise for disorder defects.
	The first point is that, unlike order defects, the two-dimensional degrees of freedom are not added by hand, instead arising as {\it edge modes} of the four-dimensional gauge field.  To see this, consider the boundary term in the variation of the 4d CS action \eqref{eq:delta4DCS}
		\begin{align}\label{eq:bounddisord}
			\int_{ \Sigma \times C}\dd\omega\wedge\left<A,\delta A\right>\,.
		\end{align}
		Although $\omega$ is meromorphic, its exterior derivative is nonzero as a distribution. Writing $\varphi(z)$ in terms of its poles \eqref{eq:principalpart} and using \eqref{eq:dzbonz}, one finds
		\begin{equation}
			\dd \omega=2\pi i\sum_{r=1}^{N}\sum_{k=0}^{m_r-1}\frac{(-1)^{k+1}\ell_{r,k}}{k!}\partial^k_z\delta^{(2)}(z-p_r)\,\dd z\wedge\dd \bar{z}\,,
		\end{equation}
		leading to the boundary contribution in the variation of the action localized
		to the poles $\mathbb{P}=\left\{p_r\right\}_{r=1}^N$
		\begin{equation}
			\frac{i}{4\pi}\int_{\Sigma\times C}\dd\omega\wedge\left<A,\delta A\right>=\frac{1}{4}\sum_{r=1}^N\sum_{k=0}^{m_r-1}\frac{\ell_{r,k}}{k!}\int_{\Sigma\times \{p_r\}}\partial^k_z\left<A,\delta A\right>\,.\label{eq:boundcond}
		\end{equation}
	For the variational problem to be well-defined, we must impose boundary conditions on $A$ that make this term vanish. For a given one-form
    $\omega$ and gauge group $G$, there may be several admissible choices, and
    different choices lead to different two-dimensional theories.\footnote{We
    will see such examples in \cref{ex:etapcm}. Theories
    arising from the same $\omega$ and $G$ but from different admissible
    boundary conditions are often related by Poisson--Lie $T$-duality \cite{Vicedo:2015pna,Klimcik:2015gba,Hoare:2015gda,Sfetsos:2015nya}.} The general classification of such boundary conditions is somewhat involved; we refer to \cite{Benini:2020skc,Lacroix:2020flf} for the detailed analysis and \cite{Lacroix:2021iit} for a concise review.  For simplicity, below we focus on the specific case of $m_r=2$, although much of the discussion extends to more general cases, see \cite{Lacroix:2021iit}. In this case, a simple admissible boundary condition is
	\begin{align}\label{eq:bcgauge}
		\left. A_{\pm} \right|_{z=p_r}
		= \left. \delta A_{\pm} \right|_{z=p_r}=0 \,.
	\end{align}
	Once this condition is imposed, the four-dimensional gauge symmetry is
    restricted: allowed gauge transformations must preserve \eqref{eq:bcgauge}.
    Consequently, $A_{\bar z}$ cannot be completely gauged away at the poles $z=p_r$.
    Writing
    \begin{align}
    \left. A_{\bar z}\right|_{z=p_r}
    =
    -\partial_{\bar z}g_r\,g_r^{-1},
    \end{align}
    we identify the group-valued fields $g_r$ as edge modes. These are genuine dynamical degrees of freedom of the corresponding 2d integrable field theory.

    To extract the Lax connection, we follow the discussion around \eqref{eq:4Dgaugetr}; we use the shift gauge symmetry to set $A_z$ to zero, and perform a {\it formal} four-dimensional gauge transformation \eqref{eq:4Dgaugetr}
		\begin{equation}
			A\mapsto A^{\hat g^{-1}},
			\qquad
			\left.\hat g\right|_{z=p_r}=g_r \,,
		\end{equation}
    to set $A_{\bar{z}}=0$. This transformation does not preserve the boundary condition \eqref{eq:bcgauge}, and hence is not an allowed gauge transformation of the original four-dimensional problem.
		Nevertheless, it is a useful change of variables since it sets $A_{\bar{z}}^{\hat{g}^{-1}}=0$. We then identify the Lax connection as
		\begin{equation}
			L_\pm := A_\pm^{\hat g^{-1}} .
		\end{equation}
		The boundary condition \eqref{eq:bcgauge} translates to the following condition on the Lax connection:
		\begin{equation}\label{eq:LaxBC}
			L_{\pm}\big|_{z=p_r}= A_{\pm}^{\hat g^{-1}}\big|_{z=p_r}=g_r^{-1}\partial_{\pm} g_r\,.\end{equation} 
        Thanks to $A_{\bar{z}}^{\hat{g}^{-1}}=0$,\footnote{As mentioned, the gauge transformation $A\mapsto A^{\hat{g}^{-1}}$ violates the boundary condition at $z=p_r$, but it is a perfectly fine transformation away from the poles. Thus, the Lax equations still hold at a generic point on $C$.} the earlier derivations of the Lax equations \eqref{Lax4d} are still applicable.
		 In particular, the equation \eqref{eq:LaxMero} implies that \(L_\pm\) can have poles only at zeros of the twist function \(\varphi(z)\), with orders bounded by the multiplicities of those zeros. As mentioned before, we allow $L_{+}$ to have poles only at the zeros in $\mathbb{Z}^{+}$ and $L_{-}$ to have poles only at the zeros in $\mathbb{Z}^{-}$. This condition comes from the requirement that the resulting two-dimensional theory has a finite action, although it can be relaxed in some cases.\footnote{More general pole structures may also lead to finite two-dimensional actions. Examples include the \(T\bar T\)-deformation \cite{Smirnov:2016lqw,Cavaglia:2016oda}, the root-\(T\bar T\) deformation \cite{Babaei-Aghbolagh:2022uij,Conti:2022egv,Ferko:2022cix,Babaei-Aghbolagh:2022leo,Rodriguez:2021tcz,Borsato:2022tmu}, and integrable auxiliary-field deformations \cite{Ferko:2024ali}. Their 4d CS descriptions have been studied in \cite{Fukushima:2024nxm,Sakamoto:2025hwi,Fukushima:2025tlj}.}

        We then write the Lax connection in the form \cite{Liniado:2023uoo}
		\begin{align}\label{LfromJ}
			L_{\pm}&=\sum_{i=1}^{N_\pm}\sum_{n=0}^{m_i^{\pm}-1}\frac{J_{\pm}^{(i,n)}}{(z-z_i^\pm)^{n+1}}
			+J_{\pm}^{\infty}\,,
		\end{align}
        The coefficients $J_{\pm}^{(i,n)}$ and $J_{\pm}^\infty$ are not new degrees of freedom; they are uniquely\footnote{Even in more general cases of $m_r\neq 2$, one can verify that there are precisely the same number of boundary conditions as the number of currents $J$'s. Thus, the currents are uniquely fixed in terms of the edge modes.} determined in terms of the edge modes $g_r$ by the boundary conditions \eqref{eq:LaxBC}. After this substitution, the remaining four-dimensional equation \eqref{eq:LaxFlat} becomes the equation of motion for the edge modes.\footnote{This is
        the essential difference from the order-defect construction. For order defects,
        one begins with explicit two-dimensional fields, and their equations of motion
        imply the flatness of the Lax connection. For disorder defects, the logic is
        reversed: the edge modes arise from the restricted gauge symmetry, and the
        flatness of the Lax connection supplies their equations of motion.}
	
	\medskip
	
	Finally, substituting \eqref{LfromJ} back into the 4d CS action and integrating over $C$ gives a two-dimensional action on $\Sigma$
	\begin{equation}\label{eq:2dactionabs}
		\text{S}_{\rm 2d}[\phi_i]=\int_{\Sigma}\dd^2 x\Big(\int_{\mathbb{CP}^1}S_{\rm CS_4}[A[g_r]]\Big)\,,
	\end{equation}
	where $g_r$ are the edge modes. For the $m_r=2$ cases that we have been discussing, it was found in \cite{Delduc:2019whp} that the action \eqref{eq:2dactionabs} takes the following universal form (The case of higher-order poles ($m_r>2$) is discussed in \cite{Liniado:2023uoo}.)
	\begin{equation}
		\text{S}_{\text{2d}}\left[\left\{g_r\right\}_{r=1}^{N-1}\right]=\frac{1}{2}\int_{\Sigma}\sum_{r=1}^{N-1}\left<\text{res}\left[\omega\wedge L\right]_{z=p_r},g_r^{-1}\dd g_r\right>-\frac{1}{2}\sum_{r=1}^{N-1}\ell_{r,0}\,\text{S}_{\text{WZ}}[g_r]\,,\label{2Dfromdis}
	\end{equation}
	where $L=L_{+}\dd x^{+}+L_{-}\dd x^{-}$ and $\text{S}_{\text{WZ}}$ is the Wess-Zumino action with the residue $\ell_{r,0}$ playing the role of the corresponding level. 
    
    The equations of motion derived from \eqref{2Dfromdis} are equivalent to the flatness of the Lax connection. Moreover, the same action determines the Poisson brackets of the Lax connection. From these brackets one can read off the classical \(r\)-matrix, and thereby verify that the twist function \eqref{eq:twistfunction} is precisely the function $\varphi(z)$ appearing in the meromorphic one-form
	$\omega =\varphi(z) \dd z$. 
	
	\subsubsection*{Example: Principal chiral model}
	Let us illustrate the general construction with the PCM. We choose $\omega=\varphi(z) \dd z$ where $\varphi(z)$ is the twist function of PCM \eqref{eq:rPCM}. Explicitly,
	\begin{align}
		\omega=\frac{h^2-z^2}{2\,z^2}\dd z\,,\qquad h=\text{const}\,.\label{eq:wPCM}
	\end{align}
     This differs slightly from the general discussion above since $z=\infty$ is allowed to be a pole. We keep this convention because it is a standard one for the PCM. Equivalently, one could apply a Möbius transformation to move this pole to a finite point. without changing the underlying 2d action.
    
	The 1-form $\omega$ has two zeroes at $z=\pm h$, which we assign to $\mathbb{Z}^+$ and $\mathbb{Z}^-$ respectively. It also has double poles at $z=0$ and $z=\infty$, both with vanishing residues, at which we impose the simple boundary condition \eqref{eq:bcgauge}
    \begin{align}\label{eq:bc-pcm}
    \left.A_\pm\right|_{z=0}=A_\pm|_{z=\infty}=0\,,\qquad \Leftrightarrow \qquad L_{\pm}\big|_{z=0}=g_0^{-1}\partial_{\pm}g_0\,,\quad L_{\pm}\big|_{z=\infty}=g_{\infty}^{-1}\partial_{\pm}g_{\infty}\,.
    \end{align}
    Here  $g_0,g_{\infty}\in G$ are two corresponding edge modes, i.e.\ the degrees of freedom in $A_{\bar{z}}$ that cannot be gauged away:
    $
    \left.A_{\bar{z}}\right|_{z=0}=-\partial_{\bar{z}}g_0 \,g_0^{-1}\,,\quad\left.A_{\bar{z}}\right|_{z=\infty}=-\partial_{\bar{z}}g_\infty \,g_\infty^{-1}\,.
    $

    Following the general discussion, we write $L_{\pm}={J_{\pm}^{0}}/{(z\mp h)}+J^{\infty}_{\pm}$ and solve for $J^{0,\infty}_{\pm}$ using the boundary condition \eqref{eq:bc-pcm}. We then obtain
    \begin{align}
    L_{\pm}=\frac{h(g_{0}^{-1}\partial_{\pm}g_{0}-g_{\infty}^{-1}\partial_{\pm}g_{\infty})}{h\mp z}+g_{\infty}^{-1}\partial_{\pm}g_{\infty}\,,
    \end{align}
 	Substituting this solution into the universal disorder-defect action \eqref{2Dfromdis} gives
	\begin{align}
		\text{S}[g]
		=
		\frac{h}{2}
		\int_\Sigma
		\left\langle
		g^{-1}\partial_+g,
		g^{-1}\partial_-g
		\right\rangle
		\dd x^+\wedge \dd x^-,
	\end{align}
	where $g=g_0\,g_\infty^{-1}$.
	This is precisely the action of the PCM \eqref{eq:SCPM}.
	Note that although the construction begins with two edge modes, $g_0$ and
	$g_\infty$, the two-dimensional theory depends only on their relative value. One of them is therefore redundant and should be regarded as a residual
	two-dimensional gauge degree of freedom.\footnote{The existence of this redundancy is common to all integrable theories obtained from the 4d CS theory on $C=\mathbb{CP}^1$. For the elliptic cases, all the edge modes become physical degrees of freedom in two dimensions.} This redundancy is not the original
	four-dimensional gauge symmetry, which has already been taken care of. 
    
    The redundancy is visible at the level of the Lax connection. Performing a gauge transformation $v$: $\Sigma \to \mathfrak{g}$, the Lax connection transforms as
    \begin{align}
    \partial_\pm + L_\pm
    \longmapsto
    v^{-1}
    \left(
        \partial_\pm + L_\pm
    \right)
    v\,.\label{eq:Lv}
    \end{align}
    As can be easily verified, this shifts the value of $J_{\pm}^{\infty}$. Thus one can gauge fix this redundancy by setting $v=g_{\infty}^{-1}$ to obtain the standard PCM Lax connection
    \begin{align}
    L_\pm
    =
    \frac{
        h\,g^{-1}\partial_\pm g
    }{
        h\mp z
    } .
    \label{eq:pcm-lax}
    \end{align}
    	
	\medskip
	
	Just like for order defects, it is possible to consider deformations and generalisations of the PCM, either by changing the Riemann surface $C$, the 1-form $\omega$, or the boundary conditions, which will all be integrable by construction.
	
	\section{Time-dependent integrability from the 4d CS theory}\label{sec:timedep4dCS}
	We now extend four-dimensional Chern--Simons theory so that it also produces
    time-dependent integrable models. To keep the discussion explicit, we focus on the rational case, $C=\mathbb{CP}^1$. As we will see, the analysis of time dependence closely parallels that of the
    one-loop RG flow. We therefore expect the trigonometric and elliptic cases to admit analogous treatments, following the RG flow analyses for disorder defects in \cite{Lacroix:2024wrd,Lacroix:2025ias} and for order defects in Appendices \ref{app:ClRG} and \ref{app:elliptic}.

    The general strategy is the following. We first allow the parameters of the 4d CS theory (the location of the order defects/the meromorphic one-form) to depend on both space and time. Consistency of the four-dimensional theory then imposes constraints on this dependence. In the fully spacetime-dependent setting, these constraints still leave a large space of admissible deformations. The situation becomes much more rigid when the physical parameters are required to depend only on time. In that case, the 4d CS structure fixes the allowed time dependence uniquely, and the result coincides with the one-loop RG flow. 

    For this reason, before introducing the spacetime-dependent version of the
    4d CS theory, we first review how RG flow acts on the parameters of the 4d CS theory.
	
	\subsection{One-loop RG flow from the 4d CS theory}\label{sec:RGflow}
	\subsubsection*{RG flow of order defects}
    Let us begin with theories obtained from order defects.
	A systematic analysis of their RG flow intrinsically in the 4d CS language does not appear to have been carried out previously (see, however, \cite{AY} for recent work). However, the effective two-dimensional action \eqref{eq:2daction-order} is a current-current deformation, so its one-loop RG flow can be computed using standard methods \cite{Gerganov:2000mt}. 
    
    In Appendix \ref{app:RG_order}, we perform this computation for rational integrable order defects in full generality, and also discuss representative trigonometric and elliptic examples. Recall that, for order defects, the coupling constants of the resulting two-dimensional theories are the relative positions of chiral and antichiral defects, $z^+_i-z_j^-$. The analysis in Appendix \ref{app:RG_order} then strongly supports the following conjecture for the one-loop RG flow:
	\begin{equation}
		\frac{\dd}{\dd\tRG} (z_i^+-z_j^-)=c_G\,.\label{eq:RGorderdef}
	\end{equation}
    This indeed reproduces the RG flow of the GN model \eqref{eq:RGGN} under the identification \eqref{eq:hfromzGN}. For the elliptic case $C=\mathbb{T}_\tau$, there is one additional parameter,
    the complex structure modulus $\tau$. We conjecture that $\tau$ does not flow under the one-loop RG.
	
	\subsubsection*{RG flow of disorder defects}
    For disorder defects, the coupling constants of the resulting two-dimensional theories are encoded in the meromorphic $1$-form $\omega=\varphi(z) \dd z$. The beauty of the 4d CS formulation is that the RG flow of this data can be written in a compact geometric form \cite{Delduc:2020vxy}. This was first obtained for one-forms with at most double poles in \cite{Delduc:2020vxy} and later established in full generality in \cite{levine_universal_2023,Levine:2023wvt,Lacroix:2024wrd,Lacroix:2025ias}. The result is
	\begin{equation}
		\frac{\dd\varphi(z)}{\dd \tRG}=\frac{\dd \Psi(z)}{\dd z}\,.\label{eq:phiflow}
	\end{equation}
    $\Psi(z)$ is determined entirely by the twist function $\varphi(z)$.
    More precisely, it is the unique meromorphic function with the same pole locations and pole orders as $\varphi(z)$, and with the following behaviour at the zeros of $\varphi$
	\cite{Lacroix:2025ias}:
	\begin{equation}
		\begin{cases}\Psi(z_i^\pm)=\mp 2c_G
			\\
			\partial^k_z\Psi(z^\pm_i)=0\,, \qquad k=1,\dots,m_i^\pm-1\label{eq:Psiat0}
		\end{cases}
		\,.
	\end{equation}
    The RG flow is thus completely fixed by the data of the $1$-form $\omega$.\footnote{Precisely speaking, the boundary conditions used to cancel the boundary term \eqref{eq:boundcond} are also part of the data and must flow consistently with $\omega$. We refer to \cite{Lacroix:2025ias} for the full treatment of this point.}
    
	For later purposes, it will be useful to rewrite $\Psi$ in the form
	\begin{equation}\label{eq:Psi-fphi}
		\Psi(z)=-f(z)\varphi(z)\,,
	\end{equation}
    where $f(z)$ is regular at the poles of $\varphi$, ${\mathbb{P}}=\left\{p_r\right\}_{r=1}^N$, and has poles only at the
    zeros of $\varphi$, $\mathbb{Z}=\left\{z_i^{\pm}\right\}_{i=1}^{N_{\pm}}$.
	We further decompose $f$ into two pieces,
	\begin{align}\label{eq:fphi-fpm}
		f(z)={f_+(z)-f_{-}(z)}\,,\quad\,c_G\varphi(z)^{-1}={f_+(z)+f_{-}(z)}\,.
	\end{align}
	As we shall see, $f_\pm$ are essentially the functions that will appear in the deformed derivatives $\hat{\partial}_\pm$ in the Lax pair for the time-dependent case.
    
    For $C=\mathbb{CP}^1$, these functions can be written explicitly as
	\begin{equation}
		\label{eq:fRG}
		\begin{split}
			f_+(z)=c_G\sum_{i=1}^{N_+}\sum_{k=0}^{m_i^+-1}\frac{c_+^{(i,k)}}{(z-z^+_i)^{k+1}}+c_+^{(0,0)}+c^{(0,1)}_+z+c^{(0,2)}_+ z^2\,\\     f_-(z)=c_G\sum_{i=1}^{N_-}\sum_{k=0}^{m_i^--1}\frac{c_-^{(i,k)}}{(z-z^-_i)^{k+1}}+c_-^{(0,0)}+c^{(0,1)}_-z+c^{(0,2)}_- z^2
		\end{split} 
	\end{equation}
	where $c^{(i,k)}_\pm$ are rational functions of the locations of the poles and zeros $p_r$ and $z_i^\pm$; their exact expressions will not be needed in the following discussion.
    The polynomial terms, parametrized by $c_\pm^{(0,k)}$, encode the freedom to perform an RG-scale-dependent Möbius transformation of the spectral parameter and are therefore physically redundant.
    Expanding both sides of \eqref{eq:phiflow} near the poles and zeros of $\varphi(z)$ gives the one-loop flow of the location of the poles, zeroes, and overall normalization of the twist function; see \cite{Delduc:2020vxy} for details. In trigonometric and elliptic models, the rational kernels above are replaced by the corresponding trigonometric or elliptic functions
    \cite{Lacroix:2024wrd}.

	\subsection{Order defects}\label{sec:time-order-defect}
        
        We now consider the theories obtained by coupling the 4d CS theory to order surface defects (\ref{eq:2daction-order}), and ask how they can be made spacetime-dependent while preserving classical integrability.

		We start from the same 4d-2d action \eqref{eq:orddefS}, but allow the parameters defining the construction to vary over the two-dimensional spacetime $\Sigma$. In the order-defect setup, these parameters are the choice of the curve $C$, the meromorphic one-form $\omega$, and the defect positions $z_i^\pm$. First, the choice of the curve is fixed once we fix the theory; in the following discussions, we mostly consider rational integrable models, i.e.~$C=\mathbb{CP}^1$. Next, recall that, for $\mathbb{CP}^1$ without disorder defects, $\omega$ is required to be regular at finite $z$ and to have a double pole at $z=\infty$. The most general such one-form can be written as $\omega=\dd (a(t,x) \,z+b(t,x))$. A spacetime-dependent Möbius transformation $a(t,x)z+b(t,x)\rightarrow z$ brings this back to the standard form $\omega=\dd z$ without affecting the physics.\footnote{When $\Sigma$ is compact, such a Möbius transformation may be incompatible with the periodicity of the coordinates $(t,x)$. It is possible to extend our discussion to such cases, but we will not discuss it explicitly in what follows.} Thus, without loss of generality, we can also keep $\omega$ fixed. Therefore, all nontrivial spacetime dependence is carried by the defect positions $z^\pm_i$. 

  \begin{figure}[t]
            \centering
                  \begin{tikzpicture}
        
    \newcommand{\R}{0.125\textwidth}   
    
    \filldraw[color=gray!70, fill=gray!15, very thick] ({-4*\R},-\R) rectangle ({-2*\R},\R);

    \draw[red,->, very thick] ({-3*\R},{-0.75*\R})--({-3.35*\R},{-0.4*\R});
    \draw[green,->, very thick] ({-3*\R},{-0.75*\R})--({-2.65*\R},{-0.4*\R});
    \node[red,anchor=east] at ({-3.17*\R},{-0.65*\R}) {$x^-$};
    \node[green,anchor=west] at ({-2.83*\R},{-0.65*\R}) {$x^+$};
    \node[gray!70,font=\Huge] at ({-1.5*\R},0){$\times$};
    
    \filldraw[color=gray!70,fill=gray!15,thick] (0,0) circle (\R);
    
    \draw[gray!70, dashed] (-\R,0) arc [start angle=180, end angle=0, x radius=\R, y radius={0.3*\R}];
    
    \draw[gray!70] (-\R,0) arc [start angle=180, end angle=360, x radius=\R, y radius={0.3*\R}];
    
    
    \foreach \x/\y in {
        0/1
    } {
        \fill[blue] (\x*\R, \y*\R) circle (1.5pt);
    }
    
    \foreach \x/\y in {{0.94}/{0.34},
        {-0.94}/{0.34}%
    } {
        \draw[orange, thick] ({\x*\R - 0.03*\R}, {\y*\R - 0.03*\R}) -- ({\x*\R + 0.03*\R}, {\y*\R + 0.03*\R});
        \draw[orange, thick] ({\x*\R - 0.03*\R}, {\y*\R + 0.03*\R}) -- ({\x*\R + 0.03*\R}, {\y*\R - 0.03*\R});
    }
    
    \draw[green,thick,->] ({1.02*\R*cos(22)},{1.02*\R*sin(22)}) arc [start angle=22, end angle=40, radius={1.02*\R}];
    \draw[red,thick,->] ({-1.02*\R*cos(22)},{1.02*\R*sin(22)}) arc [start angle=158, end angle=140, radius={1.02*\R}];

    \node[left, orange] at ({-1.05*\R*cos(45)},{\R*sin(45)}) {\footnotesize$\hat{z}^-\sim ax^-+b^-$};
    \node[right, orange] at ({1.05*\R*cos(45)+0.1*\R},{\R*sin(45)}) {\footnotesize$\hat{z}^-\sim ax^-+b^-$};
    \node[above, blue] at (0,\R) {$\infty $};
    
    \node[] at ({-3*\R},1.5*\R) {\LARGE $\Sigma$};
    \node[] at (0,1.5*\R) {\LARGE $C$};
    \node[] at ({-1.5*\R},1.5*\R) {\LARGE $\times$};
\end{tikzpicture}
            \caption{Time-dependent order defect. This figure shows the case with one chiral defect located at $\hat{z}^+$ and one antichiral defect at $\hat{z}^-$. The location of the chiral order defect is a function of $x^+$, while the location of the antichiral order defect is a function of $x^-$.}
            \label{fig:ordertime}
        \end{figure}
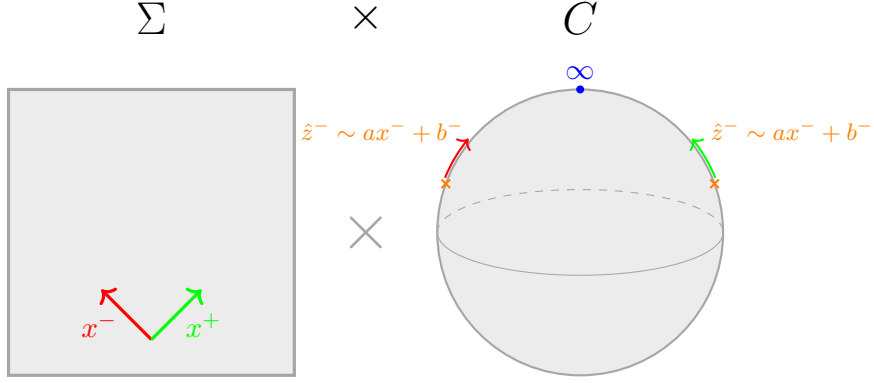
	
        Making $z^{\pm}_i\mapsto \hat{z}^\pm_i(t,x)$ spacetime-dependent, the action  now takes the form
		\begin{equation}
			\begin{split}
				\hat{\text{S}}_{\text{4d}}[A,\phi_\pm^i]=&\int_{\Sigma\times C} \dd z\wedge \text{CS}[A]
				\\
				+
				\sum_{i=1}^{N_+} 
				&\int_{\Sigma \times C}\Big(\mathcal{L}_+^i\left[\phi_{+}^i\right]\dd x^+\wedge \dd x^- 
				+\left< A, J_{+}^i\,\dd x^+\,\right>
				\Big)\wedge \updelta\Big(z-\hat{z}_i^+\Big)\\
				+\sum_{i=1}^{N_-} 
				&\int_{\Sigma \times C} 
				\Big(\mathcal{L}^{i}_- 
				\left[ \phi^{i}_{-} \right]\dd x^+\wedge \dd x^-
				+ \left<A, J_-^{i}\,\dd x^-\,\right>
				\Big)\wedge\updelta\Big(z-\hat{z}_i^-\Big)\,,\label{eq:timeorddefS}
			\end{split}
		\end{equation}
        where $\mathcal{L}^i$ are the two-dimensional defect Lagrangians and $\updelta(z-\hat{z}_{i}^{\pm})$ denotes a distribution\footnote{This distribution is the \textit{Poincaré dual} of the 2d surface defined by $z-\hat{z}_i=0$. More generally, the dual of the surface $(f(x),\bar{f}(x))=(0,0)$, is given by $\updelta\Big(f(x)\Big)=\delta^{(2)}\Big(f(x)\Big)\dd f\wedge\dd\bar{f}$.} localised at the location of the spacetime-dependent surface defect, defined by
		\begin{equation}
			\updelta\Big(z-\hat{z}_i^\pm(t,x)\Big)=\delta^{(2)}\Big(z-\hat{z}_i^\pm(t,x)\Big)\Big[\dd(z-\hat{z}_i^\pm(t,x))\wedge \dd(\bar{z}-\hat{\bar{z}}_i^\pm(t,x))\Big]\,.\label{eq:deltaform}
		\end{equation}
		 The main constraint on the allowed spacetime-dependence comes from the shift gauge symmetry of the 4d CS theory,
       $A \longmapsto A+\chi(x,z)\,\dd z $. For fixed defects, this symmetry is automatic. For spacetime-dependent defects, however, the $\updelta(z-\hat{z}^{\pm}_i)$ contains derivatives of $\hat z_i^\pm$, and the action changes by
      \begin{equation}
			\begin{split}
				\hat{\text{S}}_{\text{4d}}[A+\chi\,\dd z,\phi_\pm^i]
                &= \hat{\text{S}}_{\text{4d}}[A,\phi^i_\pm]
				\\
				+\sum_{i=1}^{N_-}&\int_{\Sigma\times C}\partial_- \hat{z}_i^+\left<\chi,J_+^i\right>\delta^{(2)}(z-\hat{z}_i^+)\dd z\wedge \dd x^+\wedge\dd x^-\wedge \dd\bar{z}
				\\
				+\sum_{i=1}^{N_-}&\int_{\Sigma\times C}\partial_+ \hat{z}_i^-\left<\chi,J_-^i\right>\delta^{(2)}(z-\hat{z}_i^-)\dd z\wedge \dd x^-\wedge\dd x^+\wedge \dd\bar{z}\,.
			\end{split}
		\end{equation}
        Thus, in order to preserve the shift gauge symmetry, we need to impose
		\begin{equation}
			\partial_\pm \hat{z}_i^\mp=0\,.\label{eq:OrderTimeReq}
		\end{equation}
		In other words, chiral defect positions can depend only on $x^+$, while antichiral defect positions can depend only on $x^-$. This condition is the only additional requirement needed for the spacetime-dependent generalisation of order defects, see Figure \ref{fig:ordertime}.
      		
		  When the condition \eqref{eq:OrderTimeReq} is satisfied, one can take the axial-like gauge
		\begin{equation}
			(A_+,A_-,A_z,A_{\bar{z}})\rightarrow (\hat{L}_+,\hat{L}_-,0,0)\,,
		\end{equation}
		and set the boundary condition $\hat{L}_\pm\lvert_{z=\infty}=0$ as in the time-independent case. The equations of motion then read
		\begin{equation}
			\partial_{\bar{z}}\hat{L}_{\pm}(z)=\pm 2\pi i\sum_{i=1}^{N_{\pm}}\delta^{(2)}(z-\hat{z}_i^{\pm})J_{\pm}^i\,,
			\label{disordmodtime}
		\end{equation}
		where the only change from the time-independent case is that the arguments of $\delta$-functions are spacetime-dependent. The solution to this equation also takes the same form as in the time-independent case
		\begin{align}\label{eq:lax-order-time}
			\hat{L}_\pm^a(z,t,x)=\mp\sum_{i=1}^{N_\pm}J_{\pm}^{i,b}(t,x)\,r_{ab}(z-\hat{z}^\pm_i(x^{\pm}))\,,
		\end{align}
		and the resulting action is given by
		\begin{align}\label{eq:2daction-torder}
			\hat{\text{S}}_{\text{2d}}[\phi^i_{\pm}]&=\int_{\Sigma}~\biggl[~\sum_{i=1}^{N_+}\mathcal{L}^i_+\left[\phi^i_{+}\right]+\sum_{i=1}^{N_-}\mathcal{L}^i_-\left[\phi^i_{-}\right]-\sum_{i=1}^{N_+}\sum_{j=1}^{N_-}J_+^{i,a}r_{ab}(\hat{z}^+_i-\hat{z}^-_j)J_-^{j,b}\biggr]\dd x^+\wedge \dd x^-\,.
		\end{align}
         The effect of the deformation is thus simply to promote the couplings of the current-current interaction to differences of chiral and antichiral functions.
        Since the model still arises from the 4d CS theory, its integrable structure is expected to survive. This can also be checked directly. The Lax connection \eqref{eq:lax-order-time} satisfies the same flatness equation \eqref{eq:LaxEq}, provided the equations of motion of \eqref{eq:2daction-torder} hold. This is because the partial derivatives $\partial_\pm$ in the Lax equation only act on the currents, thanks to $\partial_\pm \hat{z}_i^\mp=0$. 
		
		Having found the most general spacetime dependence of the parameters, we can now specialize to the purely time-dependent case. Since the coupling constants are given by
		\begin{equation}
			h_{ij}(t,x)=\frac{1}{\hat{z}_i^+(x^{+})-\hat{z}_j^-(x^{-})}\,,
		\end{equation}
        we must require
		\begin{equation}
			\hat{z}^\pm_i= \pm \frac{a}{2}\,x^\pm\pm b_i^\pm\label{eq:ztimedep}
			\quad \Rightarrow \quad h_{ij}(t)=\frac{1}{a\,t+b_i^++b_j^-}\,,
		\end{equation}
        to make the couplings depend only on time. Here $a$ and $b_i^\pm$ are arbitrary constants. This is precisely the same evolution as the one-loop RG flow in \eqref{eq:RGorderdef}, up to the normalization of the time variable, and is pictured in Fig. \ref{fig:ordertime}.
		
		\subsection{Disorder defects}\label{sec:4dCS-disorder-time}
		We now turn to disorder defects. Compared with order defects, the spacetime-dependent generalization is more intricate. For disorder defects, the two-dimensional theory is not specified by defect fields and their positions, but by the geometric data of the 4d CS theory itself. With $C=\mathbb{CP}^1$ fixed, relevant couplings are encoded in the meromorphic one-form $\omega$.
		
		We therefore allow the one-form to depend on $\Sigma$, while keeping it meromorphic in the spectral parameter $z$ (that is, it has neither dependence nor legs along $\bar{z}$). The most general form of such $\hat{\omega}$ is
		\begin{align}
			\hat{\omega}&=\hat{\varphi}(z,t,x)\,\dd z+\hat{\Psi}_+(z,t,x)\,\dd x^+
            + \hat{\Psi}_-(z,t,x)\,\dd x^-\no\\
            &=\hat{\varphi}\left[\dd z-\hat{f}_+\,\dd x^++\hat{f}_-\,\dd x^-\right]\,,
		\end{align}
		where we defined $\hat{f}_\pm=\mp \hat{\varphi}^{-1}\Psi_\pm$. We choose the function $\hat\varphi$ to have the same structure of poles and zeros as in the time-independent theory \eqref{eq:disorder-omega}, but with spacetime-dependent parameters
		\begin{equation}
			\hat{\varphi}(z,t,x)=\hat{K}(t,x)\frac{\prod_{i=1}^{N_+}(z-\hat{z}_i^+(t,x))^{m_i^+}\prod_{i=1}^{N_-}(z-\hat{z}_i^-(t,x))^{m_i^-}}{\prod_{r=1}^N(z-\hat{p}_r(t,x))^{m_r}}\,.\label{eq;phitime}
		\end{equation}
        We keep the multiplicities $m_i^\pm$ and $m_r$ fixed, since they are discrete
        data.\footnote{This assumption can fail when zeros or poles collide. Such degenerations are precisely the kind of coordinate singularities discussed in Section \ref{sec:timedep}, and can often be removed by a suitable change of coordinates.} See Figure \ref{fig:4dcsdisordertime} for an example.         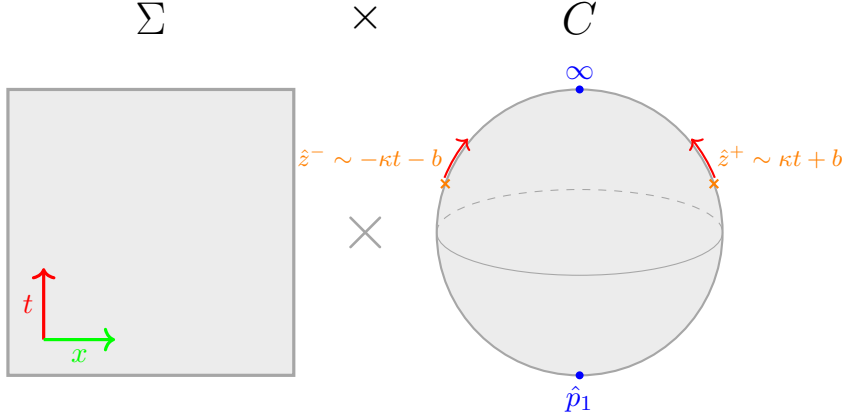
\begin{figure}[t]
            \centering
      \begin{tikzpicture}
        
    \newcommand{\R}{0.125\textwidth}   
    
    \filldraw[color=gray!70, fill=gray!15, very thick] ({-4*\R},-\R) rectangle ({-2*\R},\R);

    \draw[red,->, very thick] ({-3.75*\R},{-0.75*\R})--({-3.75*\R},{-0.25*\R});
    \draw[green,->, very thick] ({-3.75*\R},{-0.75*\R})--({-3.25*\R},{-0.75*\R});
    \node[red,anchor=east] at ({-3.75*\R},{-0.5*\R}) {$t$};
    \node[green,anchor=north] at ({-3.5*\R},{-0.75*\R}) {$x$};
    \node[gray!70,font=\Huge] at ({-1.5*\R},0){$\times$};
    
    \filldraw[color=gray!70,fill=gray!15,thick] (0,0) circle (\R);
    
    \draw[gray!70, dashed] (-\R,0) arc [start angle=180, end angle=0, x radius=\R, y radius={0.3*\R}];
    
    \draw[gray!70] (-\R,0) arc [start angle=180, end angle=360, x radius=\R, y radius={0.3*\R}];
       
    
    \foreach \x/\y in {
        0/1
        ,0/-1%
    } {
        \fill[blue] (\x*\R, \y*\R) circle (1.5pt);
    }
    
    \foreach \x/\y in {{0.94}/{0.34},
        {-0.94}/{0.34}%
    } {
        \draw[orange, thick] ({\x*\R - 0.03*\R}, {\y*\R - 0.03*\R}) -- ({\x*\R + 0.03*\R}, {\y*\R + 0.03*\R});
        \draw[orange, thick] ({\x*\R - 0.03*\R}, {\y*\R + 0.03*\R}) -- ({\x*\R + 0.03*\R}, {\y*\R - 0.03*\R});
    }
    
    \draw[red,thick,->] ({1.02*\R*cos(22)},{1.02*\R*sin(22)}) arc [start angle=22, end angle=40, radius={1.02*\R}];
    \draw[red,thick,->] ({-1.02*\R*cos(22)},{1.02*\R*sin(22)}) arc [start angle=158, end angle=140, radius={1.02*\R}];

    \node[left, orange] at ({-1.05*\R*cos(32)},{\R*sin(32)}) {\footnotesize$\hat{z}^-\sim -\kappa t-b$};
    \node[right, orange] at ({1.05*\R*cos(32)},{\R*sin(32)}) {\footnotesize$\hat{z}^+\sim \kappa t+b$};
    \node[below, blue] at (0,-\R) {$\hat{p}_1$};
    \node[above, blue] at (0,\R) {$\infty $};
    
    \node[] at ({-3*\R},1.5*\R) {\LARGE $\Sigma$};
    \node[] at (0,1.5*\R) {\LARGE $C$};
    \node[] at ({-1.5*\R},1.5*\R) {\LARGE $\times$};
\end{tikzpicture}
            \caption{Time-dependent disorder defect. This figure shows the case in which the one-form $\omega$ has two double
poles, $\hat p_1$ and $\hat p_2=\infty$, and two simple zeros, $\hat z^+$ and
$\hat z^-$. This is the configuration that gives the
time-dependent PCM \eqref{eq:SCPM}. In general, both the poles and the zeros move with time, but in the present example, the two poles can be kept fixed by a time-dependent Möbius transformation \eqref{eq:Mobius}.}
            \label{fig:4dcsdisordertime}
        \end{figure}
		
        For the moment, $\hat\Psi_\pm$, or equivalently $\hat f_\pm$, are arbitrary meromorphic functions. We will see below that the consistency of the 4d CS theory imposes various constraints on them. Starting with the spacetime-dependent action
		\begin{equation}
			\hat{\text{S}}_{\text{CS}_4}[A]=\frac{i}{4\pi}\int_{\Sigma\times\mathbb{CP}^1}\Big[\hat{\varphi}\,\dd z+\hat \Psi_+\,\dd x^++\hat \Psi_-\,\dd x^-\Big]\wedge\text{CS}[A]\,,\label{eq:disordertimeact}
		\end{equation}
		and varying it, we obtain
		\begin{equation}
			\label{eq:delta4DCStime}
			\delta \hat{\text{S}}_{\text{CS}_4}[A]=\frac{i}{2\pi}\int_{ \Sigma \times \mathbb{CP}^1}\hat{\omega}\wedge \left<\text{F}[A],\delta A\right>-\frac{i}{4\pi}\int_{ \Sigma \times \mathbb{CP}^1}\dd\hat{\omega}\wedge\left<A,\delta A\right>\,.
		\end{equation}
		As before, one needs to impose appropriate boundary conditions to make the second term vanish at the poles of $\hat{\omega}$. The new feature is that $\hat\omega$ has components along $\Sigma$, so $\dd\hat\omega$ may be nonzero even away from the poles of $\hat\varphi$. In fact, by straightforward computation, we get the following away from the poles:
		\begin{equation}
			\dd\hat{\omega}=(\partial_+\hat{\varphi}-\partial_z\hat{\Psi}_+)\dd x^+\wedge \dd z+(\partial_-\hat{\varphi}-\partial_z\hat{\Psi}_-)\dd x^-\wedge \dd z+(\partial_+\hat{\Psi}_--\partial_-\hat \Psi_+)\dd x^+\wedge\dd x^-\,.
		\end{equation}
		To make it vanish we need to impose
		\begin{equation}
			\partial_\pm\hat{\varphi}=\partial_z\hat{\Psi}_\pm\,,\qquad\partial_+\hat{\Psi}_-=\partial_-\hat{\Psi}_+\,.\label{eq:dphidpsi}
		\end{equation}
		These equations determine the allowed spacetime dependence of $\hat\varphi$, and hence the flows of $\hat K$, $\hat z_i^\pm$, and $\hat p_r$, in terms of $\hat\Psi_\pm$. In particular, using \eqref{eq;phitime} and the definition of $\hat f_\pm$, the pole positions obey
		\begin{equation}
			\frac{\dd\hat{p}_r}{\dd x^\pm}=\pm\hat{f}_\pm(\hat{p}_r)\,.\label{eq:dpdt}
		\end{equation}
        There is one further regularity requirement. The functions $\hat\Psi_\pm$ must not introduce new poles away from the poles of $\hat\varphi$, and their pole orders at these points must not exceed those of $\hat\varphi$. Otherwise, $\dd\hat\omega$ would contain additional localized contributions, forcing extra boundary conditions and overdetermining the gauge field in terms of the edge modes. Thus the singular part of $\dd\hat\omega$ should remain localized only to $\hat{\mathbb P}=\{\hat p_r\}$. With this restriction, $\dd\hat\omega$ takes the form
		\begin{equation}
			\begin{split}
				\dd\hat{\omega}=&\partial_{\bar{z}}\hat{\varphi}\,\dd \bar{z}\wedge\dd z+\partial_{\bar{z}}\hat{\Psi}_+\,\dd\bar{z}\wedge\dd x^++\partial_{\bar{z}}\hat{\Psi}_-\,\dd\bar{z}\wedge\dd x^-
				\\
				&+\partial_{+}\hat{\bar{p}}_r\frac{\dd\hat\varphi}{\dd\hat{\bar{p}}_r}\,\dd x^+\wedge \dd z+\partial_{-}\hat{\bar{p}}_r\frac{\dd\hat\varphi}{\dd\hat{\bar{p}}_r}\,\dd x^-\wedge \dd z
				\\
				&+\partial_-\hat{\bar{p}}_r\frac{\dd\hat\Psi_+}{\dd\hat{\bar{p}}_r}\,\dd x^-\wedge\dd x^++\partial_+\hat{\bar{p}}_r\frac{\dd\hat\Psi_-}{\dd\hat{\bar{p}}_r}\,\dd x^+\wedge\dd x^-\,.\label{eq:dwhat}
			\end{split}
		\end{equation}
		Let us explain the origin of the various terms in this equation. The first line is the familiar distributional contribution from taking anti-holomorphic derivatives of meromorphic functions. The remaining terms come from the fact that, when the pole positions move over $\Sigma$, $\dd\hat\omega$ must be understood as a \textit{measure derivative}; namely, derivatives with respect to $x^\pm$, ${\dd\hat \omega}/{\dd x^\pm}$, also include derivatives with respect to the complex conjugate of pole positions $\hat{\bar p}_r$, which vanish away from the poles but produce localized contributions.\footnote{Consider the function $f(z,p)=1/(z-p(\lambda))$, where $\lambda$ is an external parameter and $p$ is a holomorphic coordinate. When taking its derivative with respect to $\lambda$, we get $\frac{\dd}{\dd\lambda} f(z,p)=(\partial_\lambda p)(\partial_p f)+(\partial_\lambda \bar{p})(\partial_{\bar{p}}f)=(\partial_\lambda p)/(z-p)^2+(\partial_\lambda \bar{p})\,2\pi i\delta(z-p)$.} We refer to \cite{Lacroix:2025ias} for a detailed discussion of this measure derivative.
        
	Combining these contributions and using \eqref{eq:dpdt}, the boundary term in the variation of the action can be written compactly as
	\begin{equation}
		\begin{split}\label{eq:timebound}
			\frac{i}{4\pi}&\int_{\Sigma\times\mathbb{CP}^1}\dd\hat \omega\wedge\left<A,\delta A\right>=\frac{1}{4}\sum_{r=1}^N\sum_{k=0}^{m_r-1}\frac{\hat \ell_{r,k}}{k!}\int_{\Sigma\times\mathbb{CP}^1}\partial^k_z\left<A,\delta A\right>\wedge\updelta(z-\hat{p}_r)\,,
		\end{split}
	\end{equation}
	where $\updelta(z-\hat{p}_r)$ is defined analogously to \eqref{eq:deltaform}
	\begin{equation}\label{eq:updelta}
		\updelta(z-\hat{p}_r)=\delta^{(2)}(z-\hat{p}_r)\,\dd(z-\hat{p}_r)\wedge\dd(\bar{z}-\hat{\bar{p}}_r)
	\end{equation}
	To see explicitly that this reproduces the localized terms above, we can expand
	\begin{equation}
		\begin{split}
			\dd(z-\hat{p}_r)\wedge\dd(\bar{z}-\hat{\bar{p}}_r)
			=&\dd z\wedge\dd\bar{z}-\partial_+\hat{p}_r\,\dd x^+\wedge\dd\bar{z}-\partial_-\hat{p}_r\,\dd x^-\wedge\dd \bar{z}
			\\
			&+\partial_+\hat{\bar{p}}_r\,\dd x^+\wedge\dd z+\partial_-\hat{\bar{p}}_r\,\dd x^-\wedge\dd z
			\\
			&+\partial_+\hat{p}_r\partial_-\hat{\bar{p}}_r\,\dd x^+\wedge\dd x^-+\partial_-\hat{p}_r\partial_+\hat{\bar{p}}_r\,\dd x^-\wedge\dd x^+\,.
		\end{split}
	\end{equation}
	This matches the corresponding contributions in \eqref{eq:dwhat} term by term. As in the spacetime-independent theory, one must impose boundary conditions that make \eqref{eq:timebound} vanish. This is a delicate problem in general, now with the additional feature that the boundary term itself depends on $\Sigma$. Nevertheless, for one-forms with only double poles, we can choose the simple boundary condition 
    \begin{equation}
    A_\pm(\hat p_r)=0,
    \quad
    \text{at}\quad \forall (x^+,x^-)\in \Sigma\,.
    \end{equation}
     The examples in Section \ref{sec:examples} will use
    time-independent boundary conditions of this kind. More generally, however, the boundary conditions may themselves have to vary over $\Sigma$.\footnote{For example, they may depend explicitly on a set of spacetime-dependent parameters.} This can be analyzed in close analogy with the RG-scale-dependent boundary conditions of \cite{Lacroix:2025ias}.

    Once the boundary term has been cancelled, the bulk equations of motion are $
    \hat\omega\wedge \text{F}[A]=0 $. The action \eqref{eq:disordertimeact} is invariant under ordinary gauge transformations and under the shift
\begin{equation}
    A
    \longmapsto
    A+\chi
    \left(
        \dd z-\hat f_+\,\dd x^+
        +\hat f_-\,\dd x^-
    \right).
\end{equation}
Using these symmetries, we choose the axial gauge $(A_+,A_-,A_z,A_{\bar z})
    =
    (\hat L_+,\hat L_-,0,0)$ as in the time-independent case. In this gauge, the equations of motion become
\begin{subequations}
\begin{align}
    \hat\varphi
    \left[
        \left(\partial_+ +\hat f_+\partial_z\right)\hat L_-
        -
        \left(\partial_- -\hat f_-\partial_z\right)\hat L_+
        +
        [\hat L_+,\hat L_-]
    \right]
    &=
    0,
    \label{eq:flatness}
    \\
    \hat\varphi\,\partial_{\bar z}\hat L_\pm
    &=
    0,
    \label{eq:pzAp}
    \\
    \hat\Psi_+\,\partial_{\bar z}\hat L_-
    -
    \hat\Psi_-\,\partial_{\bar z}\hat L_+
    &=
    0 .
    \label{eq:pzA}
\end{align}
\end{subequations}

Let us spell out the meaning of these equations. First, introduce the deformed derivatives
\begin{equation}
    \hat\partial_\pm
    =
    \partial_\pm
    \pm
    \hat f_\pm\,\partial_z .
    \label{eq:dder}
\end{equation}
Then \eqref{eq:flatness} becomes
\begin{equation}
    \hat\varphi
    \left(
        \hat\partial_+\hat L_-
        -
        \hat\partial_-\hat L_+
        +
        [\hat L_+,\hat L_-]
    \right)
    =
    0 .
    \label{eq:BZ-flat}
\end{equation}
This is precisely the time-dependent Lax equation introduced in
\eqref{eq:Laxtime}. Moreover, it can be written as a flatness condition
$[\hat D_+,\hat D_-]=0$, because the deformed derivatives commute. Indeed,
\begin{equation}
\begin{split}
    [\hat\partial_+,\hat\partial_-]
    =
    \hat\varphi^{-2}
    \big[
        &
        \hat\Psi_-
        \big(
            \partial_+\hat\varphi
            -
            \partial_z\hat\Psi_+
        \big)
        -
        \hat\Psi_+
        \big(
            \partial_-\hat\varphi
            -
            \partial_z\hat\Psi_-
        \big)
        \\
        &
        +
        \hat\varphi
        \big(
            \partial_-\hat\Psi_+
            -
            \partial_+\hat\Psi_-
        \big)
    \big]\partial_z , \end{split}
\label{eq:der-comm}
\end{equation}
which vanishes precisely by \eqref{eq:dphidpsi}.

The second equation, \eqref{eq:pzAp}, has the same interpretation as in the
time-independent theory: $\hat L_\pm$ may have poles only at the zeros of
$\hat\varphi$. As before, we focus on the setups where these zeros are divided into two sets
$\hat{\mathbb Z}^\pm$ and the poles of $\hat L_\pm$ lie only in
$\hat{\mathbb Z}^\pm$ respectively. Then, general solutions to \eqref{eq:pzAp} are given by
\begin{equation}\label{LfromJtime}
			L_{\pm}(z,t,x)=\sum_{i=1}^{N_\pm}\sum_{n=0}^{m_i^{\pm}-1}\frac{J_{\pm}^{(i,n)}(t,x)}{(z-\hat z_i^\pm(t,x))^{n+1}}
			+J_{\pm}^{\infty}(t,x)\,,
		\end{equation}
        This depends explicitly on the spacetime couplings $\hat{z}_i^\pm$ and implicitly on the remaining spacetime-dependent parameters $\hat {p}_r$ and $\hat{K}$ through the solution of the boundary conditions \eqref{eq:timebound}.
        
        The third equation, \eqref{eq:pzA}, imposes additional constraints on the holomorphic dependence of $\hat\Psi_\pm $. It is then satisfied iff the following condition for $\Psi_{\pm}$ holds
		\begin{equation}
			\partial_z^k\,\hat \Psi_\pm(\hat z_i^\mp)=0\,,\quad k=0,\dots,m_i^\mp-1\,.\label{eq:Psionz}
		\end{equation}
        
		We can now integrate over the auxiliary curve to obtain the effective two-dimensional action. The reduction proceeds as in the spacetime-independent case. When $\hat\omega$ has at most double poles, the result is \begin{equation} \widehat S_{\rm 2d} \left[\{g_r\}_{r=1}^{N-1}\right] = \frac{1}{2} \int_{\Sigma} \sum_{r=1}^{N-1} \left\langle \operatorname*{res}_{z=\hat p_r} \left(\hat\omega\wedge \hat L\right), g_r^{-1}\dd g_r \right\rangle - \frac{1}{2} \sum_{r=1}^{N-1} \ell_{r,0}\, S_{\rm WZ}[g_r]\,. \label{2Dfromdist} \end{equation} Here we have used the fact that \eqref{eq:dphidpsi}, together with the rationality of $\hat\Psi_\pm$, implies that the Wess--Zumino levels $\ell_{r,0}$ are independent of spacetime. 
        
        Let us summarize the construction. We started with the most general spacetime-dependent meromorphic one-form \begin{equation} 
            \hat\omega = \hat\varphi(z,x^+,x^-)\,\dd z + \hat\Psi_+(z,x^+,x^-)\,\dd x^+ + \hat\Psi_-(z,x^+,x^-)\,\dd x^-\,. 
        \end{equation} 
        It defines a spacetime-dependent integrable field theory provided its components obey \begin{equation} \begin{split} &\partial_\pm\hat\varphi = \partial_z\hat\Psi_\pm, \qquad \partial_+\hat\Psi_- = \partial_-\hat\Psi_+, \\ & \partial_z^k\hat\varphi(\hat z_i^\pm) = 0, \qquad \partial_z^k\hat\Psi_\mp(\hat z_i^\pm) = 0, \qquad k=0,\dots,m_i^\pm-1\,. 
        \end{split} \label{eq:condonSigmadep} \end{equation} 
        The first line guarantees that $\dd\hat\omega$ has no unwanted bulk contribution, while the second line ensures that the pole structure of the Lax connection is compatible with the zeros of $\hat\varphi$. 
        
        Let us now specialize to the purely time-dependent case. The above conditions then force the one-form to take the form \begin{equation} \hat\omega = \hat\varphi(t,z)\,\dd z + \hat\Psi(t,z)\,\dd t + a\,\dd x, \end{equation} where $a$ is a constant. The time dependence of $\hat\varphi$ is then fixed by \begin{equation} \partial_t\hat\varphi = \partial_z\hat\Psi\,, \qquad \hat\Psi(\hat z_i^\pm) = \mp a\,, \qquad \partial_z^k\hat\Psi(\hat z_i^\pm) = 0\,, \qquad k=1,\dots,m_i^\pm-1 \,, \label{eq:condonpsihat} \end{equation} together with the condition that all poles of $\hat\Psi$ must lie at the poles $\hat{\mathbb P}$ of $\hat\varphi$. This is precisely the same structure that appeared in the one-loop RG flow \eqref{eq:phiflow}. There, the RG-scale dependence was governed by a meromorphic function $\Psi$ satisfying the conditions \eqref{eq:Psiat0}. The conditions on $\hat\Psi$ in \eqref{eq:condonpsihat} are identical, after the replacement
        $c_G \to a$, which can be achieved by a simple rescaling of time. This proves our central claim: the purely time-dependent deformation is governed by the same geometric data as the one-loop RG flow, with the time variable playing the role of the RG scale up to normalization. For the PCM, the time-dependent deformation is pictured in Fig. \ref{fig:4dcsdisordertime}.

       \subsection{Characterising the flow by periods}
	The conditions \eqref{eq:condonSigmadep} determine the most general spacetime dependence of the disorder-defect data. However, in terms of the positions of poles and zeros of $\omega$, it is not immediately transparent how restrictive these conditions are. It turns out that, when all zeros of $\omega$ are simple, there is a more natural set of coordinates on parameter space: the periods of $\omega$.
	
	Consider a meromorphic one-form $\omega$ with $N$ poles of arbitrary order and with $N_+ + N_-$ simple zeros, with $N_+$ of those being poles of $L_+$ and $N_-$ being poles of $L_-$. The key observation is that the positions of these poles and zeros of $\omega$, which parametrise $\omega$, can be traded for its absolute and relative {\it periods} of $\omega$ \cite{derryberry_lax_2021,Lacroix:2024wrd}, defined by
	\begin{equation}
		\Pi_{r}=\oint_{\mathcal{C}_r}\omega\,,\qquad\Pi_{i\pm}^{j\pm}=\int_{z^\pm_i}^{z^{\pm}_j}\omega\,,\label{eq:periods}
	\end{equation}
	where $\mathcal{C}_r$ is a closed contour around the pole $p_r$. 
    The signs in $\Pi_{i\pm}^{j\pm}$ may be chosen independently, so one may also consider periods between a zero in ${\mathbb Z}^+$ and a zero in ${\mathbb Z}^-$. Since $\dd\omega=0$ away from the poles, these periods are independent of continuous deformations of the integration contours (up to the addition of absolute periods).
      
	The periods are not all independent. The sum of the residues of a meromorphic one-form on $\mathbb{CP}^1$ vanishes, which gives
	\begin{equation}
		\sum_{r=1}^N\Pi_r=0\,.
	\end{equation}
	 Similarly, relative periods obey the concatenation relation
	\begin{equation}
		\Pi_{i\pm}^{j\pm}=\Pi_{i\pm}^{k\pm}+\Pi_{k\pm}^{j\pm}\,.\label{eq:concat}
	\end{equation}
	This allows us to choose $N+N_++N_--2$ independent periods, for example
	\begin{equation}
		{\Pi}_{\text{ind}}=(\Pi_r,\Pi_{1+}^{i+},\Pi_{1-}^{j-},\Pi_{1+}^{1-})\,,\quad r=2,\dots,N\,,\quad i=2,\dots,N_+\,,\quad j=2,\dots,N_-\,.\label{eq:indPi}
	\end{equation}
	These are in one-to-one correspondence with the Möbius-invariant parameters in $\omega$ \eqref{eq:disorder-omega}.
	
	The advantage of these coordinates is that they linearize the one-loop RG flow. As conjectured in \cite{derryberry_lax_2021} and later proven to be equivalent to \eqref{eq:phiflow} in \cite{Lacroix:2024wrd} for rational, trigonometric, and elliptic models, the periods evolve under the RG as
	\begin{equation}
		\frac{\dd\Pi_{r}}{\dd\tRG}=\frac{\dd\Pi_{i+}^{j+}}{\dd\tRG}=\frac{\dd\Pi_{i-}^{j-}}{\dd\tRG}=0\,,\qquad\quad\frac{\dd\Pi^{j-}_{i+}}{\dd\tRG}=-\frac{\dd\Pi^{j+}_{i-}}{\dd\tRG}=2c_G\,,\quad \forall r,i,j\,.\label{eq:periodflow}
	\end{equation}
	Equivalently, in the independent coordinates \eqref{eq:indPi}, the flow is simply
	\begin{equation}
		\frac{\dd{\Pi}_{\text{ind}}}{\dd\tRG}=(0,0,0,2c_{G})\,.
	\end{equation}
	Thus, the period parametrization turns the generally complicated one-loop $\beta$-functions of the integrable field theory into a constant linear flow. This gives a particularly transparent description of the disorder-defect RG flow.
\begin{figure}[t]
	    \centering
	    \begin{tikzpicture}
	\newcommand{\R}{0.185*\textwidth}       
	\def\tilt{10}     
	\def\lat{70}      
	\def\offset{30}
	\def\rotb{\offset}
	\def\rotg{\offset+120}
	\def\rotr{\offset+240}
	
	\pgfmathsetlengthmacro{\rxy}{\R*cos(\lat)}
	\pgfmathsetlengthmacro{\rz}{\R*cos(\lat)*sin(\tilt)}
	\pgfmathsetlengthmacro{\yc}{\R*sin(\lat)*cos(\tilt)}
	\pgfmathsetlengthmacro{\eqY}{\R*sin(\tilt)}
	
	\filldraw[color=gray!70,fill=gray!15,thick] (0,0) circle (\R);
	
	\draw[magenta!50,thick,->] ({0.95*\R},-0.03) arc [start angle=-15, end angle=-165, x radius={0.99*\R}, y radius=\eqY];
	
	\draw[cyan!50,dashed,thick] (\rxy,\yc) arc [start angle=0, end angle=180, x radius=\rxy, y radius=\rz];
	\draw[cyan!50,thick] (\rxy,\yc) arc [start angle=0, end angle=-180, x radius=\rxy, y radius=\rz];
	
	\draw[cyan!50,dashed,thick] (\rxy,-\yc) arc [start angle=0, end angle=180, x radius=\rxy, y radius=\rz];
	\draw[cyan!50,thick] (\rxy,-\yc) arc [start angle=0, end angle=-180, x radius=\rxy, y radius=\rz];
	
	\fill[blue] (0, \R) circle (1.5pt); 
	\fill[blue] (0, -\R) circle (1.5pt); 
	
    \draw[orange, thick](-\R-0.03*\R, -0.03*\R) -- (-\R+0.03*\R, 0.03*\R);
    \draw[orange, thick] (-\R-0.03*\R, 0.03*\R) -- (-\R+0.03*\R,-0.03*\R);

    \draw[orange, thick](\R-0.03*\R, -0.03*\R) -- (\R+0.03*\R, 0.03*\R);
    \draw[orange, thick] (\R-0.03*\R, 0.03*\R) -- (\R+0.03*\R,-0.03*\R);
	
	\node[left, orange] at (-\R, 0) {$z^-$};
	\node[right, orange] at (\R, 0) {$z^+$};
	
	\node[below, blue] at (0, -\R) {$p_1$}; 
	\node[above, blue] at (0, \R) {$p_2$};  
	
	\node[above,magenta] at (0,-0.3) {$\Pi_+^-\sim a\,t+b$};
	\node[above right,cyan] at (0.1*\R,\R) {$\Pi_2\sim\text{const.}$};
	\node[below right,cyan] at (0.1*\R,-\R) {$\Pi_1\sim\text{const.}$};

    \draw[red,thick,->] ({1.02*\R*cos(2)},{1.02*\R*sin(2)}) arc (2:20:{\R*1.02});
	\draw[red,thick,->] ({-1.02*\R*cos(2)},{1.02*\R*sin(2)}) arc (178:160:{\R*1.02});
\end{tikzpicture}
	    \caption{The periods \eqref{eq:periods} of a disorder defect and their time dependence. The absolute periods $\Pi_{1,2}$ are in cyan while the relative period $\Pi_{+}^{-}$ is in magenta. The latter is time-dependent, both through its dependence on $\omega$ and on the time-dependent position of the zeros $z^\pm$.}
	    \label{fig:periods}
	\end{figure}
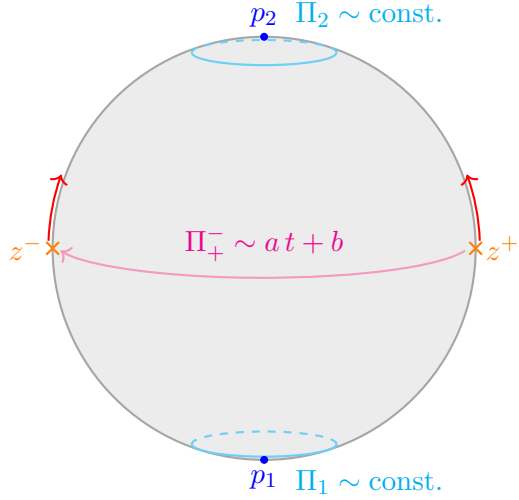
    
	Let us add two comments. First, the period description extends naturally to trigonometric and elliptic disorder defects discussed in Appendices \ref{app:ClRG} and \ref{app:elliptic}. In those cases, there are additional absolute periods, obtained by integrating $\omega$ along nontrivial cycles of $C$. Like the absolute periods above, these do not flow under the one-loop RG flow.\footnote{Unlike in the order-defect case discussed below, this does not imply that the modulus $\tau$ is fixed in the elliptic case.} Second, order defects can be studied in the same language. Their flow again takes the form \eqref{eq:periodflow}, with $z_i^\pm$ now interpreted as the positions of the order defects, or equivalently as the poles of $L_\pm$. For $\omega=\dd z$, the absolute periods vanish for $C=\mathbb{CP}^1$ while they are related to the modulus $\tau$ for $C=\mathbb{T}_\tau$. The relative periods are simply 
    \begin{equation} \Pi_{i\pm}^{j\pm} = z_j^\pm-z_i^\pm . 
    \end{equation}
    In this form, the period flow conjecture \eqref{eq:periodflow} states that $\tau$ does not flow, while the mixed relative periods flow linearly, reproducing \eqref{eq:RGorderdef} up to the normalization convention for the RG time.
	
	The simplicity of the period flow suggests that periods should also provide a useful description of the spacetime-dependent deformations (cf.\ Figure \ref{fig:periods}). From \eqref{eq:dphidpsi}, one finds
	\begin{equation}
		\partial_{\pm}\hat\Pi_{r}=0\,,\quad \partial_{+}\hat\Pi_{i\pm}^{j\pm}=\hat\Psi_+\Big(\hat z^{{\pm}}_j\Big)-\hat\Psi_+\Big(\hat z^{{\pm}}_i\Big)\,,\quad \partial_{-}\hat \Pi_{i\pm}^{j\pm}=\hat \Psi_-\Big(\hat z^{\pm}_j\Big)-\hat \Psi_-\Big(\hat z^{{\pm}}_i\Big)\,.
	\end{equation}
	Using \eqref{eq:Psionz} together with $\partial_+\hat\Psi_-=\partial_-\hat\Psi_+$, this can be simplified to
	\begin{equation}
		\partial_{\pm}\hat{\Pi}_{r}=\partial_-\hat{\Pi}_{i+}^{j_+}=\partial_+\hat{\Pi}_{i-}^{j-}=0\,,\quad \partial_+\partial_-\hat{\Pi}_{i+}^{j-}=0\,.\label{eq:periodflowtime}
	\end{equation}
	Thus the absolute periods $\hat \Pi_r$ are constants, $\hat \Pi_{i+}^{j+}(x_+)$ are chiral functions and $\hat \Pi_{i-}^{j-}(x_-)$ are antichiral while $\hat{\Pi}_{i+}^{j-}$ are the sum of a chiral and an antichiral function. The number of independent functions becomes most transparent in terms of the independent periods \eqref{eq:indPi}. They can be written as
	\begin{equation}
		\hat{{\Pi}}_{\text{ind}}(x^+,x^-)=\Big(\hat Z_r,\hat Z^+_{i}(x^+),\hat Z^-_{j}(x^-),\hat  Z^+_{1}(x^+)+\hat Z_{1}^-(x^-)\Big)\,,\label{eq:vufree}
	\end{equation}
	where $\hat Z_r$ are constants, while $\hat Z_i^+$ and $\hat Z_j^-$ are arbitrary chiral and antichiral functions, respectively. Hence, the general spacetime dependence is controlled by $N_+$ chiral functions and $N_-$ antichiral functions. This is the same pattern found for order defects, where the corresponding functions are simply the positions of the poles of $\hat L_\pm$ in \eqref{eq:OrderTimeReq}. For disorder defects, the relation is less direct: the natural variables are not the zero positions themselves, but the periods of $\omega$. If we require the theory to depend only on time and not on space, the period flow reduces to
	\begin{equation}
		\partial_t\hat \Pi_{\text{ind}}=(0,0,0,a)\label{eq:Piindflow}
	\end{equation}

    The discussions above were somewhat abstract. To illustrate it in a concrete example, let us consider the PCM, whose time-independent version corresponds to $\omega=\frac{h^2-z^2}{2z^2}\dd z$. Then the absolute periods around the poles at $z=0$ and $z=\infty$
    vanish. By \eqref{eq:periodflow}, this remains true for all values of $(x^+,x^-)$. There is a single relative period, $\hat\Pi_{1+}^{1-}=2\hat h$. As we saw in \eqref{eq:hfromZ}, this period is precisely the sum of a chiral and an antichiral function.
	
			\section{Examples of time-dependent integrable field theories} \label{sec:examples}
            In this section, we apply the 4d CS construction of the previous section to derive explicit two-dimensional actions and Lax connections for a range of time-dependent integrable field theories. The examples include both ultralocal and non-ultralocal models. In the non-ultralocal cases, we also show explicitly how the time-dependent one-form $\hat\omega$ is obtained from the time-independent one-form $\omega$.
  
            We begin with order defects. As a first example, we construct the time-dependent non-Abelian massless Thirring model from the 4d Chern--Simons theory. We then discuss several other ultralocal integrable theories and derive their time-dependent deformations in the same framework. None of these models seems to have appeared previously in the literature.

            We then turn to disorder defects. We first consider a class of models described by a meromorphic one-form with two simple zeros, two simple poles, and a double pole at $z=\infty$. By a choice of coordinates, the two zeros, which are the poles of $L_+$ and $L_-$, may be written as \begin{equation}
    z^+=-z^-=z_0 . \end{equation}
    We also fix the overall constant $K$ in \eqref{eq:disorder-omega} to be one. The one-form is then
    \begin{align}
    \omega
    =
    \varphi(z)\,\dd z
    =
    \frac{1}{2}
    \frac{z_0^2-z^2}{(z-p_1)(z-p_2)}
    \dd z .
    \label{eq:ex-omega}
    \end{align}
    The factor of $1/2$ is chosen so that the normalization of the one-loop RG flow
    \eqref{eq:phiflow} agrees with the conventions of \cite{Hoare:2020fye} for the models considered below. We also study degenerations in which the two simple poles collide to form a double pole.
    
    The time-dependent integrable field theories arising from this family of
    one-forms are summarized in Table \ref{table:disorder}. Finally, as a proof of
    concept, we go beyond the class \eqref{eq:ex-omega} and consider a coupled
    version of the $\mathrm{PCM}_k$. The resulting time-dependent deformation does
    not appear to have been discussed previously in the literature.
            \begin{table}
                \centering
				\small
				\renewcommand{\arraystretch}{1.6}
				\setlength{\tabcolsep}{4pt}
				\begin{tabularx}{\textwidth}{@{}
						>{\centering\arraybackslash}p{0.18\textwidth}
						>{\centering\arraybackslash}X
						>{\raggedright\arraybackslash}X
						@{}}
					\toprule
					Model
					&1-form $\omega$
					&
					Boundary Conditions
					\\
					\midrule 
					PCM+WZ (Sec.\,\ref{ex:pcmwz})
					&$
					\begin{aligned}
						\omega_{{\rm PCM}_k}
						&=\frac{1}{2}\frac{h^2-z^2}{(z-k)^2}\dd z
					\end{aligned}$
					&$
					\begin{aligned}
						A_\pm|_{z=k}&=A_\pm|_{z=\infty}=0
					\end{aligned}$
					\\
					$\eta$-deformed PCM (Sec.\,\ref{ex:etapcm})
					&$
					\begin{aligned}[t]
						\omega_\eta
						&=\frac{1}{2}\frac{K^2-z^2}{z^2+\eta^2K^2}\dd z
					\end{aligned}$
					&$
					\begin{aligned}[t]
						&(R+i)A_\pm|_{z=i\eta K}=(R-i)A_\pm|_{z=-i\eta K}\\
						&A_\pm|_{z=\infty}=0
					\end{aligned}$
					\\
					$\lambda$-deformed PCM (Sec.\,\ref{ex:etapcm})
					&\raisebox{0.1ex}{$
					\begin{aligned}[t]
						\omega_\lambda
						&=\frac{1}{2}\frac{k^2(\lambda^{-1}-\lambda)^2-z^2}{z^2-k^2\frac{(\lambda-1)^4}{\lambda^2}}\dd z
					\end{aligned}$}
					&
					$
					\begin{aligned}[t]
						&A_\pm|_{z=k\frac{(\lambda-1)^2}{\lambda}}=A_\pm|_{z=-k\frac{(\lambda-1)^2}{\lambda}}\,,\\
                        &A_\pm|_{z=\infty}=0
					\end{aligned}$
					\\
					\bottomrule
				\end{tabularx}
                   \caption{This table summarizes the meromorphic 1-forms $\omega$ and boundary conditions of the 4d CS action corresponding to three different 2d integrable field theories:\,(i)\,the PCM + Wess--Zumino term, (ii)\,the $\eta$-deformed PCM, and (iii)\,the $\lambda$-deformed PCM. The notation will be explained in the relevant sections.}
                \label{table:disorder}
            \end{table}
			
			The Lax integrability of the time-dependent PCM, more generally of symmetric-space coset models, was first established in \cite{Belinsky:1971nt} using the fixed-spectral formulation adopted in this
paper. It was later extended to integrable deformations in
\cite{Hoare:2020fye} using an alternative variable-spectral formulation
\cite{Breitenlohner:1986um}. Below, we show that the actions of these known
time-dependent two-dimensional integrable field theories, together with their
fixed-spectral Lax connections, arise directly from the 4d CS theory. The equivalence with
the variable-spectral formulation of \cite{Hoare:2020fye} will be demonstrated in
\cref{sec:BMtype}.
			
			\subsection{Order defect I: Massless non-Abelian Thirring model}
            Let us first consider a variation of the Gross--Neveu model \eqref{eq:SGN0}, called the massless non-Abelian Thirring model (NATM). Like the Gross--Neveu model, it contains fermions $\psi^n$, but there are now $2N$ of them, taken to be Dirac rather than Majorana fermions. The action is given by
            \begin{equation}
                \text{S}_{\text{NATM}}[\psi^n]=\int_\Sigma \dd ^2x \,i\bar{\psi}^n\slashed{\partial}\psi^n+\frac{h}{2}(\bar{\psi}^n\gamma_\mu T^a_{nm}\psi^m)(\bar{\psi}^n\gamma^\mu T^a_{nm}\psi^m)\,,\label{eq:SNATM0}
            \end{equation}
            where $T^a$ are the $N^2-1$ generators of $SU(N)$ in the $\mathbf{N}\oplus\mathbf{\bar{N}}$ representation. The theory has a chiral symmetry, so the fermions can be decomposed into $\psi_+$ and $\psi_-$. In terms of these fields, the action can be written as
            \begin{equation}
                \text{S}_{\text{NATM}}[\psi^n_\pm]\int_{\Sigma}\dd^2 x\,i(\psi_{+}^{n})^\dagger\partial_-\psi_+^n+i(\psi_-^n)^\dagger\partial_+\psi_-^n+h\left<J_+,J_-\right>\,,\label{eq:SNATM}
            \end{equation}
            where the currents $J_{\pm}$ are given by
            \begin{align}
            J_\pm^a =-i(\psi_\pm^n )^\dagger T^a_{nm}\psi_\pm^m\in\mathfrak{su}(N)\,.
            \end{align}
            
            Given the similarity of the actions, it is not surprising that the NATM can be realised from a 4d CS setup analogous to that of the GN model \cite{Costello:2019tri}. Consider the $SU(N)$ gauge field $A$, minimally coupled to chiral and antichiral order defects at $z^\pm$, so that the action takes the form
            \begin{equation}
            \begin{split}
                \text{S}_{\text{4d}}[A,\psi^n_\pm]=\int_{\Sigma\times \mathbb{CP}^1}\dd z\wedge\text{CS}[A]+&\psi_+^\dagger i(\partial_-+A_{-})\psi_+\,\dd x^+\wedge\dd x^-\wedge \updelta(z-z^+)
                \\
                +&\psi_-^\dagger i(\partial_++A_{+})\psi_-\,\dd x^+\wedge\dd x^-\wedge \updelta(z-z^-)\,.
            \end{split}
            \end{equation}
			To see the connection with the 2d action, we follow the steps outlined in Section \ref{sec:orderdefects}. Namely, we impose $A_{\bar{z}}=A_z=0$ and rename the remaining two gauge-field components $A_\pm=L_\pm$, whose equations of motion force the Lax connection to take the form
            \begin{equation}
                L_\pm(z)=\mp\frac{J_\pm}{z-z^\pm}\,.
            \end{equation}
            Integrating over $\mathbb{CP}^1$, we obtain the 2d action
            \begin{equation}
                \text{S}_{\text{NATM}}[\psi_\pm]=\int_{\Sigma}\dd^2 x\,i\psi_{+}^\dagger\partial_-\psi_++i\psi_-^\dagger\partial_+\psi_-+\frac{1}{z^+-z^-}\left<J_+,J_-\right>\,,
            \end{equation}
            which matches \eqref{eq:SNATM} upon identifying
            \begin{equation}
                h=\frac{1}{z^+-z^-}\,.
            \end{equation}
            The associated Lax connection is given by \eqref{eq:LGN}, now valued in $\mathfrak{su}(N)$.

            It is then straightforward to consider the corresponding time-dependent model. The two defect locations are given by
\begin{equation}
    \hat z^\pm=\pm \left(\frac{a}{2} x^\pm+b^\pm\right)\,,
    \label{eq:GNzpzm}
\end{equation}
where $a,b^+,b^-$ are constants. The time-dependent action is then
\begin{equation}
\begin{split}
    \hat{\text{S}}_{\text{4d}}[A,\psi^n_\pm]
    =
    \int_{\Sigma\times \mathbb{CP}^1}\dd z\wedge\text{CS}[A]
    &+i\psi_+^\dagger(\partial_-+A_{-})\psi_+\,
    \dd x^+\wedge\dd x^-\wedge
    \updelta\left(z-\frac{a}{2}x^+-b^+\right)
    \\
    &+i\psi_-^\dagger(\partial_++A_{+})\psi_-\,
    \dd x^+\wedge\dd x^-\wedge
    \updelta\left(z+\frac{a}{2}x^-+b^-\right)\,.
\end{split}
\end{equation}
Following the same steps as before, we obtain the time-dependent 2d action
\begin{equation}
   \hat{\text{S}}_{\text{NATM}}[{\psi_\pm}]
   =
   \int_{\Sigma}\dd^2 x\,
   i(\psi_{+}^{n})^\dagger\partial_-\psi_+^n
   +i(\psi_-^n)^\dagger\partial_+\psi_-^n
   +\frac{1}{a\,t+b^++b^-}\left<J_+,J_-\right>\,,
\end{equation}
whose Lax connection is
\begin{equation}
    \hat{L}_\pm(z)=
    \frac{J_\pm}{z\mp \left(\frac{a}{2}x^\pm+b^\pm\right)}\,.
\end{equation}
Indeed, one can readily check that the Lax equation for $\hat{L}$ follows from the equations of motion for $\psi_\pm$.
            
			\subsection{Order defect II: Faddeev-Reshetikhin model}
            
			The Faddeev-Reshetikhin (FR) model is a PCM-like model that does not suffer from non-ultralocality of the Poisson bracket \cite{Faddeev:1985qu}. The field content is given by a pair of $G$-valued fields $g_\pm$, controlled by the action
			\begin{equation}
				\text{S}[g_\pm]_{\text{FR}}=\int_\Sigma \dd^2 x\left<\Lambda,g_+^{-1}\partial_-g_++g_-^{-1}\partial_+g_-\right>+\frac{1}{\nu}\left<J_+,J_-\right>\,,\quad J_\pm=g_\pm \Lambda g_{\pm}^{-1}\,.\label{eq:SFR}
			\end{equation}
			Here, $\Lambda\in \mathfrak{g}$ is a fixed element of the Lie algebra and $\nu\in\mathbb{R}$ is a coupling constant. The equations of motion can be written purely in terms of $J$ and take the form
			\begin{equation}
				\partial_+J_-+\partial_-J_+=\partial_+J_--\partial_-J_++\frac{2}{\nu}[J_+,J_-]=0\,.
			\end{equation}
			Given the similarity to the Maurer-Cartan identity and equations of motion of the PCM \eqref{eq:PCMEOM}, it is not surprising that the FR-model is integrable, with essentially the same Lax connection as the PCM \eqref{eq:LPCM}
			\begin{equation}
				L_\pm(z)=\pm\frac{J_\pm}{z^\pm-z}\,,\quad \nu=z^+-z^-\,.\label{eq:nu}
			\end{equation}
			We should note, however, that there are important differences between the currents in the PCM and those of the FR-model.
			The FR-model has additional relations not obeyed by the PCM and not encoded in the flatness of the Lax connection, namely that $\Tr[(J_\pm)^{n}]$ is constant for any value of $n\in\mathbb{N}$. Furthermore, the Poisson-bracket of the currents differs, which ends up removing the non-ultralocal term in Poisson-bracket \eqref{eq:Maillet}.  

            This model can also be derived from the 4d CS theory with order defects, as shown in \cite{Caudrelier:2020xtn,Fukushima:2020tqv}. It is given by the 4d action 
            \begin{equation}
            \begin{split}
                \text{S}_{\text{4d}}[A,g_\pm]=\frac{i}{4\pi}\int_{\Sigma\times\mathbb{CP}^1}\dd z\wedge \text{CS}[A]&+\left<\Lambda,g_+^{-1}(\partial_-+A_-)g_+\right>\dd x^+\wedge \dd x^-\wedge \updelta(z-z^+)
                \\
                &+\left<\Lambda,g_-^{-1}(\partial_++A_+)g_-\right>\dd x^+\wedge \dd x^-\wedge \updelta(z-z^-)\,,
            \end{split}
            \end{equation}
            which correctly reproduces \eqref{eq:SFR} with the identification \eqref{eq:nu}. We can then generalise this to the time-dependent case, 
            such that the 4d action takes the form
           \begin{equation}
            \begin{split}
                \hat{\text{S}}_{\text{4d}}[A,g_\pm]=\frac{i}{4\pi}\int_{\Sigma\times\mathbb{CP}^1}\dd z\wedge \text{CS}[A]&+\left<\Lambda,g_+^{-1}(\partial_-+A_-)g_+\right>\dd x^+\wedge \dd x^-\wedge \updelta(z-\hat{z}^+)
                \\
                &+\left<\Lambda,g_-^{-1}(\partial_++A_+)g_-\right>\dd x^+\wedge \dd x^-\wedge \updelta(z-\hat{z}^-)\,,
            \end{split}
            \end{equation}
			where just like for the GN model and the NATM \eqref{eq:GNzpzm}, the locations of the defects are given by $\hat{z}^\pm= \pm (\frac{a}{2}\,x^\pm +b^\pm)$. To obtain the corresponding 2d action of the time-dependent FR-model, we can integrate over $\mathbb{CP}^1$, which results in the action
			\begin{equation}
				\hat{S}[g_\pm]_{\text{FR}}=\int_\Sigma \dd^2 x\,\left<\Lambda,g_+^{-1}\partial_-g_++g_-^{-1}\partial_+g_-\right>+\frac{1}{a\, t+b^++b^-}\left<J_+,J_-\right>\,.
			\end{equation}
			Thus, we see that the coupling constant $\hat{\nu}=a\,t+b^++b^-$ is linear in time.
			
			This construction can be easily generalised to the Zakharov-Mikhailov model \cite{zakharov_variational_1980}, whose time-independent action was constructed from the 4d CS theory in \cite{Caudrelier:2020xtn}.

			\subsection{Order defect III: Coupled GN model}\label{subsec:cGN}
            Lastly, we consider a more elaborate setup, namely the coupled GN model. We consider $N\times N_+$ chiral fermions and $N\times N_-$ antichiral fermions, with the action given by
			\begin{equation}
				{\text{S}}[\psi^{i,n}_\pm]_{\text{cGN}}=\sum_{i=1}^{N_+}\sum_{j=1}^{N_-}\int_{\Sigma}\dd^2x\sum_{n=1}^{N}\left(i\psi^{i,n}_+\partial_-\psi^{i,n}+i\psi_-^{j,n}\partial_+\psi_-^{j,n}\right)+h_{ij}\left<J_+^i,J_-^j\right>\,,\label{eq:cGN}
			\end{equation}
			where $J_\pm^{i,a}=-i\psi^{i,n}_\pm T^a_{nm}\psi^{i,m}_\pm$, which reduces to the regular GN-model \eqref{eq:SGN} when $N_+=N_-=1$. For arbitrary values of the couplings $h_{ij}$, the theory has an $O(N)$ symmetry. The model is known to be classically integrable, provided that the couplings $h_{ij}$ satisfy the relation \cite{Costello:2019tri}
			\begin{equation}
				\frac{1}{h_{ij}}+\frac{1}{h_{k\ell}}=\frac{1}{h_{i\ell}}+\frac{1}{h_{kj}}\,.
			\end{equation}
			When this condition is met, one can find parameters $z^+_i$ and $z^-_j$ such that\footnote{For example, one can take $z^+_i=+\frac{1}{h_{i1}}-\frac{1}{2h_{11}}$ and $z^-_j=-\frac{1}{h_{1j}}+\frac{1}{2h_{11}}$}
			\begin{equation}
				h_{ij}=\frac{1}{z_i^+-z_j^-}\,,\label{eq:hzpzm}
			\end{equation}
			and the equations of motion obtained from the action \eqref{eq:cGN} are equivalent to the flatness of the Lax connection
			\begin{equation}
				L_\pm(z)=\sum_{i=1}^{N_\pm}\frac{J^i_\pm}{z-z_\pm^i}\,.
			\end{equation}
            It is straightforward to obtain this model from the 4d CS theory. Following \eqref{fermaction1}, we add multiple order defects, each localised to a different point in $\mathbb{CP}^1$
 \begin{equation}
            \begin{split}
                \text{S}_{\text{4d}}[A,\psi^{i,n}_\pm]=\frac{i}{4\pi}\int_{\Sigma\times \mathbb{CP}^1}\dd z\wedge\text{CS}[A]+&\sum_{i=1}^{N_+}\psi_+^ii(\partial_-+A_{-})\psi^i_+\,\dd x^+\wedge\dd x^-+\wedge \updelta(z-z^+)
                \\
                +&\sum_{j=1}^{N_-}i\psi_-^j(\partial_++A_{+})\psi^j_-\,\dd x^+\wedge\dd x^-+\wedge \updelta(z-z^-)
            \end{split}
            \end{equation}
            Upon integrating over $\mathbb{CP}^1$, this correctly reproduces \eqref{eq:cGN}, with the couplings given by \eqref{eq:hzpzm}.
			
			We now generalise this to a spacetime-dependent version by promoting the locations of the chiral and antichiral defects to chiral and antichiral functions, respectively. This gives the 2d action
\begin{equation}
    \hat{\text{S}}_{\text{cGN}}\left[\psi^{i,n}_\pm\right]
    =
    \sum_{i=1}^{N_+}\sum_{j=1}^{N_-}
    \int_{\Sigma}\dd^2x\,
    \sum_{n=1}^{N}
    \left(
    i\psi^{i,n}_+\partial_-\psi^{i,n}_+
    +i\psi_-^{j,n}\partial_+\psi_-^{j,n}
    \right)
    +\hat{h}_{ij}\left<J_+^i,J_-^j\right>\,,
    \label{eq:cGN4d}
\end{equation}
where the couplings $\hat{h}_{ij}(t,x)$ are given in terms of $N_+$ chiral functions $\hat Z_i^+(x^+)$ and $N_-$ antichiral functions $\hat Z_j^-(x^-)$ by
\begin{equation}
    \hat h_{ij}(t,x)=\frac{1}{\hat Z_i^+(x^+)+\hat Z_j^-(x^-)}\,.
\end{equation}
If we require the theory to depend only on time, these functions must be restricted to
\begin{equation}
    \hat Z_i^\pm\left(x^\pm\right)=\pm \left(\frac{a}{2}\,x^\pm+ b^\pm_i\right),
\end{equation}
where $a,b^\pm_i\in\mathbb{R}$ are constants. Here $a$ is common to all defects, and the time-independent theory is recovered by taking $a\rightarrow 0$. The couplings then take the form
\begin{equation}
    \hat{h}_{ij}(t)=\frac{1}{a\,t+b^+_{i}+b^-_j}\,,
\end{equation}
so that the equations of motion follow from the flatness of the Lax connection
\begin{equation}
    \hat{L}_\pm(z)
    =
    \sum_{i=1}^{N_\pm}
    \frac{J^i_\pm}{z\mp\left(\frac{a}{2}x^\pm+ b^\pm_i\right)}\,.
\end{equation}

Taking $N=1$, this provides a realisation of the time-dependent GN-model \eqref{eq:SGNtime} from 4DCS. One can straightforwardly generalise this construction to include trigonometric and elliptic deformations as well.
			\subsection{Disorder defect I : PCM with Wess--Zumino term}\label{ex:pcmwz}
			We next turn to theories arising from disorder defects.
			We first consider the time-dependent PCM with a Wess--Zumino term at level $k$, which we denote by PCM$_k$.
			To clarify the procedure underlying our construction of time-dependent integrable field theories, we begin by reviewing the time-independent case and then explain how the construction extends to the time-dependent setting.
					
			The meromorphic 1-form $\omega_{{\rm PCM}_{k}}$ for the PCM$_k$ model is given by \cite{Costello:2019tri}
			\begin{align}
				\omega_{{\rm PCM}_{k}}=\varphi_{{\rm PCM}_{k}}(z)\dd z=\frac{1}{2}\frac{h^2-z^2}{(z-k)^2}\dd z\,,
			\end{align}
			where $h$ and $k$ are real parameters. The 1-form $\omega_{{\rm PCM}_{k}}$ has two simple zeros located at
			\begin{align}
				z = h \qquad \text{and} \qquad z = -h\,,
			\end{align}
			and it also has two double poles
			\begin{align}
				z=k \qquad \text{and} \qquad z=\infty\,.
			\end{align}
			Here, the parameter $k$ corresponds to the level of the Wess--Zumino term, and when $k=0$, the 1-form $\omega_{{\rm PCM}_{k}}$ reduces to the one for the PCM \eqref{eq:rPCM}. 
			In analogy with the PCM case, the boundary condition for PCM$_k$ is chosen to be of Dirichlet type
			\begin{align}\label{pcm-wz-bc}
				A_{\pm}|_{z=k}&=A_{\pm}|_{z=\infty}=0\,.
			\end{align}
			By solving the boundary conditions as explained in section \ref{sec:ti-disorder}, we find the Lax connection
			\begin{align}\label{pcmk:lax}
				L=\frac{h}{h-z}\left(1-\frac{k}{h}\right)j_{+}\dd x^++\frac{h}{h+ z}\left(1+\frac{k}{h}\right)j_{-}\dd x^-\,.
			\end{align}
            To obtain the 2d action corresponding to this 4d CS setup, we can plug this Lax connection into the universal action \eqref{2Dfromdis}, from where we obtain the action
			\begin{align}\label{pcmwz-action}
				\text{S}_{\text{PCM}_{k}}[g]=-\frac{h}{2}\int_{\Sigma} \dd ^2x\,\left<j_+,j_-\right>+k\,S_{\rm WZ}[g]\,,
			\end{align}
			which is precisely the action for the PCM$_k$.
					
			To introduce an appropriate time dependence into the zeros and poles of $\omega_{{\rm PCM}_k}$, we use the connection to the one-loop RG flow equation. Among the parameters of PCM$_k$, the level $k$ remains fixed, such that only the tension $h$ runs under the RG flow according to
			\begin{align}\label{pcmk-rg}
				\frac{\dd }{\dd \mathfrak{t}_{\text{RG}}}h(\mathfrak{t}_{\text{RG}})=c_{G}\left(1-\frac{k^2}{h(\mathfrak{t}_{\text{RG}})^2}\right)\,,\qquad k:\text{const}.\,.
			\end{align}
			Here we have adopted the normalization used in~\cite{Hoare:2020fye}.
			For the reader’s convenience, we briefly explain how this renormalization group equation is reproduced from the one-loop RG equation (\ref{eq:phiflow}) of the 4d CS theory. First, we write the inverse of the twist function in the form
			\begin{align}
				\varphi_{{\rm PCM}_{k}}^{-1}(z)=-\frac{(k-h)^2}{h(z-h)}+\frac{(h+k)^2}{h  (z+h)}-2\,.
			\end{align}
			From the equations (\ref{eq:Psi-fphi}) and (\ref{eq:fphi-fpm}), the scalar function $f(z)$ is taken as
			\begin{align}\label{eq:f-pcm}
				f(z)&=c_G \left(-\frac{(k-h)^2}{h(z-h)}-\frac{(h +k)^2}{h (z+h)}+\frac{2 k}{h}\right)\,,
			\end{align}
			or equivalently $f_{\pm}(z)$ are
			\begin{equation}\label{pcmk-fpm}
				f_{\pm}(z)=-\frac{c_G(z-k)(h\mp k)}{h(z\mp h)}\,.
			\end{equation}
			As already mentioned in Section \ref{sec:RGflow}, there is an ambiguity in the choices of $f(z)$ and $\varphi_{\text{PCM}_k}(z)$ due to Möbius transformations. Here, we fix this ambiguity by setting appropriate values for the parameters $c_{\pm}^{(0,1)}, c_{\pm}^{(0,2)}, c_{\pm}^{(0,3)}$ so that the RG flow equation is normalized as in \eqref{pcmk-rg}.
			Computing the product of $\varphi_{\text{PCM}_k}(z)$ and $f(z)$, we obtain
			\begin{align}\label{phi-pcmk}
				\Psi(z)=-f(z)\varphi_{\text{PCM}_k}(z)=\frac{c_G \left(k z-h ^2\right)}{h (z-k)}\,.
			\end{align}
			We then find that the flow equation (\ref{eq:phiflow}) with (\ref{phi-pcmk}) reproduces (\ref{pcmk-rg}).
			
			We now derive the action and the Lax connection of the time-dependent PCM$_k$.
			In our prescription for constructing the corresponding 4d CS action, we replace the constant parameter $h$ in the twist function with a time-dependent function $\hat{h}(t)$, and require it to satisfy the same differential equation as the RG flow, namely\footnote{In contrast, the parameter $k$ must be time-independent.}
			\begin{align}\label{eq:Rgflow-pcmk}
				\frac{\dd }{\dd t}\hat{h}(t)=a\left(1-\frac{k^2}{\hat{h}(t)^2}\right)\,.
			\end{align}
            Integrating this RG flow equation gives that the tension $\hat{h}(t)$ is a function satisfying an algebraic curve,
            \begin{align}
              \hat{h}(t)-k\tanh^{-1}\left(\frac{\hat{h}(t)}{k}\right) =a\,t+b\,,\qquad b:\text{const}.\,,\label{eq:hkflow}
            \end{align}
            which in the $k\to 0$ limit reduces to the PCM result $\hat{h}(t)=a t+b$. Then the 1-form $\hat{\omega}_{\text{PCM}_k}$ corresponding to the time-dependent $\text{PCM}_k$ is given by
			\begin{equation}\label{omega2-tPCMk}
				\begin{split}
                \hat{\omega}_{\text{PCM}_{k}}&=\hat{\varphi}_{\text{PCM}_{k}}(z,t)\,\dd z+\hat{\Psi}_{\text{PCM}_k}\,\dd t+a\,\dd x
                \\
					\hat{\varphi}_{\text{PCM}_{k}}(z,t) &=\frac{1}{2}\frac{\hat{h}(t)^2-z^2}{(z-k)^2}\,,\qquad\hat{\Psi}_{\text{PCM}_k}(z,t)=\frac{a(kz-h^2)}{h(z-k)}\,.
				\end{split}
			\end{equation}
			The remaining steps are carried out in parallel with the time-independent case. After imposing the same boundary condition (\ref{pcm-wz-bc}) and solving the equations of motion of the 4D CS theory with (\ref{omega2-tPCMk}), we obtain the Lax connection
			\begin{align}\label{pcmk:bz}
				\hat{L}=\frac{\hat{h}(t)}{\hat{h}(t)-z}\left(1-\frac{k}{\hat{h}(t)}\right)j_{+}\dd x^++\frac{\hat{h}(t)}{\hat{h}(t)+ z}\left(1+\frac{k}{\hat{h}(t)}\right)j_{-}\dd x^-\,.
			\end{align}
			In the present case, since the zeros of the twist function are time-dependent, the poles of the Lax connection also become time-dependent accordingly.
			Finally, substituting the Lax connection (\ref{pcmk:bz}) into the 4d CS action and then integrating out the spectral parameter space $\mathbb{CP}^1$, we obtain the time-dependent action 
			\begin{align}\label{pcmwzt-action}
				\hat{\text{S}}_{\text{PCM}_{k}}[g]=-\frac{1}{2}\int_{\Sigma} \dd ^2x\,\hat{h}(t)\,\left<j_+,j_-\right>+k\,S_{\rm WZ}\,.
			\end{align}
			When we set $k=0$, the resulting 2d action and Lax connection reduce to those in (\ref{eq:SCPM}) and (\ref{eq:time-pcm-lax}), respectively, for the time-dependent PCM.
			One finds that the modified Lax equation
			\begin{equation}
            \begin{split}
				&\hat{\partial}_{+}\hat{L}_--\hat{\partial}_{-}\hat{L}_++[\hat{L}_+,\hat{L}_-]=0\,,\qquad \hat{\partial}_\pm=\partial_\pm\pm \hat f_\pm(z,t)\,\partial_z\,,
                \\
					&\hat{f}_{\pm}(z,t)=\mp\frac{1}{2}\hat{\varphi}_{\text{PCM}_k}(z,t)^{-1}\left(\hat{\Psi}_{\text{PCM}_k}(z,t)\mp a\right)=-\frac{a(z-k)(\hat{h}(t)\mp k)}{\hat{h}(t)(z\mp \hat{h}(t))}
                \end{split}
			\end{equation}
			is equivalent to the equations of motion
			\begin{align}
				\partial_+(\hat{h}(t)j_-)+k\partial_+j_-+\partial_-(\hat{h}(t)j_+)-k\partial_-j_+=0\,.\label{PCMWZ-eom}
			\end{align}
			Therefore, the time-dependent system is classically integrable.
			When the scalar function $\hat{h}(t)$ approaches $\pm k$, the equations (\ref{PCMWZ-eom}) reduce to the equations of motion for the (time-independent) WZW model.
            \subsection*{Spacetime-dependence}
            Although we have chosen to focus in this Section on models that only depend on time, the construction in Section \ref{sec:timedep4dCS} is not restricted to such cases. 
            For the PCM+WZ, 
        \eqref{eq:condonSigmadep} reduces to
\begin{equation}
    \frac{\hat h\,\partial_+\partial_- \hat h}{k+\hat h}+\frac{k\,\partial_+ \hat h\,\partial_- \hat h}{(k+\hat h)^2}=\frac{\hat h\,\partial_+\partial_- \hat h}{k-\hat h}+\frac{k\,\partial_+ \hat h\,\partial_- \hat h}{(k-\hat h)^2}\,.
\end{equation}
It is straightforward to see that this reduces to $\partial_+\partial_-\hat h=0$ in the limit $k\rightarrow 0$. The solution to this differential equation is given by the curve
\begin{equation}
    \hat h-k\tanh^{-1}\left(\frac{\hat h}{k}\right)=f(x^+)+g(x^-)\,,
\end{equation}
which is simply a coordinate change of \eqref{eq:hkflow}. However, this is due to the relative simplicity of the model form 4DCS: had we considered a model with more poles in the twist function, like the one considered in Section \ref{sec:PCMk}, this would not have been the case.

			\subsection{Disorder defect II : \texorpdfstring{$\eta$- and $\lambda$-deformations of PCM}{eta- and lambda-deformations of PCM}}\label{ex:etapcm}
            We next consider two well-known one-parameter integrable deformations of the PCM: the $\eta$- and $\lambda$-deformations. The $\eta$-model is a Yang--Baxter deformation defined in terms of a skew-symmetric $R$-operator satisfying the modified classical Yang--Baxter equation \cite{Klimcik:2002zj,Klimcik:2008eq}. The $\lambda$-model, on the other hand, is an integrable sigma model that interpolates between the WZW model and the non-abelian $T$-dual of the PCM \cite{Sfetsos:2013wia}.

Although these models are constructed differently, they are related by Poisson--Lie $T$-duality and analytic continuation \cite{Vicedo:2015pna,Klimcik:2015gba,Hoare:2015gda,Sfetsos:2015nya}. In the 4d CS formulation, this relation is reflected in the fact that they share a one-form $\omega$ with the same analytic structure, while obeying different boundary conditions. This motivates treating their time-dependent versions in parallel.

First, let us briefly review the time-independent actions of these two integrable deformations. The action of the $\eta$-deformed PCM is given by \cite{Klimcik:2002zj,Klimcik:2008eq}
\begin{align}
    \text{S}_{\eta}[g]
    &=
    -\frac{K(1+\eta^2)}{2}
    \int_{\Sigma}\dd^2x\,
    \left<
    j_+,\frac{1}{1-\eta\,R_g}j_-
    \right>\,,
    \label{etaPCM}
    \\
    R_g
    &=
    {\rm Ad}_g^{-1}\circ R\circ{\rm Ad}_g\,.
\end{align}
This is a one-parameter deformation with deformation parameter $\eta \in \mathbb{R}$, and it reduces to the PCM in the limit $\eta \to 0$. Here, ${\rm Ad}_g$ denotes the adjoint action on $\mathfrak{g}$, defined by ${\rm Ad}_g(x)=g\,x\,g^{-1}$ for $x \in \mathfrak{g}$. The deformation is specified by an $\mathbb{R}$-linear operator $R:\mathfrak{g}\to \mathfrak{g}$. For the resulting deformation to be classically integrable, this $R$-operator must be skew-symmetric,
\begin{align}
    \left<x,R(y)\right>=-\left<R(x),y\right>\,,
\end{align}
and satisfy the modified classical Yang--Baxter equation
\begin{align}
    [R(X),R(Y)]
    -R([R(X),Y]+[X,R(Y)])
    =
    -c^2[X,Y]\,,
    \qquad X,Y\in \mathfrak{g}\,.
    \label{eq:mCYBE}
\end{align}
Here, the constant $c$ takes the value $c=1$ or $c=i$, and the $\eta$-model corresponds to $c=i$.\footnote{The deformation associated with the homogeneous classical Yang--Baxter equation, corresponding to $c=0$, also describes an integrable deformation \cite{Matsumoto:2015jja}. From the perspective of the 4d CS theory, the corresponding meromorphic one-form $\omega$ is the same as that of the undeformed PCM, and the deformation is again realised by imposing different boundary conditions \cite{Fukushima:2020dcp}.}

The $\lambda$-deformed PCM is instead described by \cite{Sfetsos:2013wia}
\begin{align}
    \text{S}_{\lambda}[g]
    =
    \text{S}_{{\rm WZW},k}[g]
    -\frac{k}{2}
    \int_{\Sigma} \dd^2x\,
    \left<
    g^{-1}\partial_+g,
    \frac{1}{\lambda^{-1}-{\rm Ad}_g}
    \partial_-gg^{-1}
    \right>\,,
\end{align}
where the WZW action $\text{S}_{{\rm WZW},k}[g]$ is given by
\begin{align}
    \text{S}_{{\rm WZW},k}[g]
    &=
    -\frac{k}{4}
    \int_{\Sigma}\dd^2x
    \left<
    g^{-1}\partial_+g,
    g^{-1}\partial_-g
    \right>
    -\frac{k}{2}\,\text{S}_{\rm WZ}[g]\,,
    \\
    \text{S}_{\rm WZ}[g]
    &=
    \frac{1}{3}
    \int_{\Sigma\times [0,1]}
    \left<
    \tilde{g}^{-1}\dd \tilde{g},
    \tilde{g}^{-1}\dd \tilde{g}
    \wedge
    \tilde{g}^{-1}\dd \tilde{g}
    \right>\,.
\end{align}
For $\lambda \ll 1$, this model can be viewed as the WZW model perturbed by a current-current interaction. To first order in $\lambda$, it reduces to a bosonised version of the non-Abelian Thirring model \eqref{eq:SNATM} \cite{Dashen:1973nhu,Dashen:1974hp}. In the opposite limit $\lambda\to 1$, it is effectively described by the non-Abelian $T$-dual of the PCM \cite{delaOssa:1992vci}.

        The $\eta$- and $\lambda$-deformations exhibit closely related quantum-group symmetries: the former involves a $q$-deformation of the Lie algebra $\mathfrak g$ \cite{Kawaguchi:2011pf,Delduc:2013fga}, the latter is associated with a corresponding quantum-group structure at a root of unity, with $q=\exp(i\pi/k)$ fixed by the WZW level \cite{Hollowood:2015dpa}.  For details, we refer the reader to reviews, e.g. \cite{Thompson:2019ipl,Hoare:2021dix}.

			\subsubsection*{Time-independent models from the 4d CS theory}
            Both integrable deformations admit a description within the 4d CS framework.
            The corresponding meromorphic one-forms are obtained as a one-parameter deformation of the meromorphic one-form associated with the PCM \cite{Delduc:2019whp}:
			\begin{align}
				\omega=\varphi(z)\dd z
				=\frac{1}{2}\frac{K^2-z^2}{z^2-c^2\eta^2K^2}\dd z\,,
			\end{align}
			where $K$ and $\eta$ are real parameters, and $c$ is chosen to be either $1$ or $i$.
			Under this deformation, the double pole at $z=\infty$ remains unchanged, while the double pole of the PCM twist function at $z=0$ splits into two simple poles located at $z=c\eta K$ and $z=-c\eta K$.
			At $z=\infty$, we impose a Dirichlet-type boundary condition, as in the  PCM$_k$ case (\ref{pcm-wz-bc}). For the boundary conditions associated with the simple poles, we find the following two possible choices:
			\begin{align}
				\eta\text{-model type}&: \qquad (R+c)A_{\pm}\lvert_{z=c \eta K}=(R-c)A_{\pm}\lvert_{z=-c \eta K}\,,\label{eq:eta-bc}\\
				\lambda\text{-model type}&:\qquad   A_{\pm}\lvert_{z=c \eta K}=A_{\pm}\lvert_{z=-c \eta K}\,.\label{eq:lambda-bc}
			\end{align}
            Here, the operator $R$ satisfies the modified Yang-Baxter equation (\ref{eq:mCYBE}).
			In particular, the $\eta$-model corresponds to the meromorphic 1-form with $c=i$, whereas the $\lambda$-model is obtained by taking $c=1$. In the latter case, it is often useful to reparametrize the parameters as follows:
			\begin{align}
				K\to k(\lambda^{-1}-\lambda),
				\qquad
				c\eta \to -\frac{1-\lambda}{1+\lambda} .
				\label{eq:eta-lambda-map-summary}
			\end{align}
			More explicitly, the meromorphic 1-forms associated with the $\eta$- and $\lambda$-deformed PCMs are given by \cite{Delduc:2019whp}
			\begin{align}
				\eta\text{-model}&:\quad   \omega_{\eta}=\varphi_{\eta}(z)\dd z=\frac{1}{2}\frac{K^2-z^2}{z^2+\eta^2K^2}\dd z\quad \text{with}\,\,(\ref{eq:eta-bc})\,\,\text{at}\,\, c=i\,,\label{eq:eta-bc-def}\\
				\lambda\text{-model}&:\quad     \omega_{\la}=\varphi_{\lambda}(z)\dd z=-\frac{1}{2}\frac{(z-z^+)(z-z^-)}{(z-p^+)(z-p^-)}\dd z\quad \text{with}\,\,(\ref{eq:lambda-bc})\,\,\text{at}\,\, c=1\,.\label{eq:la-bc-def}
			\end{align}
			For the $\eta$-model, we denote the zeros and poles by
			\begin{align}
				z^\pm=\pm K\,,    \qquad
				p^{\pm}=\pm i\eta K\,.
			\end{align}
			For the $\lambda$-model, they are denoted by
			\begin{align}
				z^{\pm}=\pm k\left(\lambda^{-1}-\lambda\right)\,,
				\qquad
				p^{\pm}=\mp k\frac{(\lambda-1)^2}{\lambda}\,.
			\end{align}
			With these boundary conditions, one can construct the Lax connections and actions for both models. 
			
			At the classical level, it has been shown that the $\eta$- and $\lambda$-models are related by a combination of Poisson–Lie $T$-duality \cite{Klimcik:1995jn,Klimcik:1995dy,Klimcik:1995ux} and analytic continuation \cite{Vicedo:2015pna,Klimcik:2015gba,Hoare:2015gda,Sfetsos:2015nya}. Under this relation, the parameters are mapped as
			\begin{align}
				K \longleftrightarrow k(\lambda^{-1}-\lambda)\,,
				\qquad
				\eta \longleftrightarrow i\,\frac{1-\lambda}{1+\lambda}\,.
				\label{eq:eta-lambda-map}
			\end{align}
			From the 4d CS viewpoint, this relation can be interpreted as an exchange of boundary conditions together with the analytic continuation above. 
			
			\subsubsection*{One-loop RG flow}
			
			Before extending the discussion to the time-dependent case, we write down the one-loop RG flow equations for these two models. For these models, we choose the normalization of the functions $f_{\pm}(z)$ as follows,
			\begin{align}
				\eta\text{-model}:& \quad &&f_{\pm}(z)=-\frac{a\left(z \pm K \eta^2\right)}{z\mp K}\,,\\
				\la\text{-model}:& \quad  && f_{\pm}(z)=-\frac{a \left(2\la (1+\la)z \mp k(1-\la)^3\right)}{2\la(1+\la)(z-z^{\pm})}\,.
			\end{align}
			With these normalization, the one-loop RG flow equation (\ref{eq:phiflow}) gives
			\begin{align}
				\eta\text{-model}:&\qquad     \frac{\dd \eta(\mathfrak{t}_{\text{RG}})}{\dd \mathfrak{t}_{\text{RG}}}=\frac{a \eta(\tRG)(1+\eta(\tRG)^2)}{K(\tRG)}\,,\qquad
				\frac{\dd K(\tRG)}{\dd \mathfrak{t}_{\text{RG}}}=a(1-\eta(\tRG)^2)\,,\label{eq:eta-rg-d}\\
				\la\text{-model}:&\qquad   \frac{\dd \la(\tRG)}{\dd \tRG}=-\frac{2a}{k}\frac{\la(\tRG)^2}{(1+\la(\tRG))^2}\,,\qquad \frac{\dd k}{\dd \tRG}=0\,.     \label{eq:la-rg-d}
			\end{align}
			These flow equations are consistent with the known results for the $\eta$-model \cite{Squellari:2014jfa}\footnote{The one-loop RG flow of the $\eta$-deformed PCM with a Wess-Zumino term was originally computed for $SU(2)$ in \cite{Kawaguchi:2011mz}, and was later extended to arbitrary groups in \cite{Demulder:2017zhz}.} and the $\la$-model \cite{Itsios:2014lca} (see also \cite{Kutasov:1989dt}).
			From the flow equations for the $\eta$-model, we obtain the RG invariant quantity 
			\begin{align}\label{eq:eta-rginv}
				\frac{1}{\nu}=K(\tRG)\frac{1+\eta(\tRG)^2}{\eta(\tRG)}=\text{const}.\,.
			\end{align}
			The RG invariance of this combination can be understood intuitively from the fact that, under the map (\ref{eq:eta-lambda-map}), it is mapped to the level $k$ in the $\lambda$-deformed model and $1/\nu$ is proportional to a residue in the period parametrisation, which is RG-invariant \eqref{eq:periodflow}. 
			
			\subsubsection*{Time-dependent models from the 4d CS theory}
			
			We now derive the time-dependent action and the corresponding Lax connection from the 4d CS theory.
			As in the PCM$_k$ case, we introduce the generalized meromorphic 1-forms for each deformed model. For the $\eta$-deformed PCM, we take
			\begin{align}
				\hat{f}_{\pm}(z,t)=-\frac{a\left(z \pm \hat{K}(t)\hat{\eta}(t)^2\right)}{z\mp \hat{K}(t)}\,,\label{eq:eta-f}
			\end{align}
			and the time dependences of parameters are determined by the differential equations
			\begin{align}
				\frac{\dd \hat{\eta}(t)}{\dd t}=\frac{a \hat{\eta}(t)(1+\hat{\eta}(t)^2)}{\hat{K}(t)}\,,\qquad
				\frac{\dd \hat{K}(t)}{\dd t}=a(1-\hat{\eta}(t)^2)\,.
			\end{align}
			On the other hand, for the $\lambda$-deformed PCM, the time-dependent scalar function $\hat{f}_{\pm}(z,t)$ is taken as
			\begin{align}
				\hat{f}_{\pm}(z,t)=-\frac{a \left(2\hat{\la} (1+\hat{\la})z \mp k(1-\hat{\la})^3\right)}{2\hat{\la}(1+\hat{\la})(z-\hat{z}^{\pm})}\,,\label{eq:lambda-fpm-t}
			\end{align}
			where the level $k$ remains constant, and the zeros and poles are defined as
			\begin{align}
				\hat{z}^{\pm}=\pm k\left(\frac{1}{\hat{\la}}-\hat{\la}\right)\,,\qquad \hat{p}^{\pm}=\mp k\frac{(\hat{\la}-1)^2}{\hat{\la}}\,,
			\end{align}
			with the time-dependent function $\hat{\la}(t)$ satisfies
			\begin{align}
				\frac{\dd \hat{\la}(t)}{\dd t}=-\frac{2a}{k}\frac{\hat{\la}(t)^2}{(1+\hat{\la}(t))^2}\,.   
			\end{align}
			We again impose the same boundary conditions (\ref{eq:eta-bc}) and (\ref{eq:lambda-bc}) for the $\eta$- and $\lambda$-models, respectively.
			These boundary conditions determine the explicit form of the ansatz (\ref{LfromJtime}), which in turn yields the Lax connections
			\begin{align}\label{pcmeta:bz}
				\eta\text{-model}:&\qquad\hat{L}=\frac{1}{\hat{K}(t)-z}\frac{\hat{K}(t)(1+\hat{\eta}^2)}{1+\hat{\eta} R_{g}}j_{+}\dd x^++\frac{1}{\hat{K}(t)+ z}\frac{\hat{K}(t)(1+\hat{\eta}^2)}{1-\hat{\eta} R_{g}}j_{-}\dd x^-\,,\\
				\la\text{-model}:&\qquad   \hat{L}=\frac{\hat{z}^+}{\hat{z}^+-z}\frac{2}{1+\hat{\la}}J_+\dd x^+
				+\frac{\hat{z}^-}{\hat{z}^--z}\frac{2}{1+\hat{\la}}J_-\dd x^-\,,\label{lambda-BZ-lax}
			\end{align}
			where we introduced the deformed current
			\begin{align}
				J_+=\frac{1}{\hat{\la}^{-1}-{\rm Ad}_g^{-1}}g^{-1}\partial_+g\,,\qquad 
				J_-=\frac{1}{\hat{\la}^{-1}-{\rm Ad}_g^{-1}}\partial_-gg^{-1}\,.
			\end{align}
			Using these Lax connections, we reduce the 4d CS action to 2D actions by integrating over the spectral parameter space $C$, and obtain the 2D actions
			\begin{align}\label{tetaPCM}
				\eta\text{-model}:&\qquad   \hat{\text{S}}_{\eta}[g]=-\frac{1}{2}\int_{\Sigma} \dd^2x\,\hat{h}(t)\,\left
				<j_+,\frac{1}{1-\hat{\eta}(t)\,R_g}j_-\right>\,,\\
				\la\text{-model}:&\qquad      \hat{\text{S}}_{\la}[g]=\text{S}_{{\rm WZW},k}[g]+k\int_{\Sigma} \dd^2x\,\left<g^{-1}\partial_+g,\frac{1}{\hat{\la}^{-1}-{\rm Ad}_g}\partial_-gg^{-1}\right>\,.\label{tlaPCM}
			\end{align}
			These time-dependent actions precisely reproduce those constructed in \cite{Hoare:2020fye}.
			
			\subsection{Disorder defect III :  \texorpdfstring{Coupled PCM$_k$}{Coupled PCMk}}\label{sec:PCMk}
		
            Lastly, let us consider $N$ coupled $\text{PCM}_k$ models. This is a theory of $N$ group-valued fields $g^i$, with action given by
			\begin{equation}
				\text{S}[g^i]_{\rm cPCM}=-\frac{1}{2}\int_\Sigma \dd^2 x\sum_{i,j}^Nh_{ij}\left<J^i_+,J_-^j\right>+\sum_{i=1}^N k_i\,\text{S}_{\rm WZ}[g^i]\,,\quad J_\pm^i=\big(g^i\big)^{-1}\partial_\pm g^i\,,\label{eq:cPCM}
			\end{equation}
			where, $h_{ij}$ are the coupling constants and $k_i$ are Wess-Zumino levels for each of the fields. In order for the theory to be integrable, the couplings and levels can not be arbitrary. Instead, if we introduce the functions \cite{Delduc:2018hty,Delduc:2019bcl,Delduc:2020vxy}
			\begin{equation}\label{eq:phicPCM}
				\varphi^i_{\pm}(z)=\frac{\prod_{j=1}^N(z-z_j^\pm)}{\prod_{i\neq j}^N(z-p_j)}\,,
			\end{equation}
			defined in terms of the $3N$ parameters $z_j^+,z_j^-,p_j$,
            the couplings must take the form
			\begin{equation}
				\begin{split}
					h_{ij}&=\frac{\varphi_+^i(p_i)\varphi_-^j(p_j)}{p_i-p_j}\,\quad\text{for}\quad i\neq j\,,\\
					h_{ii}&=\frac{1}{2}\Big[\varphi_-^i(p_i)\partial_z\varphi_+^i(p_i)-\varphi_+^i(p_i)\partial_z\varphi_-^i(p_i)\Big]\,,
					\\
					k_i&=\frac{1}{2}\Big[\varphi_-^i(p_i)\partial_z\varphi_+^i(p_i)+\varphi_+^i(p_i)\partial_z\varphi_-^i(p_i)\Big]\,.\label{eq:hkcpm}
				\end{split}
			\end{equation}
			Provided this form of the couplings, the associated Lax connection can be found to take the form
			\begin{equation}\label{eq:LcPCM}
				L_\pm(z)=\sum_{i=1}^N\frac{\varphi_\pm^i(p_i)}{\varphi_\pm^i(z)} J_\pm^i\,,
			\end{equation}
			which in particular has poles at $z^\pm_i$.     
            This integrable model was originally constructed from the affine Gaudin model \cite{Delduc:2018hty,Delduc:2019bcl}. The derivation from the 4d CS action was first carried out in \cite{Costello:2019tri}.
            We consider a $G$-valued gauge field $A$, whose action is given by
            \begin{equation}
                \text{S}[A]=\int_{\Sigma\times\mathbb{CP}^1}\omega\wedge\text{CS}[A]\,,
            \end{equation}
            where the 1-form $\omega$ can be written in the form
            \begin{equation}
                \omega=-\prod_{i=1}^{N}\frac{(z-z^+_i)(z-z_i^-)}{(z-p_i)^2}\,\dd z\,,
            \end{equation}
            which thus goes beyond the class of models given by \eqref{eq:ex-omega}. In this case, the 1-form has $2N$ simple zeroes and $N+1$ double poles, including one at $z=\infty$.
			To each of the finite poles, we associate one group-valued edge-mode $g_i$.\footnote{The field at the double pole at $z=\infty$ can be removed by a similar procedure to \eqref{eq:Lv}.} We should similarly impose boundary conditions on $A_\pm$, which are given by
            \begin{equation}
                A_\pm|_{z=p_i}=0\,.
            \end{equation}
            With these conditions fixed, we can construct the Lax connection, which precisely takes the form \eqref{eq:LcPCM}. It is also straightforward to obtain the 2d action, since it follows directly from plugging \eqref{eq:LcPCM} into \eqref{2Dfromdis}.

            To obtain the time-dependent version of this theory, we can use that its one-loop RG flow has already been determined \cite{Delduc:2020vxy}.\footnote{For the $N=3$ case, namely the so-called integrable $G \times G$ model, the two-loop RG flow was analyzed in \cite{Levine:2021fof}. It was shown there that, in a particular regularization scheme, the conditions imposed on the sigma-model couplings by classical integrability are preserved under the two-loop RG flow.} For the time-dependent model, it implies that $\hat p_i$ and $\hat z_i^\pm$ are promoted to functions of time, with their dependence related to the functions $\hat f_\pm$ appearing in the modified Lax equation \eqref{eq:Laxtime}. The latter are given by
			\begin{equation}
				\hat{f}_\pm(z)=\sum_{i=1}^{N}\frac{a}{\hat\varphi'(\hat z_i^\pm)}\frac{1}{z-\hat z_i^\pm}-\frac{a}{2}\,,\label{eq:fPCM}
			\end{equation}
			and following \cite{Delduc:2020vxy}, the time-dependent version of the theory is then given by 
			\begin{equation}
				\hat{\text{S}}[g^i]_{\rm cPCM}=-\frac{1}{2}\int_\Sigma \dd^2 x\sum_{i,j}^N\hat h_{ij}(t)\left<J^i_+,J_-^j\right>+\sum_{i=1}^N k_i\,\text{S}_{\rm WZ}[g^i]\,.\label{eq:cPCMtime}
			\end{equation}
			This theory is integrable provided that \eqref{eq:hkcpm} is kept intact, with the parameters on the right-hand side obeying the coupled differential equations
			\begin{equation}
				\partial_t\,\hat{p}_i= \hat f_+(\hat p_i)-\hat f_-(\hat p_i)\,,\quad \partial_t \hat z_i^\pm =\mp 2\hat f_{\mp}(\hat z_i^\pm)\,.\label{eq:pzflow}
			\end{equation}
			     In writing \eqref{eq:cPCMtime}, we used that $k_i$ is the residue at the pole $p_i$ and is thus time-independent, since it is a closed period \eqref{eq:periodflow}.
                 
                 Indeed, the period-parametrisation can also allow us to separate the complicated coupled differential equations given by \eqref{eq:fPCM} and \eqref{eq:pzflow}, since we can characterise \eqref{eq:phicPCM} by the knowledge of its residues, and the relative periods $
                 \hat \Pi_{1\pm}^{i\pm}$ and $\hat \Pi_{1+}^{1-}$. Without loss of generality, we can use $z\mapsto \alpha z+\beta$ to fix $\hat z_1^\pm=\pm 1$.
			 Explicitly, the periods are then given by 
			\begin{equation}
				\begin{split}
				&\hat \Pi_{1\pm}^{i\pm}=(\hat z_i^\pm\mp 1)+\sum_{j=1}^N\left[\frac{(\hat z_i^\pm\mp 1)\hat \varphi_-^j(\hat p_j)\hat \varphi_+^j(\hat p_j)}{(\hat p_j-\hat z_i^\pm)(\hat p_j\mp 1)}+[\hat \varphi_-^j(\hat p_j)\partial_z\hat \varphi_+^j(\hat p_j)+\hat \varphi_+^j(\hat p_j)\partial_z\hat \varphi_-^j(\hat p_j)]\log\frac{\hat p_j-\hat z_i^\pm}{\hat p_j\mp 1}\right]\,,
				\\
				&\hat \Pi_{1+}^{1-}=2+\sum_{j=1}^N\left[\frac{2\hat \varphi_-^j(\hat p_j)\hat \varphi_+^j(\hat p_j)}{\hat p_j^2-1}+[\hat \varphi_-^j(\hat p_j)\partial_z\hat \varphi_+^j(\hat p_j)+\hat \varphi_+^j(\hat p_j)\partial_z\hat \varphi_-^j(\hat p_j)]\log\frac{\hat p_j+1}{\hat p_j-1}\right]\,.
				\end{split}
			\end{equation}
			As seen in \eqref{eq:Piindflow}, this parametrisation has a distinct advantage that the only time-dependent parameter is $\hat \Pi_{1+}^{1-}$, which is linear in time.
            
			\section{Lax connection with a variable spectral parameter}\label{sec:BMtype}
In the previous sections, we used the fixed-spectral Lax connection (Belinski--Zakharov linear system). Its compatibility condition gives the modified flatness equation \eqref{eq:Laxtime}. In the literature, this formulation is known to be suited for the inverse-scattering or dressing method: solution-generating techniques that allow one to construct new solitonic solutions starting from a seed solution. In the context of dimensionally reduced gravity, it has been applied to coordinate dependent two-dimensional integrable sigma models, both for constructing higher-dimensional black holes \cite{Belinsky:1971nt,Belinsky:1979gra,Pomeransky:2005sj} and for describing soliton-like gravitational waves and nonlinear interactions of gravitational and electromagnetic waves \cite{Belinsky:1980jh,Tomimatsu:1989vw,Tomizawa:2013soa,Tomizawa:2015zva,Igata:2015oea,Igata:2015oxb}.

There exists a complementary formulation in the literature called the Breitenlohner--Maison linear system \cite{Breitenlohner:1986um}. Below, we call it the variable-spectral Lax connection. Instead of working with a fixed spectral parameter, one introduces a variable-spectral parameter $\sfz=\sfz(t,x;w)$, which depends on the two-dimensional base coordinates $(t,x)$ and on a spacetime-independent parameter $w$. When $w$ satisfies a specific algebraic equation, the equations of motion are equivalent to the standard flatness condition of the Lax connection \eqref{eq:LaxEq}. 
The price of the variable-spectral description is a more complicated analytic structure. Since $\sfz(t,x;w)$ is obtained by solving an algebraic relation, it is generically multi-valued as a function of $w$. The branch points can be identified with the zeros $\hat{z}_i$ of the twist function $\hat{\varphi}$, while the corresponding poles $\hat{p}_r$ are mapped to the local geometry around $w=\infty$. 

For the symmetric coset sigma model, the multi-sheeted structure allows us to consider monodromy matrices in the $w$-plane that connect distinct sheets, and is well suited for studying the hidden infinite-dimensional symmetry known as the Geroch group \cite{Breitenlohner:1986um}. 
It is also useful for the Riemann--Hilbert approach, which is an alternative way to construct classical solutions through the factorisation of the monodromy matrix, and has also been used in the construction of gravitational solutions \cite{Breitenlohner:1986um,Chakrabarty:2014ora,Katsimpouri:2012ky,Katsimpouri:2013wka,Sakamoto:2025xbq,Sakamoto:2025sjq}.\footnote{This approach remains less developed in this respect than the dressing method. However, recent advances have shown that it enables the construction of asymptotically AdS solutions \cite{Sakamoto:2025jtn} and extremal (multi-)black hole solutions \cite{Roy:2018ptt,Sakamoto:2026cyo}, which has not been extensively developed within the dressing method.}
For more general models, it remains unclear whether a corresponding monodromy matrix can be defined in a mathematically well-defined way. Nevertheless, constructing the associated Lax connection is expected to be useful for elucidating the underlying symmetry structure.

In this section, we derive this variable-spectral Lax connection directly from the fixed-spectral Lax connection obtained from the 4d CS theory. The construction is a direct generalization of the prescription for space-dependent symmetric coset sigma models in \cite{Figueras:2009mc}. Throughout this section, we focus on the purely time-dependent case unless otherwise stated.

			\subsection{Variable-spectral Lax connection for PCM: review}For concreteness, let us first review the variable-spectral Lax connection for the time-dependent PCM \eqref{eq:SCPM} and see how it is related to the fixed-spectral formulation used above.

\subsubsection*{Variable-spectral Lax connection}
            The variable-spectral Lax connection of the PCM has the same expression as the time-independent one \cite{Breitenlohner:1986um}
			\begin{align}
				\hat{L}_{\pm}^{\rm var}(w)=\frac{j_\pm(t,x)}{1\mp \hat{z}(t,x;w)}\,,
			\end{align}
			where the simple poles $z^\pm=\pm 1$ are fixed at constant values, and the non-trivial time dependence is instead encoded in the variable spectral parameter $\hat{z}(t,x;w)$.
            The corresponding Lax equation takes the standard form
            \begin{align}\label{eq:bm-laxeq}
            \partial_+\hat{L}_{-}^{\rm var}(w)
           -\partial_-\hat{L}_{+}^{\rm var}(w)
            +[\hat{L}_{+}^{\rm var}(w),\hat{L}_{-}^{\rm var}(w)]
           =0\,,
            \end{align}
            which is equivalent to the equations of motion for the time-dependent PCM \eqref{eq:PCMEOMtime}, provided that $\hat{z}$ satisfies
            \begin{equation}\label{eq:dz-pcm}
            \dd\hat{z}
                =
           -\frac{\hat{z}}{\hat h(t)}
           \left(
           \frac{\hat{z}+1}{\hat{z}-1}\dd x^+
           +
          \frac{\hat{z}-1}{\hat{z}+1}\dd x^-
           \right)\,.
\end{equation}
Here, the scalar function $\hat h(t)$ is given in \eqref{eq:hPCMtime}. In the following, we set $\hat{h}(t)=t$, corresponding to the choice $a=1,b=0$, for simplicity.

\begin{figure}[t]
                    \centering
           \begin{tikzpicture}
        
    \newcommand{\R}{0.185\textwidth}
    
    \filldraw[color=gray!70,fill=gray!15,thick] (0,0) circle (\R);
    
    \draw[gray!70, dashed] (-\R,0) arc [start angle=180, end angle=0, x radius=\R, y radius={0.3*\R}];
    
    \draw[gray!70] (-\R,0) arc [start angle=180, end angle=360, x radius=\R, y radius={0.3*\R}];
    
    \foreach \x/\y in {0/1, 0/-1} {
        \fill[blue] (\x*\R, \y*\R) circle (1.5pt);
    }
    \draw[orange, thick](-\R-0.03*\R, -0.03*\R) -- (-\R+0.03*\R, 0.03*\R);
    \draw[orange, thick] (-\R-0.03*\R, 0.03*\R) -- (-\R+0.03*\R,-0.03*\R);

    \draw[orange, thick](\R-0.03*\R, -0.03*\R) -- (\R+0.03*\R, 0.03*\R);
    \draw[orange, thick] (\R-0.03*\R, 0.03*\R) -- (\R+0.03*\R,-0.03*\R);
    
    \draw[red,thick,->] ({1.02*\R*cos(2)},{1.02*\R*sin(2)}) arc [start angle=2, end angle=20, radius={1.02*\R}];
    \draw[red,thick,->] ({-1.02*\R*cos(2)},{1.02*\R*sin(2)}) arc [start angle=178, end angle=160, radius={1.02*\R}];
    
    \node[left, orange] at (-\R, 0) {$\hat{z}^-$};
    \node[right, orange] at (\R, 0) {$\hat{z}^+$};
    \node[below, blue] at (0,-\R) {$\hat{p}_1$};
    \node[above, blue] at (0,\R) {$\hat{p}_2$};
      
    \draw[thick,->] ({1.35*\R},0) -- ({1.65*\R},0);
    \node[below] at ({1.5*\R},-0.1*\R) {$z=\mathsf{z}(w,t,x)$};

    \filldraw[color=gray!70,fill=gray!15,thick] ({3*\R},0) circle (\R);
    
    \draw[gray!70, dashed] ({2*\R},0) arc [start angle=180, end angle=0, x radius=\R, y radius={0.3*\R}];
    
    \draw[orange!70, decorate, decoration={zigzag, segment length=8.1pt, amplitude=2pt}, thick] 
        ({2*\R},0) arc [start angle=180, end angle=360, x radius=\R, y radius={0.3*\R}];
    
    \fill[blue] ({3*\R}, \R) circle (1.5pt);
    
    \draw[orange, thick] (2*\R+0.03*\R, 0.03*\R) -- (2*\R-0.03*\R,-0.03*\R);
    \draw[orange, thick] (2*\R-0.03*\R, 0.03*\R) -- (2*\R+0.03*\R,-0.03*\R);

    \draw[orange, thick](4*\R-0.03*\R, -0.03*\R) -- (4*\R+0.03*\R, 0.03*\R);
    \draw[orange, thick] (4*\R-0.03*\R, 0.03*\R) -- (4*\R+0.03*\R,-0.03*\R);
    
    \draw[red,thick,->] ({3*\R + 1.02*\R*cos(2)},{1.02*\R*sin(2)}) arc [start angle=2, end angle=20, radius={1.02*\R}];
    \draw[red,thick,->] ({3*\R - 1.02*\R*cos(2)},{1.02*\R*sin(2)}) arc [start angle=178, end angle=160, radius={1.02*\R}];
    
    \node[left, orange] at ({2*\R},0) {$\hat{w}^-$};
    \node[right, orange] at ({4*\R},0) {$\hat{w}^+$};
    \node[above, blue] at ({3*\R},\R) {$\infty$};
    
    \node[] at (0,1.5*\R) {\LARGE $C:(z,\bar{z})$};
    \node[] at ({3*\R},1.5*\R) {\LARGE $C:(w,\bar{w})$};
        
\end{tikzpicture}
                    \caption{As shown in \eqref{eq:ztow}, the map \eqref{gene-replace} maps the original disorder defect system with two double poles and two time-dependent zeros $\hat{z}^\pm$ to a system with one double pole at $w=\infty$ and a branch cut from $\hat w^+=w(z^+)$ to $\hat w^-=w(z^-)$.}
                    \label{fig:ztow}
                \end{figure}
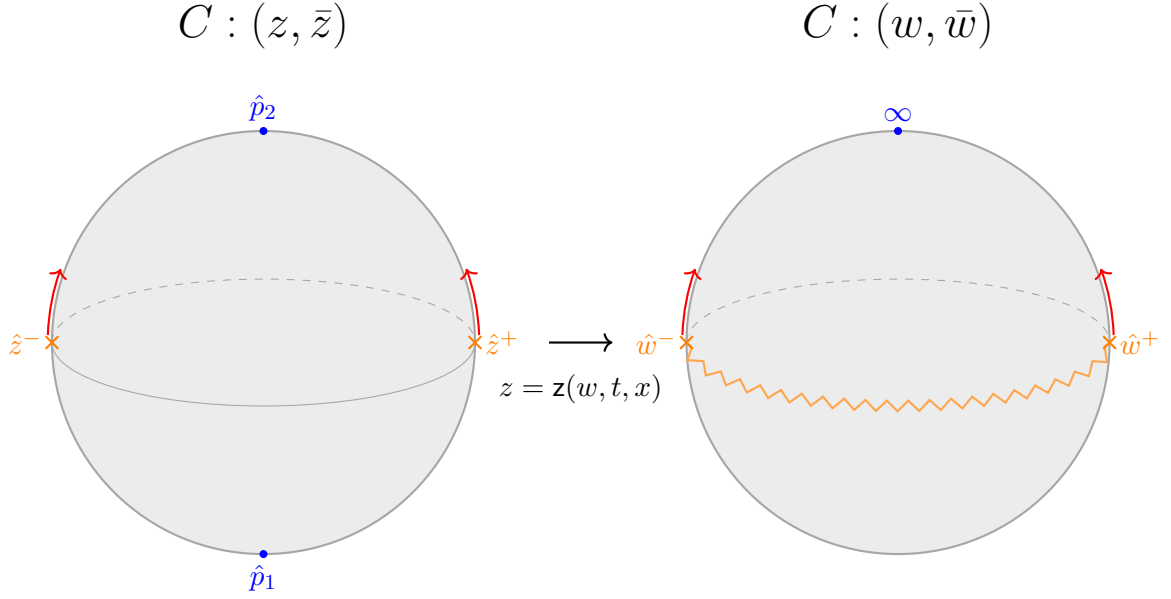

			The solutions to the equation (\ref{eq:dz-pcm}) are characterized by the algebraic relation
			\begin{align}\label{eq:algebra-eq}
				w=-x-\frac{t}{2}\left(\frac{1}{\hat{z}}+\hat{z}\right)\,,
			\end{align}
			where $w\in\mathbb{CP}^1$ is an integration constant.
			The algebraic relation \eqref{eq:algebra-eq} admits two solutions 
			\begin{align}\label{zla-rel1}
				\hat{z}=-\frac{1}{t}\left(w+x\pm \sqrt{(w+x)^2-t^2}\right)\,,
			\end{align}
			which have two branch points located at $w=-t-x$ and $w=t-x$.
			This implies that while the fixed-spectral parameter $z$ is valued in $\mathbb{CP}^1$, the function $\hat{z}(w)$ lives in the genus zero Riemann surface defined by (\ref{eq:algebra-eq}) which is a
			twofold covering of the $w$-plane with two branch points at $w=-t-x$ and $w=t-x$. See Figure \ref{fig:ztow}.
            The Lax connection $\hat{L}_{\pm}^{\rm var}$ defines a standard auxiliary linear system
			\begin{align}
				\partial_{\pm}\Phi(t,x;w)=-\hat{L}_{\pm}^{\rm var}(w)\Phi(t,x;w)\,,
			\end{align}
			which is known as the Breitenlohner--Maison linear system \cite{Breitenlohner:1986um}.
			In this formulation, the complex parameter $w$, rather than $\hat z$, plays the role of the spectral parameter.

\subsubsection*{Relation to the fixed-spectral formulation}
We now compare the variable-spectral connection with the fixed-spectral Lax
connection $\hat L_\pm^{\rm fix}(z)$ in \eqref{eq:time-pcm-lax}. Following the
spectral-parameter transformation discussed in \cite{Figueras:2009mc}, the two
descriptions are related by
\begin{equation}
    z \mapsto \sfz = t\,\hat z(t,x;w)\,,
    \label{eq:pcm-bm-bz}
\end{equation}
where $\hat z(t,x;w)$ is given in \eqref{zla-rel1}. Under this replacement, the two Lax connections coincide
\begin{align}
    \hat L_\pm^{\rm fix}(z,t,x)
    \mapsto
    \hat L_\pm^{\rm fix}(\sfz,t,x)
    =
    \hat L_\pm^{\rm var}(w,t,x)\,.
    \label{eq:bz-bm-map}
\end{align}
Thus the variable-spectral connection $\hat{L}_{\pm}^{\rm var}$ is obtained from the fixed-spectral connection $\hat{L}_{\pm}^{\rm fix}$ by evaluating the fixed spectral parameter at the coordinate-dependent value $\sfz$.

This also explains how the two Lax equations are related, namely that the modified
flatness equation \eqref{eq:BZ-flat} for $\hat{L}_{\pm}^{\rm fix}$ is mapped to the
standard flatness equation \eqref{eq:bm-laxeq} for $\hat{L}_{\pm}^{\rm var}$. Indeed,
using \eqref{zla-rel1}, one finds
			\begin{align}\label{eq:dsfz-w}
				\partial_+\sfz=\hat{f}_+(\sfz,t)\,,\qquad \partial_-\sfz=-\hat{f}_-(\sfz,t)\,,
			\end{align}
			where we recall that, for the time-dependent PCM, $f_{\pm}$ are given by \eqref{eq:fpmPCM}\eqref{eq:fPCM}
            \begin{equation}
				\hat f_{\pm}(z,t)=\pm\frac{z}{t\mp z}\,.
			\end{equation}
            This identity shows that for any function $F(w,t,x)$, regarded equivalently as $F(\sfz,t,x)$, the ordinary light-cone derivatives act as
            \begin{align}
				\partial_{\pm}F(w,t,x)&=\partial_{\pm}F(\sfz,t,x)\lvert_{\sfz:\text{fixed}}+\partial_{\pm}\sfz\,\partial_{\sfz}F(\sfz,t,x)\no\\
                &=\left(\partial_{\pm}\pm \hat f_\pm(\sfz,t)\partial_{\sfz}\right)F(\sfz,t,x)=\hat{\partial}_{\pm}F(z,t,x)\lvert_{z=\sfz}\,,\label{eq:dF}
			\end{align}
			where we used the definition (\ref{eq:dder}) of the twisted derivatives $\hat{\partial}_{\pm}$.
			In particular, we can also identify the solutions $\Phi^{\rm fix}(t,x;z)$ and $\Phi^{\rm var}(t,x,;w)$ of the two linear systems  under the replacement (\ref{eq:pcm-bm-bz}). 
					
			\subsection{Variable-spectral Lax from fixed-spectral Lax: general prescription 
            }
            We now show that a fixed-spectral Lax connection with modified derivatives can always be rewritten as a variable-spectral Lax connection with ordinary derivatives. This gives a general map between the two formulations.
            
            Consider a time-dependent integrable field theory with fixed-spectral Lax
connection $\hat L_\pm^{\rm fix}(z,t,x)$ satisfying the modified flatness
equation \eqref{eq:Laxtime} with the modified derivatives $\hat\partial_\pm
    =
    \partial_\pm
    \pm
    \hat f_\pm(z)\partial_z$. We introduce a coordinate-dependent spectral parameter
			\begin{equation}\label{gene-replace}
				z \mapsto \sfz=\sfz(w,t,x)\,,
			\end{equation}
			where $w\in C$ is constant and $\sfz$ satisfies
			\begin{align}\label{eq:bm-diffeq}
				\partial_+\sfz=\hat{f}_+(\sfz,t)\,,\qquad \partial_-\sfz=-\hat{f}_-(\sfz,t)\,.
			\end{align}
            The variable-spectral Lax connection is then defined by evaluating the fixed-spectral connection at this coordinate-dependent value of the spectral parameter:
            \begin{align}\label{eq:bz-bm-map2}
				\hat L_\pm^{\rm var}(w,t,x)
    =
    \hat L_\pm^{\rm fix}(\sfz,t,x)\,.
			\end{align}
            With this definition, the modified derivatives acting on the fixed-spectral
connection become ordinary derivatives acting on the variable-spectral
connection. Indeed, by the chain rule and \eqref{eq:bm-diffeq},
\begin{equation}
    \partial_\pm\hat L_\mp^{\rm var}(w,t,x)=\hat\partial_\pm\hat L_\mp^{\rm fix}(z,t,x)\big|_{z=\sfz}\,.
\end{equation}
Therefore the modified flatness equation for $\hat L_\pm^{\rm fix}$ is mapped
directly to the standard flatness equation
\begin{equation}
    \partial_+\hat L_-^{\rm var}-\partial_-\hat L_+^{\rm var}+[\hat L_+^{\rm var},\hat L_-^{\rm var}]=0\,.
\end{equation}
For completeness, let us check that a variable spectral parameter $\sfz$
satisfying \eqref{eq:bm-diffeq} exists. The required integrability condition is the commutativity of mixed derivatives. Differentiating \eqref{eq:bm-diffeq}
gives
\begin{align}
    \partial_-\partial_+\sfz
    &=
    \partial_-\hat f_+(\sfz,t)
    \notag\\
    &=
    \partial_-\hat f_+(\sfz,t)\big|_{\sfz:{\rm fixed}}
    +
    \partial_-\sfz\,\partial_\sfz\hat f_+(\sfz,t)
    \notag\\
    &=
    \partial_-\hat f_+(\sfz,t)\big|_{\sfz: {\rm fixed}}
    -
    \hat f_-(\sfz,t)\,\partial_\sfz\hat f_+(\sfz,t)\,,
    \\
    \partial_+\partial_-\sfz
    &=
    -\partial_+\hat f_-(\sfz,t)
    \notag\\
    &=
    -\partial_+\hat f_-(\sfz,t)\big|_{\sfz:{\rm fixed}}
    -
    \partial_+\sfz\,\partial_\sfz\hat f_-(\sfz,t)
    \notag\\
    &=
    -\partial_+\hat f_-(\sfz,t)\big|_{\sfz:{\rm fixed}}
    -
    \hat f_+(\sfz,t)\,\partial_\sfz\hat f_-(\sfz,t)\,.
\end{align}
Using $\hat f_\pm=\mp\hat\varphi^{-1}\hat\Psi_\pm$, their difference becomes
\begin{align}
    [\partial_-,\partial_+]\sfz
    =
    \hat\varphi^{-2}
    \biggl[
        \hat\Psi_-
        \bigl(
            \partial_+\hat\varphi
            -
            \partial_z\hat\Psi_+
        \bigr)
        -
        \hat\Psi_+
        \bigl(
            \partial_-\hat\varphi
            -
            \partial_z\hat\Psi_-
        \bigr)
        +
        \hat\varphi
        \bigl(
            \partial_-\hat\Psi_+
            -
            \partial_+\hat\Psi_-
        \bigr)
    \biggr]\bigg|_{z=\sfz}\,.
\end{align}
This vanishes precisely by the flow equations \eqref{eq:dphidpsi}. Hence
$[\partial_-,\partial_+]\sfz=0$, and the system \eqref{eq:bm-diffeq} is locally integrable and the coordinate-dependent spectral parameter $\sfz$ exists, at least locally. 

Although solving the differential equations \eqref{eq:bm-diffeq} explicitly is generally difficult, one can characterize the variable spectral parameter $\sfz$ algebraically (see the following subsection for concrete examples). In terms of $t$ and $x$, the equations
\eqref{eq:bm-diffeq} are equivalent to
\begin{subequations}
			\begin{align}
				\partial_t\sfz&=\hat{f}_+(\sfz,t)-\hat{f}_-(\sfz,t)=\hat{f}(\sfz,t)\,,\label{eq:sfz-t}\\
				\partial_x\sfz&=\hat{f}_+(\sfz,t)+\hat{f}_-(\sfz,t)=a\,\hat{\varphi}(\sfz,t)^{-1}\,.\label{eq:sfz-x}
			\end{align}
    \end{subequations}
			Integrating the second equation with respect to $x$ gives
	      \begin{equation}\label{eq:bm-eq}
				\log \hat{\cP}(\sfz,t)+\hat{\cP}_0(t)=a(x+w)\,,
			\end{equation}
			where $w$ is an integration constant, and the potential function $\log \hat{\cP}(z,t)$ is introduced such that it satisfies
		\begin{equation}\label{eq:def-logP}
				\partial_z\log \hat{\cP}(z,t)=\hat{\varphi}(z,t)\,.
			\end{equation}
            The function $\hat{\cP}_0(t)$ is then fixed by requiring consistency with the
$t$-evolution \eqref{eq:sfz-t}.
One simple requirement that $\hat{\cP}_0(t)$ needs to satisfy can be derived by differentiating the algebraic equation (\ref{eq:bm-eq}) with respect to the time $t$
            \begin{align}\label{eq:dert-P0}
                \partial_t\hat{\cP}_0(t)&=-\partial_t\log \hat{\cP}(\sfz,t)\no\\
                &=-\partial_t\sfz\,\hat{\varphi}(\sfz,t)-\partial_t\log \hat{\cP}(\sfz,t)\lvert_{\sfz:\text{fixed}}\no\\
                &=\hat{\Psi}(\sfz,t)-\partial_t\log \hat{\cP}(\sfz,t)\lvert_{\sfz:\text{fixed}}\,,
            \end{align}
            In the last step, we used \eqref{eq:sfz-t} and
\eqref{eq:def-logP}. The right-hand side is in fact a function only of $t$.
Indeed, rewriting the flow equation \eqref{eq:dphidpsi} in terms of
$\hat{\cP}(z,t)$ gives
            \begin{align}
                \partial_z(\partial_t \log\hat{\cP}(z,t) )=\partial_z\hat{\Psi}(z,t)\,,
            \end{align}
            and therefore the difference
$\partial_t\log\hat{\cP}(z,t)-\hat\Psi(z,t)$ is independent of $z$.
   
			Finally, differentiating \eqref{eq:bm-eq} with respect to the integration
constant $w$ gives
			\begin{align}\label{der-sfz}
				\partial_{w}\sfz=a\,\hat{\varphi}(\sfz,t)^{-1}\,.
			\end{align}
            This identity has a useful geometric interpretation. Under the change of
spectral parameter \eqref{gene-replace}, the fixed-spectral one-form is
locally trivialized
			\begin{align}
				\hat{\omega}\mapsto \hat{\omega}\lvert_{z=\sfz}&=\hat{\varphi}(\sfz,t)\left(\dd \sfz-\hat{f}_+(\sfz,t)\dd x^++\hat{f}_{-}(\sfz,t)\dd x^-\right)\no\\
                &=\hat{\varphi}(\sfz,t)\left(\dd \sfz-\partial_+\sfz\,\dd x^+-\partial_-\sfz\,\dd x^-\right)\no\\
				&=\hat{\varphi}(\sfz,t)\frac{\partial \sfz}{\partial w}\dd w=a\,\dd w\,.\label{eq:ztow}
			\end{align}
	           For the PCM, this observation was made in \cite{Cole:2024skp}, where it was
used to construct the 4d CS one-form leading to the time-dependent PCM action
and its variable-spectral Lax connection.

Note that the local trivialization of the one-form \eqref{eq:ztow} does not mean that the theory is trivial.
Although $a\,\dd w$ is holomorphic in the variable $w$, the map
$z=\sfz(t,x;w)$ changes the global structure. As in the PCM example
\eqref{eq:algebra-eq}, the poles of $\hat\varphi(z,t)$ are mapped to the pole
of $a\,\dd w$ at $w=\infty$, where boundary conditions on $A_\pm(w)$ must be
imposed. Similarly, the zeros of $\hat\varphi(z,t)$ are mapped to branch cuts
in the $w$-plane (cf.\ Figure \ref{fig:ztow}).

\subsection{Examples}

We now apply the general prescription above to the examples of
\cref{sec:examples}. Starting from the fixed-spectral Lax connections derived
there, we obtain the corresponding variable-spectral Lax connections by the
replacement $z\mapsto\sfz(t,x;w)$. The resulting expressions agree with those
of \cite{Hoare:2020fye}.

\subsubsection*{PCM with WZ term}

We first consider the time-dependent PCM with the WZ term. The variable-spectral
Lax connection is obtained from the fixed-spectral connection \eqref{pcmk:bz}
by replacing the coordinate-independent spectral parameter $z$ with the
coordinate-dependent one $\sfz$. It is useful to write
\begin{equation}
    \sfz
    =
    \hat h(t)\hat z ,
\end{equation}
so that the poles of the variable-spectral Lax connection are fixed. This gives
\begin{align}
    \hat L^{\rm var}
    =
    \frac{1}{1-\hat z}
    \left(
        1-\frac{k}{\hat h(t)}
    \right)
    j_+\,\dd x^+
    +
    \frac{1}{1+\hat z}
    \left(
        1+\frac{k}{\hat h(t)}
    \right)
    j_-\,\dd x^- .
    \label{pcmk:bm}
\end{align}

It remains to determine the algebraic equation obeyed by $\hat z$. Since
$\sfz=\hat h(t)\hat z$, the defining equations for the variable spectral
parameter become
\begin{align}
    \partial_+\bigl(\hat h(t)\hat z\bigr)
    =
    \hat f_+\bigl(\hat h(t)\hat z,t\bigr),
    \qquad
    \partial_-\bigl(\hat h(t)\hat z\bigr)
    =
    -
    \hat f_-\bigl(\hat h(t)\hat z,t\bigr).
\end{align}
Equivalently, one may use the general algebraic relation
\eqref{eq:bm-eq}. For the present model, consistency with the time evolution is
obtained by choosing
\begin{align}
    \hat{\cP}_0(t)
    =
    \frac{k}{2}
    \log
    \left(
        1-\frac{\hat h(t)^2}{k^2}
    \right).
\end{align}
Substituting this into \eqref{eq:bm-eq} gives
\begin{align}
    0
    &=
    a(w+x)
    +
    \frac{k}{2}
    +
    \frac{1}{2}
    \frac{
        \hat h(t)
        \bigl(
            2k\hat z-\hat h(t)(1+\hat z^2)
        \bigr)
    }{
        k-\hat h(t)\hat z
    }
    \notag\\
    &\quad
    +
    k\log
    \left[
        -\hat h(t)
        \left(
            \frac{k}{\hat h(t)}-\hat z
        \right)
    \right]
    -
    \frac{k}{2}
    \log
    \left(
        1-\frac{\hat h(t)^2}{k^2}
    \right).
    \label{PCMWZ-la-const}
\end{align}
This equation implicitly determines the variable spectral parameter $\hat z$.
Using \eqref{PCMWZ-la-const}, one checks that
$\partial_w(\hat h(t)\hat z)$ satisfies \eqref{der-sfz}, and therefore the corresponding variable-spectral flatness equation is equivalent to the
equations of motion \eqref{PCMWZ-eom}. The same algebraic equation is also
equivalent to the constraint on the variable spectral parameter given in
\cite{Hoare:2020fye}.
			
			\subsubsection*{$\eta$-deformed PCM}

We next consider the $\eta$-deformed PCM. The variable-spectral Lax connection is given by
\begin{equation}
    \hat L^{\rm var}
    =
    \frac{1}{1-\hat z}
    \frac{1+\hat\eta^2}{1+\hat\eta R_g}
    g^{-1}\partial_+ g\,\dd x^+
    +
    \frac{1}{1+\hat z}
    \frac{1+\hat\eta^2}{1-\hat\eta R_g}
    g^{-1}\partial_- g\,\dd x^- .
    \label{pcmeta:bm}
\end{equation}

The algebraic equation for $\hat z$ follows from the general relation
\eqref{eq:bm-eq}. For the functions $\hat f_\pm(z,t)$ in \eqref{eq:eta-f},
one may take $\hat{\cP}_0(t)=0$. After writing $\sfz=\hat K(t)\hat z$, one
finds
\begin{align}
    a(w+x)
    +
    \hat K(t)
    \left[
        \hat z
        -
        \left(
            \hat\eta(t)
            +
            \frac{1}{\hat\eta(t)}
        \right)
        \tan^{-1}
        \left(
            \frac{\hat z}{\hat\eta(t)}
        \right)
    \right]
    =
    0 .
    \label{zw-eta}
\end{align}
Using \eqref{eq:eta-rginv} to eliminate $\hat K(t)$, this equation agrees,
up to a Möbius transformation of the spectral parameter, with the constraint
found in \cite{Hoare:2020fye}.

\subsubsection*{$\lambda$-deformed PCM}

We now turn to the $\lambda$-deformed PCM. Replacing $z$ by
$\sfz=\hat z^+\hat z$ in the fixed-spectral Lax connection
\eqref{lambda-BZ-lax} gives the variable-spectral connection
\cite{Hoare:2020fye}
\begin{align}
    \hat L^{\rm var}
    =
    \frac{1}{1-\hat z}
    \frac{2}{1+\hat\lambda}
    J_+\,\dd x^+
    +
    \frac{1}{1+\hat z}
    \frac{2}{1+\hat\lambda}
    J_-\,\dd x^- .
\end{align}

The corresponding algebraic equation is again obtained from
\eqref{eq:bm-eq}. For the choice of functions $\hat f_\pm(z,t)$ in
\eqref{eq:lambda-fpm-t}, one can set $\hat{\cP}_0(t)=0$, giving
\begin{align}
    -
    \frac{2a}{k}(w+x)
    +
    \hat z
    \left(
        \hat\lambda-\hat\lambda^{-1}
    \right)
    +
    2
    \log
    \left[
        -
        \frac{
            \hat z-\frac{1-\hat\lambda}{1+\hat\lambda}
        }{
            \hat z+\frac{1-\hat\lambda}{1+\hat\lambda}
        }
    \right]
    =
    0 .
\end{align}
This agrees with the result of \cite{Hoare:2020fye}, up to an additive
constant.

\subsubsection*{Coupled $\text{PCM}_k$}

Finally, we briefly discuss the coupled $\mathrm{PCM}_k$. By the general
construction, the variable-spectral Lax connection is obtained from the
fixed-spectral one \eqref{eq:LcPCM} by replacing $z$ with $\sfz$:
\begin{align}
    \hat L^{\rm var}
    =
    \sum_{i=1}^N
    \frac{\varphi_+^i(\hat p_i)}{\varphi_+^i(\sfz)}
    J_+^i\,\dd x^+
    +
    \sum_{i=1}^N
    \frac{\varphi_-^i(\hat p_i)}{\varphi_-^i(\sfz)}
    J_-^i\,\dd x^- .
\end{align}
The algebraic relation for $\sfz$ follows from \eqref{eq:bm-eq} and takes the
form
\begin{align}
    -
    \sfz
    +
    \sum_{j=1}^N
    \left[
        \frac{
            \varphi_+^j(\hat p_j)\varphi_-^j(\hat p_j)
        }{
            \sfz-\hat p_j
        }
        -
        2k_j\log(\sfz-\hat p_j)
    \right]
    =
    a(x+w),
\end{align}
where the levels $k_j$ are the time-independent constants defined in
\eqref{eq:hkcpm}.

Here we have chosen $\hat{\cP}_0(t)=0$. This is justified as follows. For the
functions $\hat f_\pm(z)$ in \eqref{eq:fPCM}, the difference
$\partial_t\log\hat{\cP}(z,t)-\hat\Psi(z,t)$ vanishes as $z\to\infty$.
Together with \eqref{eq:dert-P0}, this implies that $\hat{\cP}_0(t)$ is
independent of $t$. Since a constant shift of $\hat{\cP}_0$ can be absorbed
into a redefinition of $w$, we may set it to zero without loss of generality.

Unlike in the previous examples, it is no possible to redefine the
variable spectral parameter in a way that makes all poles of the
variable-spectral Lax connection constant.
            
			\section{Reformulation as dilaton gravity}\label{sec:dilaton}
			As noted in the introduction, some of the earliest integrable systems with explicit spacetime dependence arose from dimensionally-reduced Einstein gravity. In particular, imposing axial symmetry and reducing Einstein’s equations to two dimensions leads to a special coset sigma model coupled to dilaton gravity. After gauge fixing, this becomes a time-dependent coset sigma model \cite{Belinsky:1971nt,Belinsky:1979gra}.
            
            In this section, we point out that general time-dependent integrable field theories constructed above admit an analogous interpretation. By a straightforward extension of the standard arguments, they can be recast as integrable matter theories coupled to two-dimensional dilaton gravity.
            \subsection{Time-dependent integrable field theories as dilaton gravity}
            We start from the general dilaton gravity action
            \begin{align}\label{eq:dilatonaction}
            {\rm S}_{\rm dilaton}=\int \dd^2 x \,\sqrt{-g}\Big(\Phi R+\mathcal{L}_{\rm matter}(\psi, g_{\mu\nu}|\Phi)\Big)\,,
            \end{align}
            where $g_{\mu\nu}$ is the two-dimensional metric, $R$ its Ricci scalar, $\Phi$ the dilaton, $\psi$ the two-dimensional quantum fields, and $\mathcal{L}_{\rm matter}$ the Lagrangian of an integrable field theory coupled covariantly to $g_{\mu\nu}$. The coupling to the dilaton in $\mathcal{L}_{\rm matter}$ is left unspecified for now; we will determine it so that the action reproduces the time-dependent integrable field theory. One important point that will be relevant below is that the integrable field theories considered in this paper are classically conformal, even though they are asymptotically
            free quantum mechanically. Thus, at the classical level, the matter action is
            invariant under Weyl rescalings of the metric.

            We now fix the two-dimensional diffeomorphism symmetry by choosing conformal
            gauge, 
            \begin{equation}
            \dd s^2 =e^{\varphi}\eta_{\mu\nu}\dd x^{\mu}\dd x^{\nu}\,,
            \end{equation}
            where $\varphi$ is the conformal mode of the metric and $\eta_{\mu\nu}={\rm diag}(-1,1)$. In this gauge, the action becomes
            \begin{equation}\label{eq:dilatonaction2}
            {\rm S}_{\rm dilaton}=\int \dd^2 x \Big(-\Phi \partial_{\mu}\partial^{\mu}\varphi+\mathcal{L}_{\rm matter}(\psi|\Phi)\Big)\,.
            \end{equation}
            Because the matter theory is classically conformal, $\varphi$ drops out of
            $\mathcal L_{\rm matter}$. Therefore, the conformal mode appears only through
            the dilaton-gravity term as a ``Lagrange multiplier". Varying $\varphi$ gives
            \begin{equation}
        \partial_{+}\partial_{-}\Phi=0\,,
            \end{equation}
            so the dilaton is locally the sum of a holomorphic and an anti-holomorphic function, $\Phi=f(x^{+})+g(x^{-})$. Performing further holomorphic coordinate transformations, we can identify the dilaton with the time direction $\Phi =t$. This means that by choosing the coupling of the dilaton to the matter appropriately, one can reproduce the action of time-dependent integrable field theories in \cref{sec:timedep4dCS} and \cref{sec:examples}.

            For example, in the coupled Gross--Neveu model discussed in \cref{subsec:cGN}, the required dilaton coupling is obtained by replacing the time-dependent couplings by functions of $\Phi$. This gives
            \begin{equation}
            \mathcal{L}_{\rm cGN}=\sum_{i=1}^{N_+}\sum_{j=1}^{N_-}\left(\sum_{n=1}^{N}\psi^{i,n}_+i\partial_-\psi^{i,n}+\psi_-^{j,n}i\partial_+\psi_-^{j,n}\right)+\frac{\left<J_+^i,J_-^j\right>}{a\Phi+b_i^{+}+b_j^{-}}\,,
            \end{equation}
            where $a$ and $b_{i}^{\pm}$ are arbitrary constants. For the principal chiral model, the corresponding coupling is simpler, being given by
            \begin{equation}
            \mathcal{L}_{\rm PCM}=\frac{a\Phi+b}{2}\langle j_{\mu}\,,j^{\mu}\rangle\,,
            \end{equation}
            with arbitrary constants $a$ and $b$. Setting $a=1$ and $b=0$ reproduces the standard dilaton-coupled PCM studied in the literature \cite{Belinsky:1971nt,Belinsky:1979gra} in relation to dimensional reduction of Einstein gravity.
            The same prescription applies to all time-dependent integrable field theories constructed in this paper: the time-dependent parameters are promoted to functions of the dilaton. However, for more general theories arising from disorder defects, the dilaton dependence can be considerably more involved.

\medskip

            Let us add two comments. First, the relation to dilaton gravity appears to hold only in the purely time-dependent case. As discussed above, once the couplings are allowed to depend on both space and time, there is a much larger space of admissible deformations. Such freedom cannot, in general, be reproduced by coupling the theory to a single dilaton field. This mirrors the relation to the one-loop RG flow, which also emerges only in the purely time-dependent setting. It would be interesting to understand whether more general spacetime-dependent deformations can be described using several dilaton fields, or whether dilaton gravity offers a deeper explanation of the connection between time dependence and one-loop RG flow. Second, the dilaton couplings obtained here are somewhat artificial from the intrinsic two-dimensional point of view. It would be useful to know whether they can arise more naturally from dimensional reductions of higher-dimensional gravitational systems.            \subsection{Connection to string worldsheet theory} There is another useful interpretation of the time-dependent integrable field
            theories, following \cite{Hoare:2020fye} (For earlier works, see \cite{Tseytlin:1992pq,Tseytlin:1992va,Tseytlin:1992ee} ). One embeds the theory into a
            string worldsheet theory by introducing two additional light-cone target-space
            coordinates $X^\pm$. In conformal gauge, the worldsheet action takes the form
            \begin{align}
             S_{\rm ws}=\int \dd^2x \left(
             \partial_+X^+\,\partial_-X^- + \mathcal L_{\rm matter}(\psi\mid X^+) \right)\,. \end{align}
             Here $\mathcal L_{\rm matter}$ is the same matter Lagrangian as in the dilaton-gravity description, with the dilaton $\Phi$ replaced by the light-cone coordinate $X^+$.
             
             The role of $X^-$ is analogous to that of the conformal mode in the dilaton gravity discussion: it acts as a Lagrange multiplier. Its equation of motion imposes \begin{align} \partial_+\partial_{-} X^+=0 \,. \end{align} One may therefore choose the light-cone gauge \begin{align} X^+=p^+ t \,, \end{align} which is compatible with the equation of motion. After this gauge fixing, the worldsheet theory reduces precisely to the time-dependent integrable field theory, provided the $X^+$ dependence of $\mathcal L_{\rm matter}$ is chosen in the same way as the dilaton dependence described above.
             
             The discussion so far has been classical. To interpret the model as a genuine string worldsheet theory, two further conditions must be addressed. First, the theory must be conformal at the quantum level. Second, the total matter central charge must cancel the conformal anomaly, for instance by giving $c=26$ in the bosonic case.
            
            The key observation of \cite{Hoare:2020fye,Tseytlin:1992pq,Tseytlin:1992va,Tseytlin:1992ee} is that these conditions can be satisfied at every order in perturbation theory by choosing appropriate spacetime dependence of the dilaton and imposing the $X^+$ dependence of $\mathcal L_{\rm matter}$ to coincide with the RG flow. This is precisely the type of time dependence that arises universally in the integrable field theories constructed in this paper. It would be important to understand whether the quantization of 4d CS theory naturally leads to the same structure beyond one loop and whether the results in \cite{Hoare:2020fye,Tseytlin:1992pq,Tseytlin:1992va,Tseytlin:1992ee} can be extended beyond scalar field theories. 
            
			\section{Conclusion and discussion}\label{sec:conclusion}
			We demonstrated that the 4d CS theory provides a
            unifying framework for constructing spacetime-dependent integrable field
            theories. The consistency conditions of the four-dimensional theory
            constrain the allowed spacetime dependence, and the resulting theories admit Lax pairs despite this spacetime dependence. When the couplings depend on both space and time, we found a large family of
            admissible integrable deformations. The purely time-dependent case is much more
            rigid. In that setting, we proved in general that the allowed time dependence is
            precisely the one-loop RG flow.

            \medskip
            As emphasized in the introduction, the appeal of time-dependent integrable
            systems lies in their ability to connect different concepts that are usually
            studied separately, such as non-equilibrium dynamics,   integrability, dilaton gravity, and isomonodromic deformation. Thus, our construction opens a broad set
            of directions across mathematical and theoretical physics. We close by outlining
            some of them.

\medskip
            \paragraph{Non-equilibrium dynamics and inverse scattering.}
            All time-dependent integrable field theories constructed in this paper admit
            Lax pairs. They should therefore be amenable to inverse-scattering techniques,
            including dressing methods \cite{Belinsky:1971nt,Belinsky:1979gra,Pomeransky:2005sj} and Riemann--Hilbert approaches \cite{Breitenlohner:1986um,Chakrabarty:2014ora,Katsimpouri:2012ky,Katsimpouri:2013wka}.
            However, these methods have not yet been systematically worked out, apart from special examples arising in dimensionally-reduced gravity.
            It would be interesting to develop these techniques in this broader setting
            and to use them to construct explicit solitonic solutions. Such solutions could
            provide useful analytic probes of non-equilibrium dynamics, and may also have applications to dilaton gravity coupled to matter. 

            A related question would be to understand the dressing and Riemann--Hilbert constructions
            directly from the 4d CS perspective. Such a formulation could provide a unified framework for inverse-scattering methods and may also clarify their role in higher-dimensional gravity, where these techniques were originally developed. Indeed, dressing methods and Riemann--Hilbert approaches continue to be used in the construction of gravitational solutions, but the full class of gravity solutions accessible by these methods remains poorly understood. It would be interesting to see whether the 4d CS theory can shed light on this question.  
            
            \paragraph{Finite-gap method for time-dependent theories.} Another related problem is to generalize the finite-gap method to time-dependent
            integrable theories on compact spaces. In ordinary classical integrable field
            theory, inverse scattering is naturally suited for non-compact space, where one constructs
            solitonic solutions with prescribed asymptotic behaviour. On a space with a periodic boundary condition, the periodicity requires the more refined finite-gap construction \cite{Babelon:2003qt}. It would be interesting to extend it to the general time-dependent systems studied here, building on the related works on dimensionally-reduced gravity \cite{Korotkin:1988,Korotkin:1990,Korotkin:1991,Korotkin:1996,Neugebauer:1993,Klein:1998,Klein:2005,deLeon:2023}. Recent work clarifying the relation between finite-gap methods and inverse scattering  \cite{Beisert:2024tit} may also provide a useful starting point.

            \paragraph{Time-dependent affine Gaudin models.}
            Affine Gaudin models provide another systematic framework for constructing
            integrable field theories 
            and are known to be connected with the Hamiltonian formulation of the 4d CS theory \cite{Vicedo:2019dej}. It would therefore be interesting to develop a time-dependent generalization of this framework and to compare it with the 4d CS construction studied here.
            \paragraph{Time-dependent discrete and ultradiscrete integrable systems.}
            Classically integrable field theories with Lax pairs often admit spacetime
            discretisations \cite{Hirota:1981,Suris:2003}, leading to discrete integrable systems such as the discrete
            Toda system and the discrete KP equation. Discrete integrability also appears in
            a broad range of applied settings, including discrete Lotka--Volterra equations
            for predator--prey dynamics \cite{Suris:1996lv,Shinjo:2018dhLV} and discretized epidemic models such as discrete
            SIR systems \cite{Tanaka:2022,Nobe:2023}. Many of these systems can be further
            ``ultradiscretised" (see e.g.~\cite{Tokihiro:1996}), producing integrable cellular automata such as the
            box--ball system \cite{Takahashi:1990,Inoue:2011}, which also has deep connections to combinatorics.

            It would be interesting to discretize the time-dependent integrable field
            theories constructed in this paper and to understand how the standard structures
            of discrete and ultradiscrete integrability mentioned above are modified. One particularly
            important object is the Hirota equation, which governs many classical discrete integrable systems and also appears as the algebra of transfer matrices in quantum integrable models. Therefore, a time-dependent generalization of the Hirota equation may also provide an alternative route toward time-dependent quantum integrability.

            \paragraph{Time-dependent quantum integrability.} The key advantage of the 4d CS theory is that it
            produces not only integrable field theories, but also lattice integrable systems
            through the insertion of Wilson lines. Since these lattice systems admit rather straightforward quantization, this framework offers a natural route toward time-dependent
            quantum integrability. 
            
            A natural next step is therefore to extend the analysis in this paper to
            quantum integrable systems. This is also essential for clarifying the relation to recent work on time-dependent quantum-mechanical systems \cite{Pasnoori:2025kondo,Pasnoori:2025wavefunction,Pasnoori:2025qkz,Pasnoori:2025rgflow,Pasnoori:2026tdgn}, where a similar link between time dependence and renormalisation group flow has been observed.
            We should also note that a simpler class of quantum examples is provided by time-dependent Gaudin models studied in \cite{Fioretto:2012caux}. In close parallel with the construction discussed here, these models can be derived from a time-dependent generalization of three-dimensional BF theory \cite{Vicedo:2022mixbf}. They are also a natural arena for understanding the relation to isomonodromic deformations since Gaudin models are known to be closely connected to isomonodromy. Connections to isomonodromy were also discussed for integrable field theories coupled to gravity in \cite{Korotkin:1996vi}.\footnote{It is useful to note that the relation between the spacetime-dependent
            PCM and Painlevé equations has appeared previously in the
            general relativity literature. In particular, certain symmetry reductions of
            the spacetime-dependent PCM lead to the Painlevé III, V, and VI equations for suitable choices of parameters \cite{Leaute:1979yv,Leaute:1983ph,Wils_1989,Calvert_1996,Manojlovic:1999rr,Manojlovic:2000lla}. More recently, this connection has been revisited from the perspectives of twistor space and Riemann--Hilbert problems \cite{Ferrari:2018ubt}.}
            
            These and related topics on time-dependent quantum integrability will be addressed in our forthcoming work \cite{toappear}.
            \paragraph{One-loop time-dependent integrability.}
            Classical integrability often leaves a visible imprint on one-loop quantum
            physics. One aspect is renormalization. In many known two-dimensional
            integrable sigma models, the one-loop RG flow preserves the integrable family:
            the flow changes the couplings, but does not take the theory out of the class of 
            integrable models. This principle is well supported by many examples, and it was proven in \cite{Delduc:2020vxy,levine_universal_2023,Levine:2023wvt,Lacroix:2024wrd,Lacroix:2025ias} for a broad class of theories arising from the 4d CS theory.
            Another aspect is semiclassical quantization. In the context of AdS/CFT correspondence, for example, the classical spectral curve provides an efficient way to organize one-loop
            corrections to the spectrum \cite{Gromov:2008ec,Vicedo:2008ryn} and correlation functions of fluctuations around classical string solutions \cite{Giombi:2022anm}.

            It would be interesting to understand how these structures extend to the
            time-dependent systems studied here. One question is whether the one-loop stability of integrable families under RG flow has a direct analogue for time-dependent integrable deformations. Another is whether the time-dependent Lax connection, or an appropriate generalization of the spectral curve, can be used to compute one-loop fluctuations around classical backgrounds.

            The latter question is especially natural for dimensionally-reduced gravity. An integrable description of fluctuations could provide a useful route to quantum
            corrections around black-hole backgrounds, including corrections to black hole entropy and
            to response coefficients such as Love numbers.
            
            \paragraph{Mixed defect systems and the Kondo problem.}
            In this work, we considered setups involving either order defects or disorder
            defects. A natural next step is to study configurations in which the two types of defects coexist. Canonical examples are provided by Kondo models, which can be engineered using disorder surface defects together with order line defects \cite{Gaiotto:2020fdr,Gaiotto:2020dhf,Wu:2021jir}, and non-Abelian Toda field theories (NATFT), arising from coincident order and disorder surface defects \cite{fukushima_non-abelian_2022}. Special cases of NATFTs include the sine-Gordon theory and the Liouville theory.
            
            It would therefore be interesting to study the time-dependent generalisation of such mixed defect systems. This would give a systematic way to construct time-dependent Kondo models and NATFTs, and would connect naturally with recent work on the time-dependent Kondo problem \cite{Pasnoori:2025kondo,Pasnoori:2025rgflow}. Additionally, it would be interesting to explore the time-dependent counterpart of the ODE/IM correspondence for the Kondo models discussed in \cite{Gaiotto:2020fdr,Gaiotto:2020dhf,Wu:2021jir}.

            \paragraph{Time-dependent massive ODE/IM correspondence.}
            The ODE/IM correspondence, originally uncovered in the study of spectral determinants of Schrödinger-type ordinary differential equations and $Q$-operators in conformal field theory, provides a striking bridge between classical differential equations and quantum integrability \cite{Dorey:1998pt,Bazhanov:1998wj,Dorey:2007zx}. Its massive generalizations replace the Schrödinger problem by auxiliary linear problems associated with classical integrable field equations, such as modified sinh--Gordon or affine Toda systems, and relate their monodromy data to massive quantum integrable field theories \cite{Lukyanov:2010rn,Ito:2018eon}.
 
            Since the time-dependent integrable field theories constructed in this paper also admit Lax pairs, it is natural to ask whether their auxiliary linear problems define a time-dependent version of the massive ODE/IM correspondence. If so, it would be interesting to identify the corresponding deformation of the quantum integrable models.

            \paragraph{Symmetries and algebraic structures of time-dependent integrability.}
            In ordinary integrable systems, exact solvability is tied to hidden infinite-dimensional symmetries. These symmetries are often organized by algebraic structures such as Yangians or affine quantum groups, whose classical counterparts are the corresponding classical $r$-matrices. It is therefore natural to ask what replaces these structures in the time-dependent setting.

            One possible clue comes from dimensionally-reduced gravity, where infinite-dimensional symmetry groups, such as the Geroch group, have long been known to play an important role; see, for example, \cite{Geroch:1971nt,Geroch:1972yt,Katsimpouri:2012ky,Schwarz:1995dk,Julia:1996ep,Lu:2007cj}. It would be valuable to revisit these symmetries from the perspective of the 4d CS theory. This may reveal a common algebraic framework underlying time-dependent integrability, beyond special systems arising from gravitational systems.
            \paragraph{Isomonodromy and its physical implications.}
            Another important problem is to clarify the role of isomonodromic deformation.
            The relation between non-autonomous integrability and isomonodromy has long been studied, especially in the mathematical literature \cite{Jimbo:1981-1,Jimbo:1981-2,Miwa:1981,Harnad:1994}. Its physical implication, however, remains less understood. In time-dependent integrable systems, isomonodromy replaces the usual isospectrality of autonomous integrable models. It would be useful to understand precisely how this property constrains non-equilibrium dynamics, and what physical information is carried by the monodromy data that remain invariant under time evolution.
			\paragraph{Physical origin of the relation between time dependence and RG flow.} Finally, although we have shown that time dependence and one-loop RG flow are governed by the same structures in 4d CS theory, we do not yet understand the deeper physical reason for this coincidence. Explaining why these two apparently different evolutions had to be controlled by the same geometric data remains, in our view, one of the most important open questions raised by this work.
			\section*{Acknowledgement}
			We thank the organisers and the participants of the workshop “Exact techniques and their applications” (Tropea, Italy, September 2024), where this work was initiated. We also thank the organisers and the participants of the workshop ``Integrability and Nonequilibrium Phenomena in Spacetime-Modulated Systems" (Kavli IPMU, Japan, May 2026), during which this work was completed. We in particular thank Lewis Cole, Hosho Katsura, Nat Levine, and Chihiro Matsui for helpful discussions on various aspects of time-dependent integrable systems. We also thank Sylvain Lacroix for useful discussions.
            SK and AW thank Kavli IPMU for hospitality at the final stage of this work. The work of JS was supported by the JSPS Grant-in-Aid for Transformative Research Areas (A) “Extreme Universe” No. 21H05190.
            
			\newpage
			
			\appendix
			
            \section*{Appendix}
			
			\section{Integrability and RG flow of order defects: rational, trigonometric, and elliptic examples}\label{app:ClRG}
			
			In this appendix, we show the classical integrability and discuss the one-loop RG flow of the 2d integrable field theory (\ref{eq:2daction-order}) arising from the 4d CS action coupled with chiral order defects.
			
			\subsection{Classical integrability}\label{sec:int-order}
			
			We first show the classical integrability of the 2d action (\ref{eq:2daction-order}) by verifying the equivalence between its equations of motion and the Lax equation associated with the Lax connection (\ref{eq:lax-order}).
			Since the chiral and antichiral order defects have the equations of motion (\ref{eq:eom-order}), the equations of motion for the resulting action (\ref{eq:2daction-order}) are given by 
			\begin{align}
				\begin{split}\label{eq:order-eom}
					&\partial_-J_+^i-\sum_{j=1}^{N_-}\,[J_+^i,r(z_i^+-z_j^-)\cdot J_-^j]=0\,,\\
					&\partial_+J_-^{i}-\sum_{j=1}^{N_+}[r(z_i^--z_j^+)\cdot J_+^j, J_-^i]=0\,,
				\end{split}
			\end{align}
            where we introduced the shorthand notation $r(z)\cdot J_{\pm}^i=r_{ab}(z)J_{\pm}^{i,b}\,T^a$.
			Conversely, the Lax equation for the Lax connection (\ref{eq:lax-order}) is
			\begin{align}
				&\partial_+L_--\partial_-L_++[L_+,L_-]\no\\
				&=\sum_{i=1}^{N_-}r(z-z_i^-)\cdot \partial_+J_-^{i}+\sum_{i=1}^{N_+}r(z-z_i^+)\cdot \partial_-J_+^{i}-\sum_{i=1}^{N_+}\sum_{j=1}^{N_-}[r(z-z_i^+)\cdot J_+^i, r(z-z_j^-)\cdot J_-^j]\,.
			\end{align}
			For any simple Lie algebra $\mathfrak{g}_{\mathbb{C}}$, any non-degenerate solutions to the cYBE (\ref{eq:cybe}) have only simple poles on $C$, and the residue at $z=0$ is proportional to the quadratic Casimir \cite{Belavin:1982rtj}
			\begin{align}
				\sum_{a=1}^{\dim \mathfrak{g}}\kappa_{ab}T^a \otimes T^b\,.
			\end{align}
			Taking the residues of the Lax equation at $z=z^{\pm}_i$ gives
			\begin{align}
				z^+_i:\quad 0&=\underset{z=z_i^+}{\text{res}}r(z-z_i^+)\cdot \partial_-J_+^{i}-\sum_{j=1}^{N_-}[\underset{z=z_i^+}{\text{res}}r(z-z_i^+)\cdot J_+^i, r(z_i^+-z_j^-)\cdot J_-^j]\no\\
				&\propto \partial_-J_+^{i}-\sum_{j=1}^{N_-}[ J_+^i, r(z_i^+-z_j^-)\cdot J_-^j]\,,\\
				z^-_i:\quad 0&=\underset{z=z_i^-}{\text{res}}r(z-z_i^-)\cdot \partial_+J_-^{i}-\sum_{j=1}^{N_+}[r(z_i^--z_j^+)\cdot J_+^j, \underset{z=z_i^-}{\text{res}}r(z-z_i^-)\cdot J_-^i]\no\\
				&\propto \partial_+J_-^{i}-\sum_{j=1}^{N_+}[r(z_i^--z_j^+)\cdot J_+^j, J_-^i]\,.
			\end{align}
			Thus, the equations of motion (\ref{eq:order-eom}) are equivalent to the Lax equation for the Lax connection (\ref{eq:lax-order}).
				
			\subsection{Computation of one-loop RG flow equation }\label{app:RG_order}
			
			For self-consistency, we review a computation of the one-loop RG flow equation of the 2d integrable field theories \eqref{eq:2daction-order} obtained from order defects.
            
			We restrict our attention to order defects which are built from chiral and antichiral free CFTs, which allows us to use the techniques outlined in \cite{Kutasov:1989dt,Gerganov:2000mt} (see also \cite{Ludwig:2002fu}). To make the coupling constants of the current-bilinear terms explicit, we expand the classical $r$-matrix as
			\begin{equation}\label{eq:r-mat-ansatz}
				r(z)=-\sum_{A}\sum_{a,b=1}^{\dim \mathfrak{g}}h_{A}(z)\hat{r}_{ab}^{A}\,T^a\otimes T^b\,,
			\end{equation}
			where the constant matrix $\hat{r}_{ab}^{A}$ is independent of the spectral parameter $z$. 
            The coefficient functions $h_{A}(z)$ then play the role of the coupling constants, and the index $A$ runs over all independent couplings. We denote the current bilinear term by
			\begin{align}\label{eq:bilinear}
				\cO^{A,ij}(w,\bar{w})=\sum_{a,b=1}^{\dim \mathfrak{g}}\hat{r}_{ab}^{A}\,\cO^{ab,ij}(w,\bar{w})\,,\qquad 
				\cO^{ab,ij}(w,\bar{w})=J_{+}^{a,i}(w)J_{-}^{b,j}(\bar{w})\,,
			\end{align}
			and the (Euclidean) action of the resulting 2d effective integrable theory (\ref{eq:2daction-order}) 
            is then described as a current-current deformation of the 2d CFT\footnote{
            Along the lines of \cite{Kutasov:1989dt,Gerganov:2000mt}, we perform the perturbative computation after passing to Euclidean signature. More explicitly, we make the Wick rotation $t=-i y$ and
            and introduce the complex coordinates $w=x+i y, \bar{w}=x-i y$. With this prescription, the Lorentzian action $S_L$ and the Euclidean action $S_E$ are related by $S_{E}=-i S_L\lvert_{t=-i y}$.}
            \begin{align}\label{eq:2dcft-cc}
				\text{S}_r=\sum_{i=1}^{N_+}\text{S}_{+}^{i}+\sum_{i=1}^{N_-}\text{S}_{-}^{i}+\sum_{i=1}^{N_+}\sum_{j=1}^{N_-}\sum_{A}\int_{\Sigma}\dd^2w\,h_{A}(z_{i}^+-z_j^{-})\cO^{A,ij}(w,\bar{w})\,,
			\end{align}
			where $\text{S}_{+}^{i}$ and $\text{S}_{-}^{i}$ are the actions of chiral and antichiral free CFTs with the Kac-Moody currents $J_{+}^{i}$ and $J_{-}^{i}$ of level $k$, respectively. The Kac-Moody currents $J_{+}^{i}$ and $J_{-}^{i}$ obey the operator product expansion (OPE)
			\begin{align}\label{eq:kac-ope}
				J^{a,i}_{+}(w)J^{b,j}_{+}(0)\sim \frac{k\,\kappa^{ab}\delta^{ij}}{w^2}+\frac{\delta^{ij}}{w}f^{ab}{}_c J^{c,i}_{+}(0)\,,
			\end{align}
			and similarly for $J_{-}^{i}(\bar{w})$, while the OPE between $J_+$ and $J_-$ is regular. From this, one can read off the OPE of the current-current interactions \eqref{eq:bilinear}
            \begin{equation}
            \begin{split}
				&\cO^{ab,ij}(w_1,\bar{w}_1)\cO^{cd,lm}(w_2,\bar{w}_2)\sim \\&\delta^{il}\delta^{jm}\biggl[\frac{k^2 \kappa^{ac}\kappa^{bd}}{|w_1-w_2|^4}+\frac{k \kappa^{bd}}{|w_1-w_2|^2(\bar{w}_1-\bar{w}_2)}f^{ac}{}_e J_+^{e,i}(w_2)\\
				&+\frac{k \kappa^{ac}}{(w_1-w_2)|w_1-w_2|^2}f^{bd}{}_s J_-^{s,m}(\bar{w}_2) +\frac{1}{|w_1-w_2|^2}f^{ac}{}_{e} f^{bd}{}_{s} J_{+}^{e,i}(w_2)J_{-}^{s,m}(\bar{w}_2)\biggr]\,.\label{eq:OOOPE}
            \end{split}
			\end{equation}
			The correlation functions in the deformed theory can be computed using standard CFT techniques of the undeformed theory. The correlation function $\left<\dots\right>_{r}$ (where $\dots$ represents insertion of operators) with the 2d action (\ref{eq:2dcft-cc}) is expressed as
			\begin{align}
				\left<\dots\right>_{r}\propto\left<\dots\,\exp\left({-\sum_{i=1}^{N_+}\sum_{j=1}^{N_-}\int_{w}r_{ab}^{(0)}(z_i^+-z_j^-)\,\cO^{ab,ij}_{w} }\right)\right>_{0}\,,
			\end{align}
			where the superscript $(0)$ of $r_{ab}$ denotes bare couplings.
            The right-hand side is evaluated with respect to the undeformed action, and we introduced a shorthand notation
			\begin{equation}
				\int_{w}\cO^{ab,ij}_{w}=\int_{w}\dd^2w\,\cO^{ab,ij}(w,\bar{w})\,.
			\end{equation}
			To obtain the one-loop $\beta$-function, it is enough to consider a perturbative expansion at $\cO(r_{ab}^2)$, such that the right-hand side is expanded as
			\begin{align}\label{eq:cf-ex}
				&\left<\dots\,\right>_{0}+\left<\dots\,\left({-\sum_{i=1}^{N_+}\sum_{j=1}^{N_-}\int_{w_1}r_{ab}^{(0)}(z_i^+-z_j^-)\,\cO^{ab,ij}_{w_1} }\right)\right>_{0}\\
				&\quad+\left<\dots\,\left(\sum_{i,l=1}^{N_+}\sum_{j,m=1}^{N_-}\frac{1}{2}\int_{w_1}\int_{w_2}r_{ab}^{(0)}(z_i^+-z_j^-)r_{cd}^{(0)}(z_l^+-z_m^-)\,\cO^{ab,ij}_{w_1}\cO^{cd,lm}_{w_2}\right)\right>_{0}+\dots\,.\no
			\end{align}
            
            The second line can be evaluated using OPE of the current bilinears \eqref{eq:OOOPE}. The terms linear in the currents do not contribute. For example, the contribution from the term linear in $J_{+}^{e,i}$ can be written as
			\begin{align}
				\sum_{i=1}^{N_+}\sum_{j=1}^{N_-}\int_{w_1}\int_{w_2}\,\frac{ r_{ab}^{(0)}(z_i^+-z_j^-)r_{cd}^{(0)}(z_i^+-z_j^-)\kappa^{bd}f_e^{ac}}{|w_1-w_2|^2(\bar{w}_1-\bar{w}_2)}J_+^{e,i}(w_2)\,,
			\end{align}
			which vanishes due to the relation
			\begin{align}
				r_{ab}^{(0)}(z_i^+-z_j^-)r_{cd}^{(0)}(z_i^+-z_j^-)\kappa^{bd}f^{ac}{}_e=0\,,
			\end{align}
            which follows from the symmetry of $r_{ab}\kappa^{bd}r_{cd}$ in $a$ and $c$, while $f^{ac}{}_{e}$ is antisymmetric.
            
			A potential divergence arises from the remaining integrals.
			To handle this, we employ a regularization scheme which introduces a UV cutoff $\epsilon$ and an IR cutoff $R$. We then split $w_2$ into two points $u,v$ such that $\epsilon<|u-v|\ll R$ and integrate $w_1$ over the region $\Omega_{R,\epsilon}$, defined by
            \begin{align}
				\Omega_{R,\epsilon}=D_R(0)\backslash \big(D_{\epsilon}(u)\cup D_{\epsilon}(v)\big)\,,\qquad D_{\ell}(y)=\{z\in \mathbb{C}:|z-y|<\ell\}\,,
			\end{align}
            i.e.\ a disk of radius $R$ from which smaller disks of radius $\epsilon$ around $u$ and $v$ have been removed. Within this prescription, the relevant integrals over $w_1$ are given by (see, for example,  \cite{Georgiou:2016iom})
			\begin{align}
				I_1 =\int_{\Omega_{R,\epsilon}} \dd^2w_1\,\frac{1}{(w_1-u)(\bar{w}_1-\bar{v})}&=-\pi\log \frac{|u-v|^2}{R^2-u\bar{v}}\,,\label{int-1}\\
				I_2=\int_{\Omega_{R,\epsilon}} \dd^2w_1\,\frac{1}{(w_1-u)^2(\bar{w}_1-\bar{v})^2}&=\pi^2\delta^{(2)}(u-v)-\pi\frac{R^2}{(R^2-u\bar{v})^2}\,.
			\end{align}
			one observes that the second integral $I_{2}$ is obtained by differentiating $I_{1}$ with respect to $u$ and $\bar{v}$.\footnote{While we impose $|u-v|>\epsilon$, we keep the delta function term formally.} 
			Taking the large-$R$ limit and sending $u,v\to w_{2}$ in the integrals $I_{1,2}$, we obtain 
			\begin{align}
				I_1=\int_{\Omega_{R,\epsilon}}\frac{\dd^2w_1}{|w_1-w_2|^2}=-\pi \ln\frac{\epsilon^2}{R^2}\,,\qquad I_2=\int_{\Omega_{R,\epsilon}}\frac{\dd^2w_1}{|w_1-w_2|^4}=0\,.
			\end{align}
			Thus, the logarithmic divergence that contributes to the renormalization of the coupling constants comes only from the last term in the OPE \eqref{eq:OOOPE}
			\begin{align}
				\left<\dots\,\left(\sum_{i=1}^{N_+}\sum_{j=1}^{N_-}\sum_{A,B}h_{A,ij}^{(0)}h_{B,ij}^{(0)}\left(-\frac{\pi}{2} \ln \frac{\epsilon^2}{R^2}\right)\int_{w_2} \hat{r}_{ab}^{A}\hat{r}_{cd}^{B}\,f^{ac}{}_{e} f^{bd}{}_{s} J_{+}^{e,i}(w_2)J_{-}^{s,j}(\bar{w}_2) \right)\right>_{0}\,,
			\end{align}
			where we introduced
			\begin{equation}
				h_{A,ij}^{(0)}=h_A^{(0)}(z_i^+-z_j^-)
			\end{equation}
			
			This indicates that in general, the model \eqref{eq:2dcft-cc} is not renormalizable at one loop for arbitrary choices of the matrices $r^{ab}$. Renormalizability requires that the set of perturbing operators be closed under the OPE, i.e., the $r$-matrix must satisfy the relation \cite{Gerganov:2000mt}
			\begin{equation}\label{reno-con}
				\hat{r}_{ab}^{A}\hat{r}_{cd}^{B}\,f^{ac}{}_{e} f^{bd}{}_{s}
				=\sum_{C}\cC^{AB}{}_C\,\hat{r}_{es}^{C}\,,
			\end{equation}
			where $\cC^{AB}{}_{C}$ are constants.
            When the matrices $\hat{r}^{A}_{ab}$ satisfy this relation, the the OPE (\ref{eq:OOOPE}) of the current-bilinear operators $\cO^{A}(w,\bar{w})$ closes within this set
			\begin{equation}\label{eq:OPE-closed}
				\cO^{A}(w,\bar{w})\,\cO^{B}(0,0)\ \supset\ \frac{\cC^{AB}{}_C}{|w|^{2}}\,\cO^{C}(0,0)\,+\cdots\,,
			\end{equation}
            and then the constants $\cC^{AB}{}_{C}$ play the role of the structure constants appearing in the OPE.
			Absorbing the logarithmic divergence into the couplings, one finds
			\begin{align}
				h_{A}(z^+_{i}-z^-_{j})
				=h_{A}^{(0)}(z^+_{i}-z^-_{j})
				+\pi \ln \frac{\epsilon}{R}\sum_{B,C}\cC_A{}^{BC}\,
				h_{B}^{(0)}(z^+_{i}-z^-_{j})\,h_{C}^{(0)}(z^+_{i}-z^-_{j})
				+\cdots\,.
			\end{align}
			which leads to the one-loop RG flow equation
			\begin{align}\label{eq:morder-RGeq}
				\frac{\dd }{\dd \tRG}\,h_{A}(z^+_{i}-z^-_{j})
				=-\frac{1}{2}\,\cC^{BC}{}_{A}\,
				h_{B}(z^+_{i}-z^-_{j})\,h_{C}(z^+_{i}-z^-_{j})\,.
			\end{align}
			where the RG time $\tRG$ is defined by 
			\begin{equation}
					\tRG=\frac{1}{4\pi}\log \mu=-2\pi\log \epsilon\,.
			\end{equation}
			
			\subsection{RG flow as linear time evolution of the defect positions}

            We examine how the positions of order defects evolve with the RG flow time under the one-loop RG flow (\ref{eq:morder-RGeq}) derived above, focusing on various explicit classical $r$-matrices. Interestingly, their RG time evolutions are found to be linear, independently of the class of the underlying integrable system, i.e.\ 
			\begin{align}\label{eq:order-RG}
				z^+_{i}-z^-_{j}=-c_{G}\,\tRG+c_{ij}\,.
			\end{align}
            Here, the limits $\tRG\to \infty$ and $\tRG\to 0$ correspond to the UV and IR regimes, respectively.
			Since $c_G$ is positive, the 2d system (\ref{eq:2dcft-cc}) reduces to a free theory in the UV limit. This behaviour can be understood as the chiral and antichiral defects separating from each other on the Riemann surface $C$ as the energy increases, and thus their mutual interaction becomes small in the UV regime.
                
            This linear dependence on the RG time was already observed in the trigonometric case for $\mathfrak{su}(2)$ in \cite{Bykov:2020nal,Bykov:2021dbk}. Below, we examine the behaviour of this RG flow in three cases: the rational integrable theories for an arbitrary semisimple Lie algebra, and the trigonometric and elliptic theories for $\mathfrak{g}=\mathfrak{su}(2)$, postponing the 
            elliptic $\mathfrak{sl}(N)$ theory for the next section.
			
			\subsubsection*{Rational theories}
			
			We begin with the rational $r$-matrices \eqref{eq:Yang} i.e.\ $r(z)=z^{-1}\,\sum_{a}T^a\otimes T_a $.
			For a given pair of chiral and antichiral defects, there is only one independent coupling constant. The coupling constant and the associated tensor basis are taken to be\footnote{Since the Killing form $\kappa_{ab}$ is negative definite for a compact, semisimple Lie algebra $\mathfrak{g}$ in our conversion, this choice makes $\hat{r}^1_{ab}$ positive definite. }
			\begin{equation}\label{eq:rational-choice}
				h_{1}(z^+_{i}-z^-_{j})=\frac{1}{z^+_i-z^-_j}\,,\qquad \hat{r}_{ab}^{1}=-\kappa_{ab}\,.
			\end{equation}
            and the only OPE coefficient is given by $\mathcal{C}=2c_G$. One-loop renormalizability (\ref{reno-con}), is then ensured by
			\begin{equation}
				\hat{r}_{ab}^{1}\hat{r}_{cd}^{1}\,f^{ac}{}_{e} f^{bd}{}_{s} =\kappa_{ab}\kappa_{cd}\,f^{ac}{}_e f^{bd}{}_s=2c_G\,\hat{r}_{es}^{C}\,.
			\end{equation}
			Which implies
			\begin{equation}\label{eq:order-RGeq}
				\frac{\dd }{\dd \tRG}h_{1}(z^+_{i}-z^-_{j})=-c_{G}\,h_{1}(z^+_{i}-z^-_{j})^2\Rightarrow \frac{\dd }{\dd \tRG}(z^+_i-z^-_j)=c_{G}\,.
			\end{equation}
			Thus, one finds that the distance between any pair of defects evolves linearly with the RG flow time.
			
			\subsubsection*{Trigonometric $\mathfrak{su}(2)$}
			The explicit expression of the trigonometric classical $r$-matrix for $\mathfrak{su}(2)$ is given by
			\begin{align}
				r(z)&=\frac{\nu}{\sinh(\nu z)}\left(T^1\otimes T^1+T^2\otimes T^2\right)+\frac{\nu}{\tanh(\nu z)}\,T^3\otimes T^3\,,\label{eq:r-tri}
			\end{align}
			where the $\mathfrak{su}(2)$ generators $T_a\,(a=1,2,3)$ are defined as
			\begin{align}
				T_a=-\frac{i}{2}\sigma^a\,,\qquad \kappa_{ab}=-\frac{1}{2}\delta_{ab}\,,\qquad c_G=2\,.
			\end{align}
			We denote the corresponding coupling constants and $r$-matrix components by
			\begin{align}
				h_1(z)=\frac{\nu}{\sinh(\nu z)}\,,\quad h_2(z)=\frac{\nu}{\tanh(\nu z)}\,,\quad \hat{r}^1_{ab}=\frac{1}{2}(\delta_{a1}\delta_{b1}+\delta_{a2}\delta_{b2})\,,\qquad 
				\hat{r}^2_{ab}=\frac{1}{2}\delta_{a3}\delta_{b3}\,,\label{eq:tri-coupling}
			\end{align}
            with the corresponding OPE coefficients $\mathcal{C}_2{}^{11}=2\mathcal{C}_1{}^{21}=4$.
			The corresponding one-loop RG flow equations are \cite{Gerganov:2000mt}
			\begin{align}
				\begin{split}\label{eq:trig-RGflow}
					\frac{\dd }{\dd \tRG}h_{1}(z^+_{i}-z^-_{j})&=-2h_1(z^+_{i}-z^-_{j})h_2(z^+_{i}-z^-_{j})\,,\\
					\frac{\dd }{\dd \tRG}h_{2}(z^+_{i}-z^-_{j})&=-2h_1(z^+_{i}-z^-_{j})^2\,.
				\end{split}
			\end{align}
			This RG flow equation has also been obtained in the trigonometric deformations of the $\mathbb{CP}^1$ sigma models and the bosonic Thirring/Gross-Neveu-type models \cite{Bykov:2020nal,Bykov:2021dbk}. In particular, it was pointed out in \cite{Bykov:2020nal,Bykov:2021dbk} that the separation between mutually ordered defects in the spectral parameter space depends linearly on the RG time.
			Indeed, assuming a linear RG-time dependence of the defect positions,
			\begin{align}
				z^+_{i}-z^-_{j}=2\tRG+c_{ij}\,,\label{eq:trigflow}
			\end{align}
			it is straightforward verify that the coupling constants (\ref{eq:tri-coupling}) satisfy the RG flow equations (\ref{eq:trig-RGflow}), with $\nu$ appearing in \eqref{eq:tri-coupling} being constant.
			
			\subsubsection*{Elliptic $\mathfrak{su}(2)$}

            Finally, we consider the elliptic case. For simplicity, we study the one-loop RG flow in the $\mathfrak{su}(2)$ case, which was studied in \cite{LeClair:2004ps}.
 The elliptic classical $r$-matrix of $\mathfrak{su}(2)$ can be expressed in terms of elliptic functions as 
			\begin{align}
				r(z)&=\frac{\nu}{\text{sc}(\nu z;m)}T^1\otimes T^1+\frac{\nu}{\text{sd}(\nu z;m)}T^2\otimes T^2+\frac{\nu}{\text{sn}(\nu z;m)}T^3\otimes T^3\,.\label{eq:r-ell}
			\end{align}
			We define the corresponding coupling constants and $r$-matrix components as 
			\begin{align}
				h_1(z)=\frac{\nu}{\text{sc}(\nu z;m)}\,,\quad h_2(z)=\frac{\nu}{\text{sd}(\nu z;m)}\,,\quad h_3(z)=\frac{\nu}{\text{sn}(\nu z;m)}\,,\quad\hat{r}^A_{ab}=\frac{1}{2}\delta_{aA}\delta_{bA}\,.\label{eq:ell-coupling}
			\end{align}
			Then, the one-loop RG flow equation is written as \cite{LeClair:2004ps}
			\begin{align}
				\frac{\dd }{\dd \tRG}h_{A}(z^+_{i}-z^-_{j})=-|\epsilon_{ABC}|\,h_B(z^+_{i}-z^-_{j})h_C(z^+_{i}-z^-_{j})\,,
			\end{align}
			where $\epsilon_{ABC}\,(A,B,C=1,2,3)$ is a totally antisymmetric tensor normalized as $\epsilon_{123}=1$.
			It is straightforward to verify that the RG flow still takes the form \eqref{eq:trigflow}, with the additional parameters $\nu$ and $m$ both remaining constant.
			Since this RG flow is described in terms of elliptic functions, the couplings become periodic as functions of the scale, and the RG flow exhibits limit cycles.
            For discussions of the physical implications of RG limit cycles, see Ref.~\cite{LeClair:2004ps} and the references therein.

\section{RG flow of elliptic \texorpdfstring{$\mathfrak{sl}(N)$}{sl(N)} order defects}\label{app:elliptic}

In the previous section, we analyzed the one-loop RG flow of current-current deformations described by the 4d CS theory coupled to multiple chiral and antichiral order defects. Based on several examples, we conjectured that the separation between each chiral defect and antichiral defect on the spectral parameter space depends linearly on the RG time. In this section, we show that this also holds for the Belavin $\mathfrak{sl}(N)$ $r$-matrix \cite{belavin_discrete_1981}. This serves as a strong check on the conjecture of linear flow.

\subsection{Belavin elliptic \texorpdfstring{$r$-matrix}{r-matrix}}

We first collect the functions and useful identities needed for the analysis. Unless otherwise stated, we follow the notation of \cite{Lacroix:2023qlz} throughout this section.

\subsubsection{Quasi-elliptic functions}\label{sec:elliptic-function}
We take $C=\mathbb{T}(2\ell_1,2\ell_2)$, where $2\ell_i$ are the periods of the torus, which are parameters of the theory. This allows us to define the elliptic Weierstrass $\wp$-function, which satisfies
\begin{equation}
    \wp(z)=\frac{1}{z^2}+O(z^2)\,,\quad \wp(z+2\ell_i)=\wp(z)\,.
\end{equation}
From this, one can define two pseudo-elliptic functions $\zeta$ and $\sigma$, which satisfy
\begin{equation}
    \frac{\dd\zeta(z)}{\dd z}=-\wp(z)\,,\quad \zeta(z)=\frac{1}{z}+O(z)\,,\quad\frac{\dd\log(\sigma(z))}{\dd z}=\zeta(z)\,,\quad \sigma(z)=z+O(z^2)\,.
\end{equation}
 They inherit pseudo-elliptic properties from $\wp$ under shifts by $2\ell_i$, given by
\begin{equation}
\zeta(z+2\boldsymbol{\ell}_i)
= \zeta(z) + 2\boldsymbol{L}_i\,,\quad 
\sigma(z+2\boldsymbol{\ell}_i)
= -\exp\!\left(2\boldsymbol{L}_i[z+\boldsymbol{\ell}_i]\right)\sigma(z)\,,\label{eq;zetasigmashift}
\end{equation}
where $\boldsymbol{L} = (\boldsymbol{L}_1,\boldsymbol{L}_2) \equiv \bigl(\zeta(\boldsymbol{\ell}_1),\zeta(\boldsymbol{\ell}_2)\bigr)$. Introducing the two-dimensional scalar-valued ``cross-product''
\begin{equation}
(a_1,a_2) \times (b_1,b_2) \equiv a_1 b_2 - b_1 a_2\,,
\end{equation}
we can define fractional shifts $q_\alpha$ and $Q_\alpha$ for $\alpha \in \mathbb{C}^2$ as
\begin{equation}
q_\alpha \equiv \frac{2}{N}\,\alpha \times \boldsymbol{\ell}\,,
\qquad
Q_\alpha \equiv \frac{2}{N}\,\alpha \times \boldsymbol{L}\,,
\end{equation}
where, $N=\text{rk}\,\mathfrak{sl}(N)+1$.
Using $q_\alpha$ and $Q_\alpha$, we can consider the following combination of $\sigma$-functions
\begin{align}
r^\alpha(z) \equiv
\exp(-Q_\alpha z)
\frac{\sigma(z+q_\alpha)}{\sigma(z)\sigma(q_\alpha)}\,,\label{eq:ralpha}
\end{align}
which is well defined for $q_\alpha \notin \Gamma$.
The functions $r^{\alpha}(z)$ are defined such that they satisfy the Fay identity
\begin{align}\label{fay-id}
r^{\alpha}(z_1)r^{\beta}(z_2)
=r^{\alpha+\beta}(z_1)r^{\beta}(z_2-z_1)
+r^{\alpha}(z_1-z_2)r^{\alpha+\beta}(z_2)\,.
\end{align}
They are not periodic under shifts by $2\ell_i$, since it is built out of the non-elliptic $\sigma$-function. Instead, (\ref{eq;zetasigmashift}) implies that
\begin{equation}
r^\alpha(z+2\boldsymbol{n}\cdot\boldsymbol{\ell})
=\xi^{\boldsymbol{n}\cdot \alpha}
\,r^\alpha(z)\,,\quad \xi=\exp\left(\frac{2\pi i}{N}\right)\,.\label{eq:ralphapseudo}
\end{equation}
\subsubsection{The Belavin basis of $\mathfrak{sl}(N)$}

Here, we introduce the Belavin basis of $\mathfrak{sl}(N)$.
To do this, we first define the following two $N \times N$ matrices $\Xi_1$ and $\Xi_2$:\footnote{Equivalently, the $N \times N$ matrix $\Xi_1$ is defined as
\begin{align}
    \left(\Xi_1\right)_{jk}=
\begin{cases}
1, & j+1\equiv k \pmod N\\
0, & \text{otherwise}
\end{cases}\,.
\end{align}
}
	\begin{equation}
		\Xi_1\equiv\begin{pmatrix}
			0&1&&&\\
			0&0&\ddots&&\\
			&\ddots&\ddots&\ddots&\\
			&&\ddots&0&1\\
			1&&&0&0
		\end{pmatrix},\qquad
		\Xi_2\equiv\begin{pmatrix}
			1&0&&&\\
			0&\xi&\ddots&&\\
			&\ddots&\ddots&\ddots&\\
			\phantom{\xi^{N-2}}&\phantom{\xi^{N-2}}&\ddots&\xi^{N-2}&0\\
			&&&0&\xi^{N-1}
		\end{pmatrix},\label{eq:defofXi}
	\end{equation}
 which satisfy the algebraic relations
\begin{align}
\Xi_1^N = \Xi_2^N = I\,,
\qquad
\Xi_1\,\Xi_2 = \xi\,\Xi_2\,\Xi_1\,.
\end{align}
In particular, both $\Xi_i$'s are cyclic of order $N$. This implies that for $\alpha = (\alpha_1,\alpha_2) \in \mathbb{A}_0 = \mathbb{Z}_N \times \mathbb{Z}_N$, the following matrices are well defined:
\begin{align}
T_\alpha \equiv \frac{1}{\sqrt{N}}\,
\Xi_1^{-\alpha_2}\Xi_2^{\alpha_1}\,,
\end{align}
where we have $T_{(0,0)}=I$.
For $\alpha\neq 0$ the matrix $T_\alpha$ is traceless.
Hence, the $N^2-1$ matrices $\{T_\alpha\}_{\alpha\in \mathbb{A}}$ with $\mathbb{A} = \mathbb{A}_0 \setminus \{(0,0)\}$ span $\mathfrak{sl}_{\mathbb{C}}(N)$, and we call this the Belavin basis. 

Using this matrix representation, one can verify that the generators
$T_{\alpha}$ of the Belavin basis have an inner product and commutation relations given by
\begin{equation}\label{eq:elliptic-inner}
\left<T_\alpha,T_\beta\right>=-\delta_{\alpha+\beta,0}\xi^{-\beta_1\beta_2}\,,\quad [T_\alpha,T_\beta]
=
f_{\alpha,\beta}T_{\alpha+\beta}\,,
\qquad
f_{\alpha,\beta}
=
\frac{1}{\sqrt{N}}
\left(
\xi^{\alpha_1\beta_2}
-
\xi^{\beta_1\alpha_2}
\right)\,.
\end{equation}
Furthermore, the structure constants satisfy the relation
\begin{align}\label{eq:blavin-dual-c}
    \sum_{\beta,\alpha \in \mathbb{A}}\xi^{\beta_1\beta_2}f_{\beta,\alpha}f_{-\beta,\beta+\alpha}=2c_G\,,
\end{align}
where $c_G=N$ is the dual Coxeter number of $\mathfrak{sl}(N)$.

\subsubsection{Belavin’s elliptic $r$-matrix}
One can then combine the functions \eqref{eq:ralpha} and the basis \eqref{eq:defofXi} to constructs Belavin’s $\mathfrak{sl}(N)$ elliptic $r$-matrix 
\begin{equation}\label{eq:Belavinr}
r(z)=\sum_{\alpha\in\mathbb{A}} r^\alpha(z)\,T^\alpha\otimes T_{\alpha}\,,
\end{equation}
where $T^\alpha$ is the dual of $T_\alpha$ with respect to the inner product \eqref{eq:elliptic-inner}.
$r(z)$ satisfies the classical Yang--Baxter equation \eqref{eq:cybe}, which follows from the commutation relates \eqref{eq:elliptic-inner} and the Fay identity \eqref{fay-id}.
\subsection{Derivation of 2d action of elliptic order defects}

We construct the action and Lax connection of an integrable field theory obtained by deforming a 2d CFT by current-current interactions characterized by the elliptic $r$-matrix \eqref{eq:Belavinr}, from the 4d CS theory coupled to order defects.
For the $\mathfrak{sl}(2)$ case, the corresponding two-dimensional action and Lax connection were constructed from the 4d CS theory in \cite{Costello:2019tri}, and its classical integrability was established. Here, we extend it to the $\mathfrak{sl}(N)$ case and present an explicit derivation, specifying in particular the boundary conditions used to construct the elliptic deformation of the PCM \cite{Lacroix:2023qlz}.

We take the spectral parameter plane to be
\begin{align}
  C = \mathbb{T}(2\ell_1,2\ell_2)
\end{align}
and choose the meromorphic one-form to be
\begin{equation}
  \omega = \dd z \, ,
\end{equation}
which is without poles on $\mathbb{T}$.
The corresponding 4d--2d action is given by
\begin{align} 
\begin{split}
  \text{S}_{\textrm{4d--2d}}[A,\{\phi_{\pm}^i\}]
  =
  \text{S}_{\text{CS}_4}[A]
  &+
  \sum_{i=1}^{N_+}
  \int_{\Sigma \times \{z_i^+\}}
  \left(
    \mathcal{L}^i_+\!\left[\phi^i_{+}\right]
    +
    \left< A_{-}, J_{+}^i \right>
  \right)
  \dd x^+ \wedge \dd x^- 
  \\
  &+
  \sum_{i=1}^{N_-}
  \int_{\Sigma \times \{z_i^-\}}
  \left(
    \mathcal{L}^{i}_-\!\left[\phi^{i}_{-}\right]
    +
    \left\langle A_{+}, J_-^{i} \right\rangle
  \right)
  \dd x^+ \wedge \dd x^-\,,
\end{split}
\end{align}
where the inner product $ \left\langle\cdot,\cdot \right\rangle$ is defined in (\ref{eq:elliptic-inner}) and the defect currents are expanded as 
\begin{align}
    J_{\pm}^i = \sum_{\alpha \in \mathbb{A}} J_{\pm}^{i,\alpha} T_\alpha= \sum_{\alpha \in \mathbb{A}} J_{\pm,\alpha}^{i} T^\alpha\,,\qquad J_{\pm,\alpha}^{i}=-J_{\pm}^{i,-\alpha}\xi^{\alpha_1\alpha_2}\,.
\end{align}

By analogy to the disorder defect case considered in \cite{Lacroix:2023qlz}, we will require that the theory transforms \textit{equivariantly} under $\mathbb{A}_0=\mathbb{Z}_N\times\mathbb{Z}_N$. An element  $\alpha=(\alpha_1,\alpha_2)\in \mathbb{A}_0$ acts on the torus $\mathbb{T}$ by the translation
\begin{align}
    z\mapsto z+2\alpha\cdot \ell\,.
\end{align}
On the other hand, \(\alpha\) also acts on the Lie algebra $\mathfrak{sl}(N)$ by the automorphism
\begin{equation}
    X \mapsto
\operatorname{Ad}_{\Xi_1^{\alpha_1}}
\operatorname{Ad}_{\Xi_2^{\alpha_2}}X\,,\qquad X\in \mathfrak{sl}(N)\,.
\end{equation}
We then require that the gauge field $A$ is equivariant with respect to the action of $\mathbb{A}_0$, which imposes
\begin{align}\label{eq:A-eq-con}
    A(z+2\ell_i)={\rm Ad}_{\Xi_i}A(z)\,,
\end{align}
such that $A$ is not an elliptic function.
We also impose the same equivalence condition on the position of the order defect in the spectral-parameter space.

We now construct the Lax pair and the corresponding 2d action.
Following the arguments of \cite{Lacroix:2023qlz}, one can find that it is still possible to impose the gauge condition
\begin{align}
    A_{\bar{z}}=0\,,
\end{align}
and we can replace $A_{\pm}$ by $L_{\pm}$.
The equations of motion are then written as
	\begin{align}\label{eq:eom-lax-el}
		\partial_{\bar{z}}L_{\pm}(z)=\pm 2\pi i\sum_{i=1}^{N_{\pm}}\delta^{(2)}(z-z_i^{\pm})J_{\pm}^i\,.
	\end{align}
To solve this differential equation, we can use the functions $r^\alpha$. In particular, one can find that the Lax connection takes the form
\begin{equation}\label{eq:elliptic-lax}
L_{\pm}
=\mp\sum_{i=1}^{N_\pm}
\sum_{\alpha \in \mathbb{A}}
J_{\pm}^{i,\alpha}\, r^\alpha(z-z_i^{\pm})\, T_\alpha\,,
\end{equation}
where the equavariance condition \eqref{eq:A-eq-con} follows from the pesudo-elliptic behaviour of $r^\alpha$ \eqref{eq:ralphapseudo}.
Finally, substituting the Lax connection (\ref{eq:elliptic-lax}) into the 4d CS action gives the effective 2d action in the Belavin basis 
\begin{align}
S
&=
\sum_{i=1}^{N_+} S_i^+
+
\sum_{j=1}^{N_-} S_j^-
-
\sum_{i=1}^{N_+}
\sum_{j=1}^{N_-}
\int_{\Sigma} \dd x^+\wedge \dd x^-
\sum_{\alpha \in \mathbb{A}}
h_\alpha(z_{ij})
J^{i}_{+,\alpha}
J^{j,\alpha}_{-}\,,
\end{align}
where we adopt the notations
\begin{align}
z_{ij}=z_i^+ - z_j^-\,,\qquad h_\alpha(z_{ij})=-r^\alpha(z_{ij})\,.
\end{align}
The sign convention in $h_\alpha(z_{ij})$ is chosen to match the rational case, where $h(z)=-1/z$. 

\subsection{One-loop RG equation}

We now study the one-loop RG flow of the 2d CFT perturbed by a current-current interaction with the elliptic $r$-matrix.
Defining the perturbing operator by
\begin{align}
\mathcal{O}^{ij}_{\alpha}(w,\bar{w})
&=-J^{i}_{+,\alpha}(w)
J^{j,\alpha}_{-}(\bar{w})\,,
\end{align}
the interaction term is
\begin{align}
S_{\mathrm{int}}
&=
\sum_{i,j}
\sum_{\alpha \in \mathbb{A}}
\int_{\Sigma} \dd^2 w\,
h_\alpha(z_{ij})
\mathcal{O}^{ij}_{\alpha}(w,\bar{w})\,.
\end{align}
As in the previous cases, we need to know the OPE of the operators $\mathcal{O}^{ij}_{\alpha}$. In the Belavin basis, this takes the form
\begin{align}
J^{i}_{+,\alpha}(w)J^{l}_{+,\beta}(0)
&\sim
\delta_{il}
\left(
\frac{k \langle T_\alpha,T_\beta\rangle}{w^2}
+
\frac{f_{\alpha,\beta} J^{i}_{+,\alpha+\beta}(0)}{w}
\right)\,.
\end{align}
On the antichiral side, we use the commutator of the dual basis elements $T^\alpha$ with the structure constants $f^{\alpha,\beta}$, and then we write
\begin{align}
J^{j,\alpha}_{-}(\bar{w})J^{m,\beta}_{-}(0)
&\sim
\delta_{jm}
\left(\frac{k \langle T^\alpha,T^\beta\rangle}{\bar{w}^2}
+
\frac{f^{\alpha,\beta}
J_{-}^{j,\alpha+\beta}(0)}{\bar{w}}
\right)\,.
\end{align}
Hence, the relevant OPE of the operators $\mathcal{O}^{ij}_{\alpha}$ is written as
\begin{align}
\mathcal{O}^{ij}_{\alpha}(w_1,\bar{w}_1)
\mathcal{O}^{lm}_{\beta}(w_2,\bar{w}_2)
\supset\
\delta_{il}\delta_{jm}
\frac{C^{\alpha+\beta}_{\alpha,\beta}}{|w_{12}|^2}
\mathcal{O}^{ij}_{\alpha+\beta}(w_2,\bar{w}_2)\,,
\end{align}
where we have defined the structure constants
\begin{align}\label{eq:structure-ell}
C^\gamma_{\alpha,\beta}
&=\delta_{\alpha+\beta,\gamma}
f_{\alpha,\beta}
f^{\alpha,\beta}\,.
\end{align}
If $\alpha+\beta=0$, the operator $\cO_0$ is not in the $\mathfrak{sl}(N)$ elements, but it does not appear in the OPE channel because $f_{\alpha,-\alpha}=0$.
Hence, the logarithmically divergent current-current term is only obtained for
\begin{align}
\alpha,\beta,\alpha+\beta\in\mathbb{A}\,.
\end{align}
Thus, we obtain the one-loop RG flow
\begin{align}\label{eq:elliptic-rg}
\frac{\dd h_\gamma}{\dd\tRG}
=\frac{1}{2}\sum_{\substack{\alpha,\beta\in\mathbb{A}\\
\alpha+\beta=\gamma}}C_{\alpha,\beta}^{\gamma}
\,h_\alpha h_\beta\,.
\end{align}
Next, we show explicitly that the one-loop RG equation for the elliptic case is equivalent to
\begin{align}
\frac{\dd\ell_i}{\dd\tRG}=0\,,\quad \frac{\dd z_{ij}}{\dd\tRG}=c_G\,,\label{eq:dldz}
\end{align}
To establish this statement, let us discuss what this implies for the functions $r^\alpha(z)$. Suppose that $z=z(\tRG)$. Then, applying the chain rule to the left-hand side of the RG flow equation (\ref{eq:elliptic-rg}), we obtain
\begin{equation}
\frac{\dd h_\gamma}{\dd\tRG}
=
-\frac{\dd}{\dd\tRG}r^\gamma(z(\tRG))
=
-\frac{\dd z}{\dd\tRG}\,\partial_zr^\gamma(z)\,.
\end{equation}
Thus, if $\dd z/\dd\tRG=c_G$, consistency with the RG flow equation requires $r^\alpha(z)$ to satisfy
\begin{align}
c_G\,\partial_zr^\gamma(z)
=-\frac{1}{2}
\sum_{\substack{\alpha,\beta\in\mathbb{A}\\
\alpha+\beta=\gamma}}
C_{\alpha,\beta}^{\gamma}
r^\alpha(z)r^\beta(z)\,.
\label{eq:delli-identity}
\end{align}
In what follows, we prove this identity.

\subsection{Proof of derivative formula for \texorpdfstring{$r^{\alpha}(z)$}{ralpha(z)}}

For this purpose, it is useful to employ the identity
\begin{align}\label{eq:fay-der-id}
r^\alpha(z)r^\beta(z)
&=-\partial_z r^\gamma(z)
+(\rho_\alpha+\rho_\beta)r^\gamma(z)\,,
\qquad \gamma=\alpha+\beta\,.
\end{align}
Here we have introduced
\begin{equation}\label{eq:rho-ref}
    \rho^\alpha=\zeta(q_{\alpha})-Q_{\alpha}\,,\quad \rho^\alpha=-\rho^{-\alpha}\,.
\end{equation}
This identity follows from the Fay identity (\ref{fay-id}), together with the expansion of $r^{\alpha}(z)$ for small $z=\varepsilon\ll 1$,
\begin{equation}
r^\alpha(\varepsilon)=
\frac{1}{\varepsilon}+\rho_{\alpha}+\cO(\varepsilon)\,.
\end{equation}
For further details, see Appendix D of \cite{Lacroix:2024wrd}.
We then multiply the identity (\ref{eq:fay-der-id}) by $C^\gamma_{\alpha,\beta}$ and sum over $\alpha,\beta$ satisfying $\alpha+\beta=\gamma$. This gives
\begin{align}\label{eq:sum-rr}
\frac{1}{2}
\sum_{\substack{\alpha,\beta\in\mathbb{A}\\
\alpha+\beta=\gamma}}
C^\gamma_{\alpha,\beta}
r^\alpha(z) r^\beta(z)
&=
\frac{1}{2}
\sum_{\substack{\alpha,\beta\in\mathbb{A}\\
\alpha+\beta=\gamma}}
C^\gamma_{\alpha,\beta}
\left[-\partial_z r^\gamma(z)
+(\rho^\alpha+\rho^\beta) r^\gamma(z)\right] \\
&=
-\frac{1}{2}
\biggl(
\sum_{\substack{\alpha,\beta\in\mathbb{A}\\
\alpha+\beta=\gamma}}
C^\gamma_{\alpha,\beta}
\biggr)
\partial_z r^\gamma(z)
+
\frac{1}{2}
r^\gamma(z)
\biggl(
\sum_{\substack{\alpha,\beta\in\mathbb{A}\\
\alpha+\beta=\gamma}}
C^\gamma_{\alpha,\beta}
(\rho_\alpha+\rho_\beta)
\biggr)\,.\no
\end{align}
We can simplify the first sum 
due to \eqref{eq:blavin-dual-c}, from where we get
\begin{equation}
    \sum_{\substack{\alpha,\beta\in\mathbb{A}\\
\alpha+\beta=\gamma}}
C^\gamma_{\alpha,\beta}
=2c_G
\end{equation}
For the second sum, we can show the identity 
\begin{align}\label{eq:c-sum-rho}
\sum_{\substack{\alpha,\beta\in\mathbb{A}\\
\alpha+\beta=\gamma}}
C^\gamma_{\alpha,\beta}
(\rho_\alpha+\rho_\beta)&=0\,.
\end{align}
To see this, we use the following identities for the structure constants:
\begin{align}\label{eq:c-rel}
    C_{\gamma-\alpha,\alpha}^{\gamma}=C_{\alpha,\gamma-\alpha}^{\gamma}\,,\qquad 
       C_{-\alpha,\gamma+\alpha}^{\gamma}=C_{\alpha,\gamma-\alpha}^{\gamma}\,.
\end{align}
These relations follow directly from the definition  (\ref{eq:structure-ell}) of the structure constants.
We first rewrite the summation in (\ref{eq:c-sum-rho}) as
\begin{align}
\sum_{\substack{\alpha,\beta\in\mathbb{A}\\
\alpha+\beta=\gamma}}
C^\gamma_{\alpha,\beta}
(\rho_\alpha+\rho_\beta)&=\sum_{\substack{\alpha\in\mathbb{A}\\
\gamma-\alpha\in\mathbb{A}}}
C^\gamma_{\alpha,\gamma-\alpha}
(\rho_\alpha+\rho_{\gamma-\alpha})
=2\sum_{\substack{\alpha\in\mathbb{A}\\
\gamma-\alpha\in\mathbb{A}}}
C^\gamma_{\alpha,\gamma-\alpha}
\rho_\alpha\,.
\end{align}
In the second equality, we used the symmetric property in (\ref{eq:c-rel}). Indeed,
\begin{align}
    \sum_{\substack{\alpha\in\mathbb{A}\\
\gamma-\delta\in\mathbb{A}}}C^\gamma_{\alpha,\gamma-\alpha}\rho_{\gamma-\alpha}=\sum_{\substack{\delta\in\mathbb{A}\\
\gamma-\delta\in\mathbb{A}}}C^\gamma_{\gamma-\delta,\delta}\rho_{\delta}=\sum_{\substack{\delta\in\mathbb{A}\\
\gamma-\delta\in\mathbb{A}}}C^\gamma_{\delta,\gamma-\delta}\rho_{\delta}\,.
\end{align}
Using the second identity in (\ref{eq:c-rel}), together with the reflection property (\ref{eq:rho-ref}) of $\rho_\alpha$, we find
\begin{align}
C^\gamma_{\alpha,\gamma-\alpha}
\rho_\alpha+C^\gamma_{-\alpha,\gamma+\alpha}
\rho_{-\alpha}=C^\gamma_{\alpha,\gamma-\alpha}
\rho_\alpha-C^\gamma_{\alpha,\gamma-\alpha}
\rho_{\alpha}=0\,.
\end{align}
Thus, the terms in the summation cancel pairwise under the involution
$\delta\mapsto -\delta$.
Therefore, we obtain
\begin{equation}
    \sum_{\substack{\delta\in\mathbb{A}\\
\gamma-\delta\in\mathbb{A}}}C^\gamma_{\delta,\gamma-\delta}\rho_{\delta}=0\,,
\end{equation}
which proves the identity (\ref{eq:c-sum-rho}).

\medskip

Combining these results with (\ref{eq:sum-rr}), we find
\begin{align}
\frac{1}{2}
\sum_{\alpha+\beta=\gamma}
C^\gamma_{\alpha,\beta}
r^\alpha(z) r^\beta(z)
&=
-\frac{1}{2}
\biggl(
\sum_{\substack{\alpha,\beta\in\mathbb{A}\\
\alpha+\beta=\gamma}}
C^\gamma_{\alpha,\beta}
\biggr)
\partial_z r^\gamma(z)
+
\frac{1}{2}
r^\gamma(z)
\biggl(
\sum_{\substack{\alpha,\beta\in\mathbb{A}\\
\alpha+\beta=\gamma}}
C^\gamma_{\alpha,\beta}
(\rho_\alpha+\rho_\beta)
\biggr)\no\\
&=-c_G\,\partial_z r_\gamma(z)\,.
\end{align}
This is precisely the identity (\ref{eq:delli-identity}). Thus, integrability of the elliptic order defect is preserved at one-loop order with the RG flow given by \eqref{eq:dldz}. By taking the limit $\ell_1\rightarrow \infty$ or $\ell_2\rightarrow \infty$, this further proves the conjectured flow of order defects for a number of trigonometric theories.

	\newpage

	\bibliographystyle{utphys}
	\bibliography{4DCS}
			
\end{document}